\documentclass[12pt,preprint]{elsarticle}
\usepackage{hyperref,lineno,natbib,amsmath,amssymb,graphicx,multirow,algorithmic,color,microtype,fullpage}
\usepackage[skip=4pt]{caption}
\usepackage[title]{appendix}
\usepackage{longtable}
\usepackage{graphicx}% http://ctan.org/pkg/graphicx
\usepackage{array}% http://ctan.org/pkg/array
\usepackage{siunitx}
\usepackage{subfigure}
\usepackage{bm}

\usepackage{tikz}

\usepackage[utf8]{inputenc}
\usepackage[T1]{fontenc}

\usepackage{amssymb, amsmath, amsfonts}
\usepackage{outlines}

\usepackage{enumitem}
\setlist[enumerate,1]{label=\textit{\Roman*}}
\setlist[enumerate,2]{label=\textit{\roman*}}
\setlist[enumerate,3]{label=\textit{\alph*}}

\journal{Computer Physics Communications}
\usepackage{xcolor}
\usepackage{listings}

\colorlet{mygray}{black!30}
\colorlet{mygreen}{green!60!blue}
\colorlet{mymauve}{red!60!blue}

\definecolor{webgreen}{rgb}{0,.35,0}
\definecolor{webbrown}{rgb}{.6,0,0}
\definecolor{RoyalBlue}{rgb}{0,0,0.9}
\definecolor{mywhite}{rgb}{1.0,1.0,1.0}

\definecolor{purp}{rgb}{0.4,0.2,0.8}

\hypersetup{
   colorlinks=true, linktocpage=true, pdfstartpage=3, pdfstartview=FitV,
   breaklinks=true, pdfpagemode=UseNone, pageanchor=true, pdfpagemode=UseOutlines,
   plainpages=false, bookmarksnumbered, bookmarksopen=true, bookmarksopenlevel=1,
   hypertexnames=true, pdfhighlight=/O,
   urlcolor=webbrown, linkcolor=RoyalBlue, citecolor=webgreen,
   pdfauthor={Jiayin Lu and Chris H. Rycroft},
   pdfsubject={Multi-threaded geometry meshing using the Delaunay Triangulation},
   pdfkeywords={Delaunay triangulation, computational geometry, multithreaded programming},
   pdfcreator={pdfLaTeX},
   pdfproducer={LaTeX with hyperref}
}

\newcommand{\R}{\mathbb{R}}

\newcommand{\co}[1]{\texttt{#1}}
\newcommand{\vpp}{\textsc{Voro++}}
\renewcommand{\vec}[1]{\mathbf{#1}}
\newcommand{\vp}{\vec{p}}
\newcommand{\vx}{\vec{x}}
\newcommand{\vy}{\vec{y}}
\newcommand{\vF}{\vec{F}}

\newcommand{\trans}{\mathsf{T}}
\DeclareMathOperator{\sdf}{sdf}
\DeclareMathOperator{\lfs}{lfs}

\newcommand{\geps}{\textit{geps}}
\newcommand{\deps}{\textit{deps}}

\definecolor{colrev}{rgb}{0.9,0,0}

\usepackage{algorithm,algorithmic}

\newlength\myindent
\makeatletter
\newenvironment{procedure}[1][htb]{%
    \renewcommand{\ALG@name}{Procedure}% Update algorithm name
   \begin{algorithm}[#1]%
  }{\end{algorithm}}
\makeatother

\newlist{steps}{enumerate}{1}
\setlist[steps, 1]{label = \textit{Step \arabic*}}

\begin{document}
\begin{frontmatter}

  \title{\textsc{TriMe++}: Multi-threaded triangular meshing in two dimensions}

\author[UCLA,UW]{Jiayin Lu}
\author[UW,LBL]{Chris H. Rycroft}
\address[UCLA]{Department of Mathematics, University of California, Los Angeles, Los Angeles, CA 90095, United States}
\address[UW]{Department of Mathematics, University of Wisconsin--Madison, Madison, WI 53706, United States}
\address[LBL]{Mathematics Group, Lawrence Berkeley Laboratory,
Berkeley, CA 94720, United States}

\begin{abstract}
  We present \textsc{TriMe++}, a multi-threaded software library designed for generating two-dimensional meshes for intricate geometric shapes using the Delaunay triangulation. Multi-threaded parallel computing is implemented throughout the meshing procedure, making it suitable for fast generation of large-scale meshes. Three iterative meshing algorithms are implemented: the DistMesh algorithm, the centroidal Voronoi diagram meshing, and a hybrid of the two. We compare the performance of the three meshing methods in \textsc{TriMe++}, and show that the hybrid method retains the advantages of the other two. The software library achieves significant parallel speedup when generating large-scale meshes containing between $10^4$ to $10^7$ points. \textsc{TriMe++} can handle complicated geometries and generates adaptive meshes of high quality.

  \section*{Program summary}

\noindent \textit{Program title:} \textsc{TriMe++} \\ \vspace{-0.3em}

\noindent \textit{Developer's repository link:} \url{https://github.com/jiayinlu19960224/TriMe} \\  \vspace{-0.3em}

\noindent \textit{Licensing provisions:} BSD 3-clause \\  \vspace{-0.3em}

\noindent \textit{Programming language:} C++ \\ \vspace{-0.3em}

\noindent \textit{External routines/libraries:} OpenMP, multi-threaded \textsc{Voro++} \\ \vspace{-0.3em}

\noindent \textit{Nature of problem:} Multi-threaded geometry meshing in two dimension using the Delaunay triangulation \\ \vspace{-0.3em}

\noindent \textit{Solution method:}
The \textsc{TriMe++} library is built around several C++ classes that follows a structured meshing pipeline. During initialization, the \texttt{shape\_2d} class reads the geometry input and generates a signed distance field using a grid-based data structure to represent the shape. The \texttt{sizing\_2d} class subsequently produces adaptive element sizing and density fields for the mesh. It uses an adaptive quad-tree data structure, enabling efficient refinement of sizing and density values in areas with complex geometries. In the meshing procedure, the \texttt{parallel\_meshing\_2d} class iteratively improves point positions in the mesh. In each meshing iteration, the multi-threaded \textsc{Voro++} library generates the Delaunay triangulation of the points. Users can select from three meshing algorithms, the DistMesh algorithm in the \texttt{mesh\_alg\_2d\_dm} class, the centroidal Voronoi diagram meshing algorithm in the \texttt{mesh\_alg\_2d\_cvd} class, and a hybrid method of the two in the \texttt{mesh\_alg\_2d\_hybrid} class Throughout this meshing workflow, we use OpenMP for multi-threaded parallel computations.

\end{abstract}
\begin{keyword}
  Delaunay triangulation \sep Voronoi tessellation \sep geometry meshing \sep multi-threaded programming \sep centroidal Voronoi tessellation \sep \vpp{} \sep DistMesh \sep computational geometry \sep geometric adaptivity
\end{keyword}
\end{frontmatter}

\section{Introduction}
\label{sec:intro}

\subsection{Delaunay triangulation and geometry meshing}
\label{sec:intro: Delaunay and meshing}
The Delaunay triangulation~\cite{delaunay1934} is a widely-used approach in computational geometry~\cite{okabe09}. For a given set of discrete points in a two-dimensional (2D) Euclidean space, the Delaunay triangulation is the unique triangulation such that no circumcircle of any triangle contains any other point. Therefore, the Delaunay triangulation preserves local proximity of points. The Delaunay triangulation can be extended to higher dimensions, for example, Delaunay tetrahedralization in three dimensions (3D)~\cite{eppstein92, shewchuk2002tet}.

\begin{figure}
  \centering
  \includegraphics[width=0.7\textwidth]{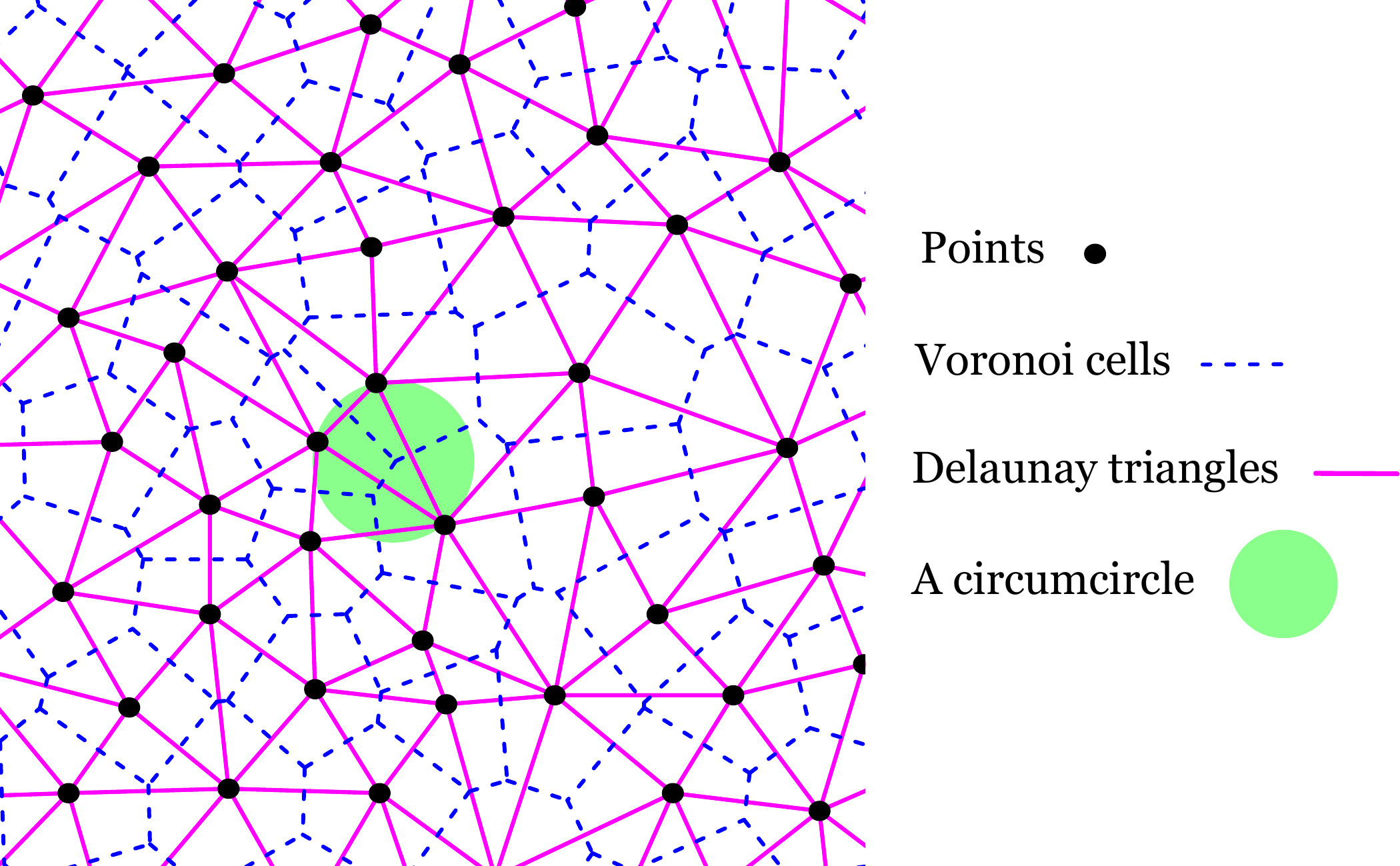}
  \caption{An example Voronoi tessellation for a set of discrete points in a two-dimensional space. The corresponding Delaunay triangulation is obtained by connecting points sharing the same Voronoi cell edges. A cricumcircle for one of the triangles is drawn in the plot. For a Delaunay triangle, no other points are in its circumcircle.\label{fig:voro_delaunay_relationship}}
\end{figure}

The Delaunay triangulation has a close relationship with the Voronoi diagram~\cite{voronoi07,okabe09}, also known as the Voronoi tessellation. For a set of discrete points in a domain, each point has a corresponding Voronoi cell, consisting of the locations in the domain that is closer to that point than any other points. In 2D Euclidean geometry, the Voronoi cells form irregular convex polyhedra that tessellate the domain. Each edge in the Voronoi tessellation is the perpendicular bisector between neighboring points. The Delaunay triangulation is the dual graph of the Voronoi tessellation. It can be obtained by first calculating the Voronoi tessellation, then connecting neighboring points sharing the same Voronoi cell edges. Figure \ref{fig:voro_delaunay_relationship} shows a Voronoi diagram for a set of points, the corresponding Delaunay triangulation, and a circumcircle for one of the triangles.

The Delaunay triangulation can be used in geometry meshing to give discrete representations of continuous shapes. It creates an unstructured mesh (i.e.~a mesh with arbitrarily varying local neighborhoods) that has the flexibility to fit arbitrarily complicated domains of geometries~\cite{eppstein92}. Geometry meshing is important in computer graphics~\cite{alliez2005, Pons07CV,Tychoniev10CV} and scientific computing for solving partial differential equations (PDEs), especially for physical simulations using the finite element method (FEM)~\cite{claes_fem, gresho2000,You2015FEM,Roarty04FEM,Shin2005FEM,lazar2022} and finite volume method (FVM)~\cite{Herbin2000,moukalled2016FVM,amaziane13FVM,ghoudi2023FVM}. In these applications, it is desirable that the mesh conforms to a given geometry shape. In addition, the mesh may have adaptive triangle element sizes, based on a density field that creates a finer mesh in regions of importance where greater resolution is required.

\begin{figure}
  \centering
  \includegraphics[width=0.3\textwidth]{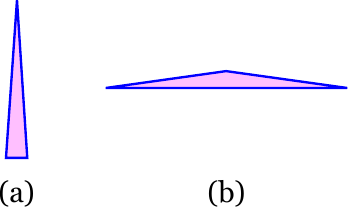}\vspace{1em}
  \caption{Two types of badly shaped tiangles in 2D: (a) a needle-type triangle, with one very small angle, and (b) a flat-type triangle with one very large angle.\label{fig:needle_flat_triangles}}
\end{figure}

Several measures of mesh quality are used, and are motivated by FEM, where the ultimate goals are speed and accuracy of the computation~\cite{eppstein92}. In some circumstances, the ideal triangle element is anisotropic (elongated and oriented in an appropriate directon)~\cite{Apel1999AnisotropicFEM,Shewchuk2002WhatIA}. The situation arises in applications where the functions being interpolated have the largest curvature in one direction far exceeding the largest curvature in another direction, or when the PDE itself has anisotropic coefficients. However, here we deal with cases in the solution isotropic space and PDE isotropic space, where equilateral triangle elements are good for interpolation and matrix conditioning. The desired triangle qualities are~\cite{eppstein92, Shewchuk2002WhatIA}:
\begin{itemize}
    \item Triangle shape: The triangles should be as close to equilateral triangles as possible. In FEM computation, the finite element formulation leads to an algebraic linear system of equations to solve, which can be written as $\vec{A} \vec{x} =\vec{b}$. The numerical condition number of the matrix $\vec{A}$ is related to the minimum angle of the mesh, where a small angle can cause poor conditioning. The interpolation error of a finite-element approximation is related to the maximum angle~\cite{Babuska1976ONTA}. In addition, the minimum height of a triangle affects the quality of a curved-surface approximation. In 2D, there are two types of badly shaped triangles: the needle type (Fig.~\ref{fig:needle_flat_triangles}(a)) and the flat type (Fig.~\ref{fig:needle_flat_triangles}(b)).
    \item Triangle size:  For an element shape in FEM, the approximation error grows with element size. We want to minimize the maximum min-containment circle of the triangles to bound the sizes of the triangles, where the min-containment circle of a triangle $t$ is the smallest circle containing $t$.
    \item Number of triangles: We want the computation to have reasonable complexity, so we cannot have too many elements in the mesh.
\end{itemize}
These qualities are often in conflict with each other. Nonetheless, if we are given a set of points and need to find a triangulation that best suit our applications, the Delaunay triangulation is a good choice for producing a unique triangulation that is optimal in several quality measures of triangulation. The Delaunay triangulation for a given set of points simultaneously optimizes the maxmin angle, minmax circumcircle, and minmax min-containment circle~\cite{eppstein92}. Its circumcircle property (no circumcircle of any triangle contains any other point) tends to favor near-equilateral triangles, as their circumcircles are more compact. The task now is to find an algorithm that generates this set of points at desired locations in the domain, so that they balance the conflicting objectives. Once we find the set of points, we can use Delaunay triangulation to generate the final geometry mesh.

\subsection{Existing Delaunay meshing algorithms and software}
\label{sec:intro: Delaunay software}
It is challenging and expensive to generate a high quality mesh on complicated shapes. There are existing popular meshing software that implement sophisticated meshing algorithms, and we discuss several examples. To obtain a quality mesh from an initial mesh, the mesh refinement algorithm by Chew~\cite{Chew93} and Rupert~\cite{Ruppert95} (pioneered in 3D by Shewchuk~\cite{Shewchuk98TetRefine}) is a popular choice. The algorithm works by adding new vertices at desired locations in the mesh and deleting unwanted vertices. In 2D, a quality mesh generator, Triangle~\cite{shewchuk96,shwechuk2002_delaunay_refinement,triangle_website}, uses this approach. It implements three efficient algorithms to generate the Delaunay triangulation for a set of points: incremental insertion algorithm~\cite{Lawson77incrementalDelaunay,green78,leonidas85incrementalDelaunay}, divide-and-conquer algorithm~\cite{leonidas85incrementalDelaunay,lee80,dwyer87divideconquerDelaunay}, and plane-sweep algorithm~\cite{fortune87}. Once an initial mesh is generated, it uses the mesh refinement algorithm to obtain a quality mesh. In 3D, a quality tetrahedral mesh generator, TetGen~\cite{Si2015TetGen,tetgen_website}, adopts this approach. Given a surface mesh, it implements the incremental insertion algorithm for Delaunay tetrahedronization. The mesh refinement algorithm is then used to obtain the quality tetrahedral mesh. Another software library, the Computational Geometry Algorithms Library (CGAL)~\cite{cgal:eb-23b,cgal_website}, also uses the mesh refinement algorithm for quality meshing of 2D and 3D shapes. It can also mesh 3D surfaces based on the work of Boissonnat and Oudot~\cite{Jean2005surfacemesh}.

The advancing front~\cite{lo85af,lohner88,lo91af,Jin93af,George1994af} approach is also a popular choice. NetGen~\cite{netgen_website,schoberl97NetGen} is a high performance FEM software with mesh generation functionality. It adopts the advancing front method to generate a mesh for 2D shapes, 3D surfaces and 3D shapes. Another software designed for FEM simulation, Gmsh~\cite{Geuzaine2009gmsh,gmsh_website}, defines shapes with parametric surfaces (e.g.~NURBS~\cite{piegl97nurbs}). It implements a modified ``metric-based'' advancing front surface meshing algorithm inspired by George and Borouchaki~\cite{george1998delaunay}, to generate surface mesh in the parametric space. Once the surface mesh is obtained, Gmsh uses TetGen and NetGen to generate the 3D tetrahedral mesh of the shape.

The above software all use quality meshing algorithms that take a local optimization approach. There are also quality meshing algorithms and software that take a global optimization approach, such as centroidal Voronoi diagram (CVD) meshing~\cite{du1999centroidal,du2003}, optimal Delaunay triangulation (ODT) meshing~\cite{Chen2004ODT,alliez2005}, and the DistMesh~\cite{Persson04, persson2005implicitgeometries} algorithm with its accompanying software.

\subsection{Contribution}
We developed a 2D multi-threaded meshing software in C++, \textsc{TriMe++}, for large-scale quality meshing with thousands of millions of points on complicated shapes. Several software packages mentioned in Sec.~\ref{sec:intro: Delaunay software} (NetGen, Gmsh and CGAL) have parallel computation implemented where applicable. In comparison, we provide an alternative multi-threaded meshing software that takes different approaches both in the choice of meshing algorithms and in the choice of Delaunay triangulation algorithm. For meshing algorithms, we implemented iterative algorithms that take a global optimization approach: DistMesh~\cite{Persson04, persson2005implicitgeometries}, CVD meshing~\cite{du1999centroidal, du2003}, and a hybrid method combining the two. We parallelize the iterative meshing algorithms used. 

For the Delaunay triangulation, we first generate the Voronoi diagram for a set of points using \vpp{}~\cite{rycroft09c, voro_website} and its recent multi-threaded extension~\cite{lu2022} for the generation of the Voronoi diagrams. From there we use the duality relationship to obtain the Delaunay triangulation. \vpp{} uses a cell-based approach~\cite{rhynsburger73,bentley80,boots83,quine84} to compute the Voronoi diagrams, which is ideal for parallelization. With multi-threaded \vpp{}, we can efficiently parallelize the generation of Delaunay triangulation from the set of points in an iteration, which is the most expensive computation in the meshing procedure.  We design the overall meshing procedure to optimize computational efficiency and parallel performance. Our software has a simple user interface, and can generate large-scale, adaptive-resolution, and high-quality meshes for complicated shapes efficiently in a reasonable timeframe.

\section{Overview}
\label{sec:key_overview}
In this section, we give an overview of the key components used in \textsc{TriMe++}: multi-threaded \vpp{}, the DistMesh algorithm, the CVD meshing algorithm, and a hybrid meshing algorithm combining the two. We also introduce the meshing pipeline used. Then in Sec.~\ref{sec: set up part} and Sec.~\ref{sec:main meshing part}, we describe the initialization and main meshing procedure of the meshing pipeline, respectively. We detail the multi-threaded implementation in Sec.~\ref{sec: parallel implementation}, and analyze potential causes of parallel inefficiency with regard to different meshing algorithms. In Sec.~\ref{sec:example code}, we show an example code that is simple and easy to use. In Sec.~\ref{sec:performance}, we provide a performance analysis, comparing the three meshing methods with respect to parallelization and mesh quality. We demonstrate that all three methods achieve high parallel efficiency with significant time speedup. We discuss some advantages and disadvantages of DistMesh and CVD, respectively. We show that in empirical testing, the hybrid method has the advantages of both while avoiding the disadvantages. Lastly, in Sec.~\ref{sec:comparison with existing software}, we compare the performance of \textsc{TriMe++} with two widely used existing software for 2D meshing, PyDistMesh~\cite{pydistmesh} and CGAL~\cite{cgal:eb-23b, cgal_website}. We show that \textsc{TriMe++} has significant advantage over PyDistMesh and CGAL in terms of computation time. A summary list of symbols used in the paper is provided in Table~\ref{tbl:meshing_parameters} in Appendix ~\ref{appendix: control parameters}.

\subsection{Multi-threaded \vpp{}}
\label{sec: intro, multi-threaded voro++}
\vpp{}~\cite{rycroft09c, voro_website} is a \textsc{C++} software library released in 2009 by Rycroft for the computation of the Voronoi tessellation in both two and three dimensions. It has been used in a wide range of scientific research projects. It has been incorporated into other research software, such as the Large Atomic/Molecular Massively Parallel Simulator (LAMMPS)~\cite{plimpton95,rycroft06a,lammps_website}, \textsc{Ovito}~\cite{stukowski10,ovito_website} for the analysis of atomistic particle systems, and \textsc{Zeo++}~\cite{willems12,pinheiro13a,zeo_website} for screening of crystalline porous materials.

\begin{figure}
  \centering
  \includegraphics[width=0.7\textwidth]{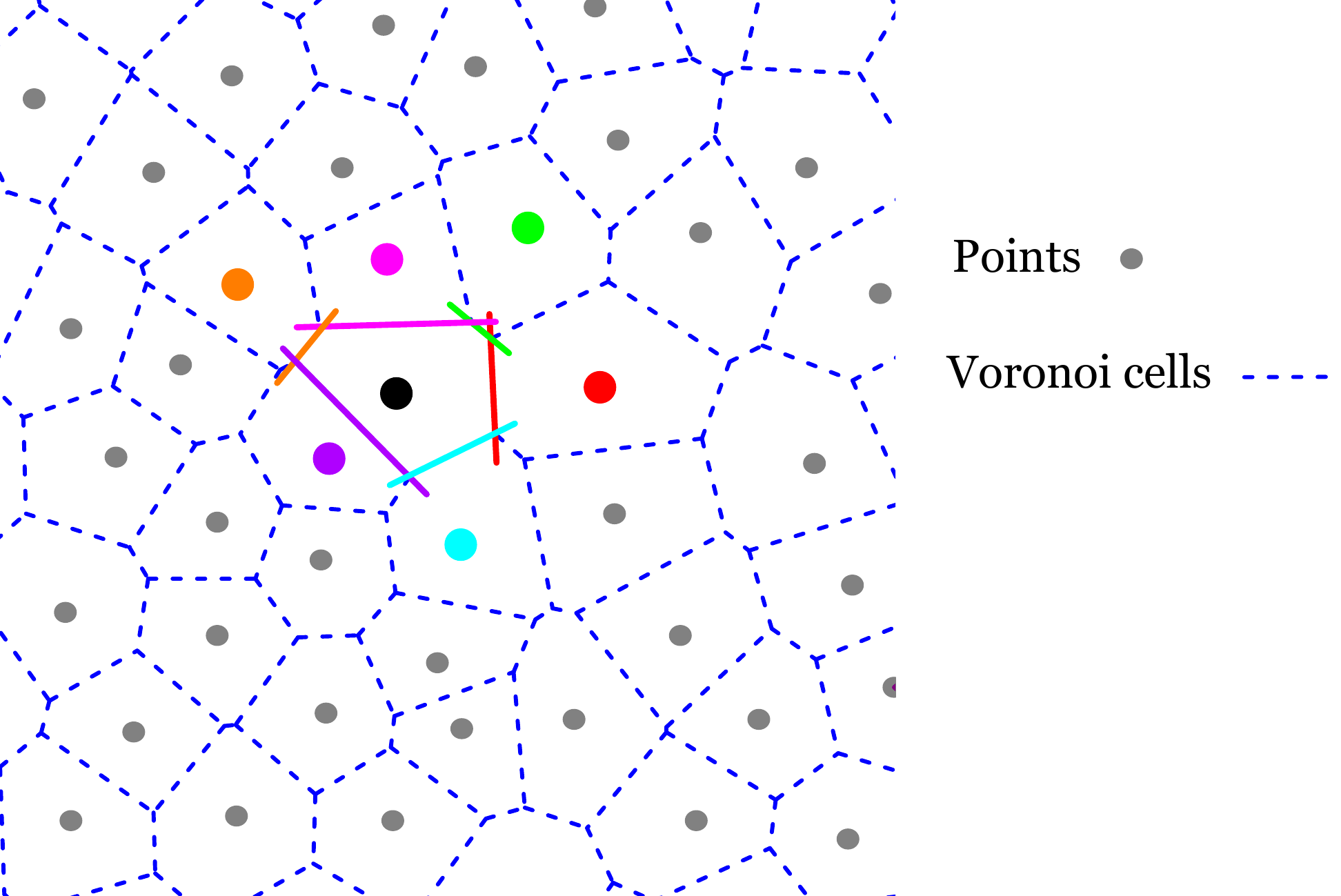}
  \caption{The cell-based approach to construct Voronoi diagram: For each point, when considering a neighboring point, we cut the point's original Voronoi cell with the associated half-space. The process is repeated until all neighboring points are found. In the end, all individual Voronoi cells are put together to obtain the Voronoi diagram.
  ~\label{fig:cell_based_approach}}
\end{figure}

\vpp{} uses a cell-based approach~\cite{rhynsburger73,bentley80,boots83,quine84}, where each Voronoi cell is computed individually as an irregular convex polygon (in 2D) or polyhedron (in 3D). As shown in Fig.~\ref{fig:cell_based_approach}, for each point, its original Voronoi cell is initialized to fill the computational domain. Each neighboring point is considered, which removes half-space from the Voronoi cell. The process is repeated, until all neighboring points are found. As discussed by Lu et al.~\cite{lu2022}, the cell-based approach has some advantages and some drawbacks. One drawback is that, because each Voronoi cell is computed independently from others, small floating point inaccuracies in one cell may lead to the formation of extra facets, leading to inconsistency in the edge and face topologies of the Voronoi cells. However, this drawback only happens rarely, typically when a vertex is equidistant from four particles. On the other hand, one significant advantage of the cell-based approach is its suitability for parallelization. Voronoi cells can be computed concurrently in parallel, as they are independent of each other, resulting in highly efficient parallel computation. Lu et al.~\cite{lu2022} introduced a multi-threaded version of \vpp{}, which distributes the Voronoi cell computations to different CPU cores. This is done using OpenMP~\cite{dagum98,openmp_website}, an application programming interface (API) for shared-memory parallel computing on multi-core computers. The multi-threaded version of \vpp{} has excellent parallel performance, achieving thread efficiencies greater than 90\% across a variety of test cases. As shown in Procedure~\ref{procedure: key meshing step: Voro, Delaunay}, we use multi-threaded \vpp{} to compute Voronoi diagrams in parallel for the meshing points, in order to obtain their Delaunay mesh using the dual graph relationship.

The two meshing methods that we employ, DistMesh and CVD, both involve iteratively updating the mesh points based on local improvements. In both methods, computing the Voronoi tessellation and/or the dual Delaunay triangulation is the most computationally expensive step. Hence, both methods can be substantially accelerated via the multi-threaded \vpp{}.

\begin{procedure}
    \caption{Voronoi tessellation and Delaunay triangulation on current points}
    \label{procedure: key meshing step: Voro, Delaunay}
    \begin{algorithmic}[1]

    \STATE Compute Voronoi diagram for points in the current ($n^{th}$) iteration, $\vec{p_n}$ \label{procedure-key meshing step-Voro}
    \STATE Obtain the Delaunay triangulation by connecting points sharing the same Voronoi cell edges \label{procedure-key meshing step-Delaunay}
    \STATE Extract valid triangles in the geometry domain, and not outside of the geometry \label{procedure-key meshing step-valid tria}

    \end{algorithmic}
\end{procedure}

\subsection{DistMesh algorithm}
\label{sec:intro-DistMesh}
\begin{figure}
  \centering
  \includegraphics[width=1\textwidth]{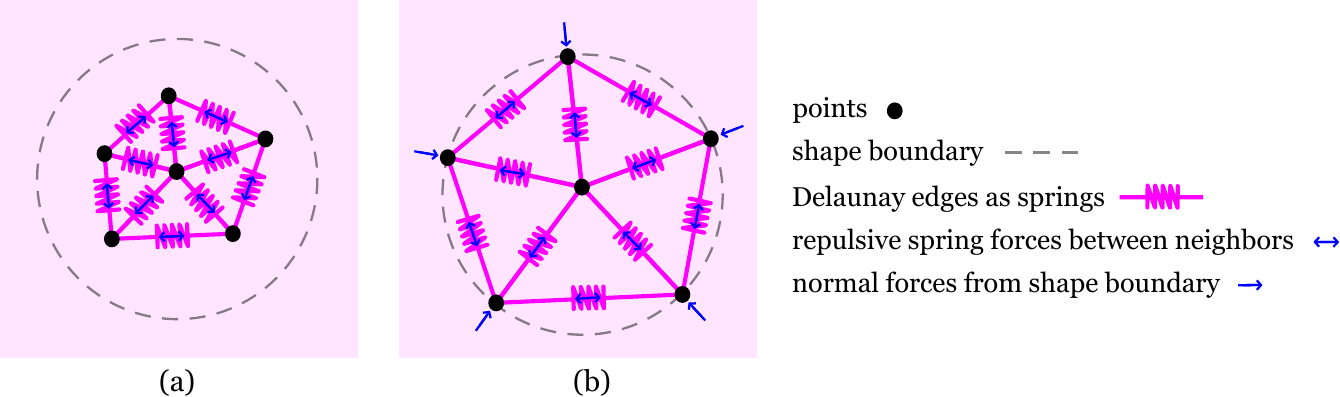}
  \caption{In DistMesh, the triangle edges that connect points are modeled as springs with repulsive forces, which push points outwards to fill out the geometry domain. A point experiences internal spring forces acting on it by its neighbors. In addition, the geometry boundaries exert normal forces pushing points inwards, so that points cannot be outside of the geometry.\label{fig:point_spring_system}}
\end{figure}

DistMesh~\cite{Persson04, persson2005implicitgeometries} is a popular meshing algorithm developed by Persson and Strang, with an accompanying software written in MATLAB. Using the physical analogy between a simplex mesh and a truss structure, the main principle is to model points connecting each other via the Delaunay triangulation as a large spring system. As shown in Fig.~\ref{fig:point_spring_system}, the springs have repulsive forces only. In this way, points are forced to keep pushing each other outwards, and fill out the geometry domain. Geometry boundaries impose constraints, where they exert inward pushing forces normal to the boundaries, so that points can only be inside or on the edges of the geometry, but cannot be outside of the geometry. The point positions are iteratively updated via force balancing, where each point moves in the net force direction of acting forces from its neighbors. When the spring system reaches an equilibrium state, the corresponding Delaunay triangulation is extracted as the final mesh.

An ordinary differential equation system is used to describe the motion of the points under the action of the springs. Consider $N$ discrete points with positions $\vp_i=(x_i,y_i)$ where $i=1,2,\ldots,N$. The net force on each point is written as $\vF(\vp_i)$ for $i=1,2,\ldots, N$. The point positions can be written in matrix form as $\vp=(\vx,\vy)$ where $\vp\in \R^{N\times 2}$ and $\vx,\vy\in \R^N$. The net forces can then be written as
\begin{equation}
  \label{eq:DistMesh-net force balancing}
  \vF(\vp)=\bigl[ \vF_{\text{int},x}(\vp) \quad \vF_{\text{int},y}(\vp)\bigr] + \bigl[ \vF_{\text{ext},x}(\vp) \quad \vF_{\text{ext},y}(\vp) \bigr],
\end{equation}
where $\vF_\text{int}(\vp)$ represents the net internal repulsive forces from the springs, and $\vF_\text{ext}(\vp)$ the net external forces from the geometry boundaries. The internal repulsive spring force for a triangle edge with length $l$ is given by
\begin{equation}
  \label{eq:DistMesh-spring force}
  f(l,l_0)=\begin{cases}
    k(l_0-l) & \text{if $l<l_0$,} \\
    0 & \text{otherwise,}
  \end{cases}
\end{equation}
where $k$ is the spring stiffness parameter, and $l_0$ is the desired edge length. The force is zero when the edge length is larger than or equal to $l_0$, and therefore the formulation only generates repulsive forces. When the edge length is much smaller than $l_0$, the repulsive force is large. When $l$ is close to $l_0$, the repulsive force becomes small. In the implementation, $l_0$ is chosen to be a factor $\textit{fac}_F=1.2$ larger than the edge length we actually desire, so that there is repulsive force when $l$ is near the desired length, to help the points spread out across the domain. This is incorporated in Eq.~\eqref{eq:DistMesh-size scaling} below, of the calculation of the desired element sizes corresponding to a mesh.

The desired edge length $l_0$ is calculated from the element size function, $\mu(\vec{x})$, which gives the relative edge length distribution over the domain, regardless of the number of points and the actual element sizes. For example, $\mu(x,y)=1+x$ for a unit square corresponds to a mesh where the edge lengths close to the left boundary ($x=0$) are about half the edge lengths close to the right boundary ($x=1$). Suppose a mesh has $M$ edges. Depending on how fine the mesh is, the actual desired edge lengths for $i=1,\dots , M$ edges are calculated via the scaling
\begin{equation}
  \label{eq:DistMesh-size scaling}
  l_{0,i}=\mu(\vec{m}_i)\cdot \textit{fac}_F\cdot\textit{fac}_{\mu}, \quad\text{where}\quad \textit{fac}_{\mu}=\left(\frac{\sum_{i=1}^{M} l_i^2}{\sum_{i=1}^{M}{\mu(\vec{m}_i)^2}}\right)^\frac{1}{2},
\end{equation}
where $\vec{m}_i$ are the edge midpoints.
We now consider $\vp$ to be a function of time $t$, and introduce the ordinary differential equation (ODE) system
\begin{equation}
  \label{eq:DistMesh-spring ODE}
  \frac{d\vec{p}}{dt}=\vec{F(p)},
\end{equation}
where $\vp(0)=\vp^0$ as the initial condition. The ODE system is integrated in time to find a stationary solution for the point positions, where $\vF(\vp)=\vec{0}$. The forward Euler method can be used to find a solution, using
\begin{equation}
  \label{eq:DistMesh-Forward Euler}
  \vec{p}^{n+1}=\vec{p}^n+\Delta t \, \vec{F}(\vec{p}^n),
\end{equation}
where $\Delta t$ is the step size and the superscript $n$ indicates the timestep index.

Algorithm~\ref{algorithm:DistMesh meshing} shows the algorithm outline for DistMesh. The point positions are iteratively improved, until they reach some termination criteria that signal the positions are close to the stationary solution. In the algorithm, the Voronoi tessellation only happens intermittently when a large point movement occurs in an iteration. This saves time, since the Voronoi tessellation is computationally expensive.

\setcounter{algorithm}{0}
\begin{algorithm}
    \caption{DistMesh meshing}
    \label{algorithm:DistMesh meshing}
    \begin{algorithmic}[1]

      \STATE Delaunay triangulation on $N$ initial points, $\vp^0$; set \texttt{re-triangulation} to \texttt{false} \label{algorithm-DistMesh-tria0}

    \WHILE {termination criteria not met} \label{algorithm-DistMesh-while}

    \IF {\texttt{re-triangulation}=\texttt{true}}
    \STATE Procedure~\ref{procedure: key meshing step: Voro, Delaunay}: Voronoi tessellation and Delaunay triangulation on current points, $\vp^n$ \label{algorithm-DistMesh-key step}
    \STATE Set \texttt{re-triangulation} to \texttt{false}
    \ENDIF

    \STATE Update point positions: \label{algorithm-DistMesh-udpate point position}
    \begin{ALC@g}
    \STATE For each point, $\vp^n_i$ for $i=1,\ldots,N$, calculate net internal force, $\vF_\text{int}(\vp^n_i)$, from spring forces imposed by its neighbors
    \STATE Update point positions pushed by their net internal forces, $\vp^{n+1}=\vp^n+\Delta t \, \vF_\text{int}(\vp^n)$
    \STATE Consider external forces from the geometry boundary constraint, and project any points lying outside of the geometry back to their closest points, $\vp^*$, on the geometry boundary
    \end{ALC@g}

    \STATE If any point in the current iteration has large point movement, set \texttt{re-triangulation} to \texttt{true}

    \ENDWHILE

    \end{algorithmic}
\end{algorithm}

The force balancing procedure to update point positions (Algorithm \ref{algorithm:DistMesh meshing} Line \ref{algorithm-DistMesh-udpate point position}) is computationally cheap, but it is less suitable for parallelization. The calculation of net internal force for a point is not independent from other points. It requires information about neighboring points and the connecting edges. In addition, the calculation of $\textit{fac}_{\mu}$ in Eq.~\eqref{eq:DistMesh-size scaling} requires calculating the sums $\sum l_i^2$ and $\sum \mu(\vec{m}_i)^2$, gathering information globally from all triangle edges in the system. Therefore, it cannot be done for each point independently in a single parallel loop over the points, resulting in some inefficiency. We must also avoid race conditions, such as when calculating the sums $\sum l_i^2$ and $\sum \mu(\vec{m}_i)^2$ in Eq.~\eqref{eq:DistMesh-size scaling}, or when updating the same point positions from the net forces of connecting edges. More detail on the parallelization is provided in Sec.~\ref{sec: parallel implementation}.

Moreover, while DistMesh generates meshes that have overall high quality, badly shaped triangles can exist in the mesh. In our empirical testing, we find that the needle-type triangles as illustrated in Fig.~\ref{fig:needle_flat_triangles}(a) are not of concern, while extremely flat triangle outliers as illustrated in Fig.~\ref{fig:needle_flat_triangles}(b) can exist near the shape boundary in an otherwise high-quality mesh. The DistMesh algorithm tends to eliminate needle-type triangles, due to the significant relative differences of the triangles' longest and shortest edges. Assuming a locally uniform element sizing field, the desired edge lengths of the local triangles should be similar; the shortest edge for a needle-type triangle would generate a large repulsive force compared to the forces generated by other local edges. Therefore, the shortest edge would push the other edges away and grow in length. In comparison, the flat type triangles do not have as large relative differences in their edge lengths. In empirical tests, a few flat outliers could exist near the boundaries in an otherwise high quality mesh.

\subsection{Centroidal Voronoi diagram and Lloyd's algorithm}
\label{sec:intro-CVD}
A centroidal Voronoi diagram~\cite{du1999centroidal} is a Voronoi diagram where the generator points coincide with the weighted centers of mass, or centroids, of the Voronoi cells. Consider points $\vp_i \in \R^2$ for $i=1,\ldots,N$, with corresponding Voronoi cells $V_i$. Given a density field $\rho(\vx)$, the weighted Voronoi cell centroid of point $i$ is
\begin{equation}
\label{eqn:voro_centroid}
\vec{c}_i=\frac{\int_{V_i} \vx \rho(\vx) d\vx}{\int_{V_i} \rho(\vx) d\vx}.
\end{equation}
The dual Delaunay triangulation of the CVD usually has high triangle quality~\cite{lazar2022,du2003}. The CVD minimizes the quadratic energy~\cite{alliez2005,du1999centroidal,Liu2009CVTsmooth}
\begin{equation}
\label{eqn:energy minimization CVD}
E_{\text{CVD}}=\sum_{i=1}^N \int_{V_i} \rho(\vec{x}) \| \vx-\vx_i \|^2 d\vec{x},
\end{equation}
which optimizes the compactness of the Voronoi cells. Compactness is defined as the ratio of the area of an object to the area of a circle with the same perimeter. A circle is the most compact 2D shape, since the boundary points are equidistant from the center, and the perimeter relative to the area is minimal. This means that for each Voronoi cell, all points on the cell boundary are as close as possible to the cell center, and the cell is as tightly packed as possible. As a result, for a CVD, points are isotropically distributed according to the density field. In 2D, a dual quality property exists~\cite{alliez2005, shwechuk2002_delaunay_refinement, eppstein2001}: a CVD mesh must result in nearly hexagonal Voronoi cells, and the dual Delaunay mesh must have nearly equilateral triangles. Therefore, in 2D, CVD meshing provides more theorectical support on the final mesh qualities compared to DistMesh. In our empirical testing, CVD meshing generates meshes with overall high qualities, and no extremely thin triangles (needle/flat type) exist in the meshes. Note that however, in 3D, this dual quality relationship does not exist; in 3D, CVD optimizes the compactness of the Voronoi cells, but not the compactness of the tetrahedrons in the Delaunay mesh, and degenerate tetrahedrons can exist~\cite{alliez2005,eppstein2001}.

\begin{figure}
  \centering
  \includegraphics[width=1.0\textwidth]{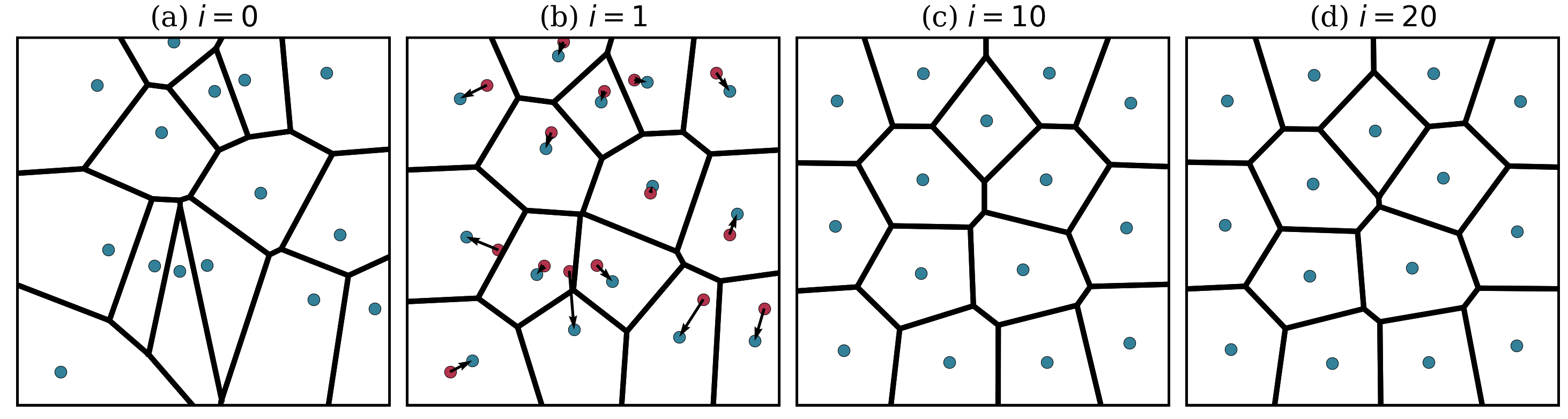}
  \caption{Lloyd's algorithm to generate a CVD: (a) Initial random points and their Voronoi diagram. (b) In an iteration $i$, points move to the locations of their Voronoi cell centroids. (c), (d) Point positions and their Voronoi diagrams after $10$ and $20$ iterations, respectively.\label{fig:lloyds_algorithm}}
\end{figure}

The CVD can be computed via Lloyd's algorithm~\cite{du1999centroidal,lloyd82}. As shown in Fig.~\ref{fig:lloyds_algorithm}, for an initial arbitrary set of points, Lloyd's algorithm iteratively moves points to their Voronoi cells' centroid positions. The step is repeated, until some termination criteria are met, to signal the points and their Voronoi diagram are close to a CVD.

Algorithm~\ref{algorithm:CVD meshing} shows our CVD meshing outline, where a geometry boundary constraint to prevent points from leaving the shape is also imposed. Compared to DistMesh, the Voronoi tessellation is required in every CVD meshing iteration. Hence, for the same number of iterations, CVD meshing is more expensive than DistMesh.

\begin{algorithm}
    \caption{CVD meshing}
    \label{algorithm:CVD meshing}
    \begin{algorithmic}[1]

    \WHILE {termination criteria not met} \label{algorithm-CVD-while}

    \STATE Procedure~\ref{procedure: key meshing step: Voro, Delaunay}: Voronoi tessellation and Delaunay triangulation on current points, $\bm{p_n}$ \label{algorithm-CVD-key step}

    \STATE Compute new point positions, $\bm{\hat{p}_{n+1}}$: \label{algorithm-CVD-compute new point positions}

    \begin{ALC@g}
    \STATE Compute the Voronoi cell centroid for each point
    \STATE Geometry boundary constraint: If the centroid lies outside of the geometry, project it back to its closest point, $\bm{p^*}$, on the geometry boundary
    \end{ALC@g}

    \STATE Update point positions, $\bm{p_{n+1}}=\bm{\hat{p}_{n+1}}$

    \ENDWHILE

    \end{algorithmic}
\end{algorithm}

In addition, in updating point positions, the Voronoi cell centroid calculation can be expensive if we have a non-constant density field $\rho(\vec{x})$. The Voronoi cell centroid calculation requires computing integrals over $V$. We can divide each Voronoi cell into a set of triangles and use an adaptive quadrature rule to compute the integrals~\cite{heath,jane2010CVTconstraints}. If $\rho(\vec{x})$ is constant, the quadrature computation is computationally cheap. We can calculate the cell area by using the generator point itself as the quadrature point, and decompose the Voronoi cell into a set of triangles by connecting the generator point to the cell vertices. However, if $\rho(\vec{x})$ varies, then during meshing, some Voronoi cells are divided into a larger number of triangles, and each triangle uses its own quadrature point(s). For example, when the local variation of $\rho(\vec{x})$ is large, to bound the centroid approximation error, we need more dividing triangles; when a point's movement is small in an iteration, a highly accurate centroid approximation is needed to update the point's position, and we need more dividing triangles. The quadrature rule to compute Voronoi cell centroids using a potentially large number of dividing triangles is more expensive than DistMesh's force balancing procedure.

Although CVD meshing is computationally more expensive than DistMesh, it is more suitable for parallelization. Since the computation of a point's Voronoi cell and its cell centroid is independent from other points, it can be distributed to threads efficiently in one single parallelized loop.

\subsection{Hybrid meshing method}
We also implement a hybrid meshing method combining DistMesh and CVD meshing. We want to achieve a low serial computation time performance like DistMesh, while having the theorectical support on triangle qualities like CVD meshing. A simple way is to use DistMesh iterations most of the time, and only switch to CVD meshing near the end of the iterations, as refinement steps for an overall high-quality mesh. In our empirical testing, the hybrid method is an effective way to avoid triangles of bad quality, and it has a serial computation time like DistMesh, much lower than that of CVD meshing.

\subsection{Meshing pipeline}
\label{sec:meshing pipeline}
Procedure~\ref{procedure: meshing pipeline} shows the meshing pipeline, which includes the initialization and the meshing procedure. In the meshing procedure, users can choose a meshing method (DistMesh, CVD, or hybrid) to iteratively improve point positions. During meshing, we use parameters to control computation efficiency and geometric adaptivity. The parameters and their recommended default values are listed in Table~\ref{tbl:meshing_parameters} in Appendix~\ref{appendix: control parameters}. We describe in detail the initialization steps in Sec.~\ref{sec: set up part}, and the meshing procedure in Sec.~\ref{sec:main meshing part}.

\setcounter{algorithm}{1}
\begin{procedure}
  \caption{Meshing pipeline} 
    \label{procedure: meshing pipeline}
    \begin{algorithmic}

    \STATE Initialization \label{procedure-set up part}

    \begin{ALC@g}
    \STATE Step 1: Geometry input and shape representation
    \STATE Step 2: Underlying geometry grid
    \STATE Step 3: Element sizing field
    \STATE Step 4: Density field
    \STATE Step 5: Point initialization
    \STATE Step 6: Point addition scheme
    \STATE Step 7: Geometric adaptivity
    \end{ALC@g}

    \STATE Meshing procedure: Iteratively improve point positions \label{procedure-Main meshing part}

    \begin{ALC@g}
    \STATE Use selected meshing algorithm: DistMesh, CVD, or the hybrid method
    \end{ALC@g}

    \end{algorithmic}
\end{procedure}

\section{Initialization}
\label{sec: set up part}

\subsection{Geometry input and shape representation}
\label{sec: Geometry input and shape representation}
The geometry is represented by a signed distance field (SDF)~\cite{Persson04, persson2005implicitgeometries}. The magnitude of the SDF represents the distance from a point to the geometry boundary. The distance is signed, with the positive and negative values representing the exterior and interior, respectively. On the geometry boundary, SDF has value $0$.

\begin{figure}
  \centering
  \includegraphics[width=1\textwidth]{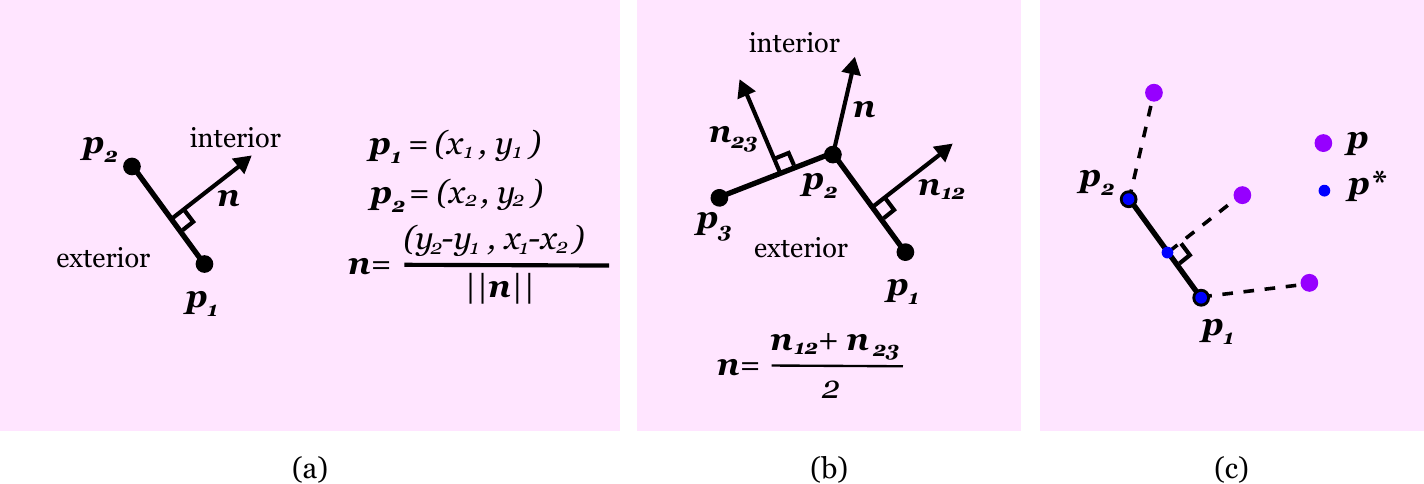}
  \caption{(a) For a geometry input line segment, $\vp_1\vp_2$, the clockwise direction is used to define its inward pointing normal of the shape, $\vec{n}$. (b) For the intersection point between two input line segments, its inward pointing normal of the shape, $\vec{n}$, is defined to be the average of the normals of its incident line segments. (c) For a point $\vp$, its closest point $\vp^*$ on a line segment $L_{12}$ between $\vp_1$ and $\vp_2$, can be either one of the endpoints, or can be a point on the line segment.~\label{fig:inward normal closest point ls}}
\end{figure}

The user may specify the SDF function directly. For instance, the function $d(x,y)=\sqrt{(x-0.5)^2+(y-0.5)^2}-r$ represents SDF of a circle with radius $r$ centered at $(0.5,0.5)$. Alternatively, users can input the geometry as line segments, and the SDF is constructed automatically in the software. The input line segments are required to be in clockwise direction and form a closed loop of the geometry. As shown in Fig.~\ref{fig:inward normal closest point ls}(a), we use the clockwise direction of $\vp_1\vp_2$, to define its inward pointing normal of the shape, $\vec{n}$. As shown in Fig.~\ref{fig:inward normal closest point ls}(b), at the intersection of two line segments, the inward pointing normal, $\vec{n}$, is the average of two connecting line segments' normals.

For a point $\vec{p}=(x,y)$ in the domain, to find its SDF magnitude, we can compute its distances to the line segments, and then take the minimum. To find the shortest distance from $\vp$ to a line segment $L_{12}$ from $\vp_1=(x_1,y_1)$ to $\vp_2=(x_2,y_2)$, we consider three possible cases as shown in Fig.~\ref{fig:inward normal closest point ls}(c). The closest point $\vp^*$ may lie at $\vp_1$, $\vp_2$, or in-between on $L_{12}$. To speed up the SDF magnitude computation, we avoid calculating the distance of $\vp$ to every line segment. We only need to check line segments close to $\vp$. A simple way is to construct a rectangular lattice grid of the domain. For each grid cell, we store all the line segments that intersect with it. When calculating the SDF magnitude of $\vp$, we first locate the grid cell $\vp$ is in. Then we loop outwards through grid cells, until finding the line segment that gives the closest distance.

\begin{figure}
  \centering
  \includegraphics[width=0.5\textwidth]{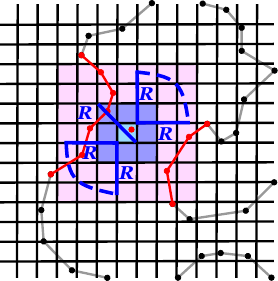}
  \caption{An efficient scheme to find closest line segment for a point. We first sort line segments in a grid, where each grid cell stores the line segments intersect with it. For a point, we locate the grid cell it lies in. We then imagine an expanding circle from the point, and loop outwards of the grids according to the expanding circle, until we find the closest line segment. With this scheme, we only need to check the (red) line segments that are close to the point, and can avoid checking the (gray) line segments that are far away from the point. ~\label{fig:bdry_pt_closest_pt_loop_outward}}
\end{figure}

The scheme is shown in Fig.~\ref{fig:bdry_pt_closest_pt_loop_outward}. Suppose each grid cell has dimensions $\Delta x$ and $\Delta y$. For the red point $\vp$ resides in the light blue grid cell, we loop grid cells outwards layer by layer, until finding a non-empty cell that has at least one line segment. We compute the distance from $\vp$ to the line segment(s). Define this dark blue layer where we first encounter line segments, $\mathcal{B}$. To avoid missing any line segments that may be closer to the point, we can define a bounding circle $\mathcal{C}$ of $\mathcal{B}$. Suppose the number of grid layers of $\mathcal{B}$ is $l$, then $\mathcal{C}$ has radius $R=\sqrt{(l\Delta x)^2+(l\Delta y)^2}$, which is the diagonal distance from the lower right corner of the grid cell the point is in, to $\mathcal{B}$'s upper left corner. Then we can define a bounding box for $\mathcal{C}$, which are the grid cells extending horizontally and vertically for a distance $R$ from the grid cell the point lies in. The bounding box corresponds to the pink region in Fig.~\ref{fig:bdry_pt_closest_pt_loop_outward}. We need to further check line segments in this region, to determine the minimum distance from $\vp$ to the line segments. All line segments outside of the pink region do not need to be considered, resulting in a substantial reduction in computation time.

Furthermore, we set the grid's dimensions to be $n_x=\lceil l_x/\bar{l}_{\text{in}}\rceil$ and $n_y=\lceil l_y/\bar{l}_{\text{in}}\rceil$, where $l_x$ and $l_y$ are the length and height of the domain, and $\bar{l}_{\text{in}}$ is the average length of the input line segments. In this way, each grid cell's edge lengths are approximately $\bar{l}_{\text{in}}$. If the grid is too fine, the procedure would loop through many empty grids before it could find the minimum distance from $\vp$ to the input line segments. If the grid is too coarse, the procedure would loop through fewer grid cells. However, for each of these cells, it would have to check many more line segments that reside in the cell.  Setting the grid's dimension in this way is a balanced choice between the two contrasting computation costs, and results in optimal timing performance in our empirical testing.

Once we know the minimum distance of $\vp$ to the geometry boundary, and the corresponding closest point $\vp^*$, the sign of the SDF is given by the sign of $(\vp -\vp^*)\cdot \vec{n}$ where $\vec{n}$ is the corresponding inward-pointing normal. Set operations (union, difference and intersection) of two geometries, $A$ and $B$, are implemented using their SDFs~\cite{Persson04, persson2005implicitgeometries}:

\begin{align*}
\textbf{Union:} \quad& \sdf_{A\cup B}(x,y)=\text{min}(\sdf_A(x,y),\sdf_{B}(x,y))\\
\textbf{Difference:} \quad& \sdf_{A\setminus B}(x,y)=\text{max}(\sdf_A(x,y),-\sdf_{B}(x,y))\\
\textbf{Intersection:} \quad& \sdf_{A\cap B}(x,y)=\text{max}(\sdf_A(x,y),\sdf_{B}(x,y))
\end{align*}

\begin{figure}
  \centering
  \includegraphics[width=1\textwidth]{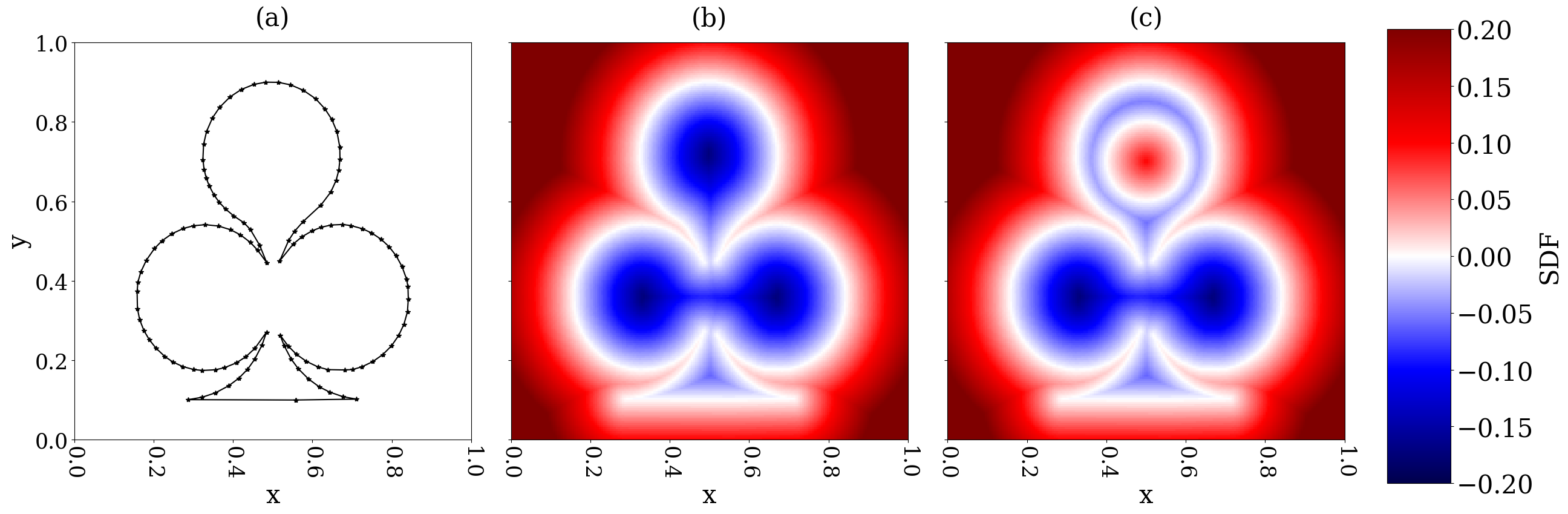}
  \caption{(a) Input geometry boundary line segments. (b) Signed distance field automatically constructed for the input geometry line segments in (a). (c) Custom shape obtained by applying a ``difference'' set operation on the input geometry shape in (a) and a circle centered at $(0.5,0.7)$ with radius $0.1$. The custom shape has challenging meshing features, such as sharp corners, concave edges, and narrow regions.\label{fig:SDF_poker_original_custom}}
\end{figure}

As an example for meshing and performance testing for the paper, we define a custom shape shown in Fig.~\ref{fig:SDF_poker_original_custom}. We use geometry contour line segment inputs in Fig.~\ref{fig:SDF_poker_original_custom}(a), to obtain the corresponding SDF in Fig.~\ref{fig:SDF_poker_original_custom}(b). We then define a circle centered at $(0.5,0.7)$ with radius $0.1$. We use the ``difference'' operation on the circle and the input shape to obtain the SDF of the custom shape in Fig.~\ref{fig:SDF_poker_original_custom}(c). This custom shape has challenging features for meshing, such as sharp corners, concave edges and narrow regions.

\subsection{Underlying geometry grid}
\label{sec: Underlying geometry grid}
An underlying geometry grid of the domain is generated based on the SDF. Each grid cell is categorized as an inner, boundary or outer cell. It is used to define geometric adaptive quantities. For example, a local geometry boundary tolerance, $\geps_i$, is defined on each boundary cell $i$. The geometry grid is also used to constrain points, so that they only lie on inner and boundary cells throughout the meshing procedure. Furthermore, the geometry grid is used to optimize computation and memory efficiency in the meshing steps.

The size of the grid depends on the number of meshing points.

Suppose we have $N_{\text{total}}$ meshing points in a rectangular domain $[a_x,b_x]\times[a_y,b_y]$, the geometry grid is set to have size $n_x^{\text{geo}}=\textit{fac}_{\text{grid}}\cdot n_x$ and $n_y^{\text{geo}}=\textit{fac}_{\text{grid}}\cdot n_y$. Here, $n_x=\lceil \lambda (b_x-a_x) \rceil$ and $n_y=\lceil \lambda (b_y-a_y) \rceil$, where $\lambda=\sqrt{N_{\text{total}}/(N_{\text{opt}}(b_x-a_x)(b_y-a_y))}$. It represents that if $N_{\text{total}}$ points are homogeneously distributed in the domain, for some grid with dimensions $n_x\times n_y$, on average $N_{\text{opt}}$ points reside in a grid cell. We then scale $n_x$ and $n_y$ by an integer factor $\textit{fac}_{\text{grid}}$ to obtain $n_x^{\text{geo}}$ and $n_y^{\text{geo}}$. If $N_{\text{opt}}$ is too small, the geometry grid is too fine, which wastes memory and increases costs to compute an unnecessary amount of geometric adaptive quantities on the grid cells. If $N_{\text{opt}}$ is too large, the geometry grid is too coarse and cannot capture local geometric adaptive features for the mesh. In our empirical testing, the default values of $N_{\text{opt}}$ and $\textit{fac}_{\text{grid}}$ shown in Table~\ref{tbl:meshing_parameters} in Appendix~\ref{appendix: control parameters} are balanced choices that give good performance.

To determine the category of each grid cell, we compute the SDF value of its midpoint $\vec{m}$, $\sdf(\vec{m})$. Grid cells within a narrow-banded distance along the geometry boundary are categorized as boundary cells. Suppose the grid cells have dimensions $\Delta x^{\text{geo}}$ and $\Delta y^{\text{geo}}$, the criteria, $\lvert \sdf(\vec{m}) \rvert \leq l_\text{diag}$, is used to define the boundary cells, where $l_\text{diag}=\sqrt{(\Delta x^{\text{geo}})^2+(\Delta y^{\text{geo}})^2}$, which is the diagonal length of a grid cell. For all other grid cells, if $\sdf(\vec{m})<0$, they are inner cells; otherwise, they are outer cells.

\begin{figure}
  \centering
  \includegraphics[width=0.6\textwidth]{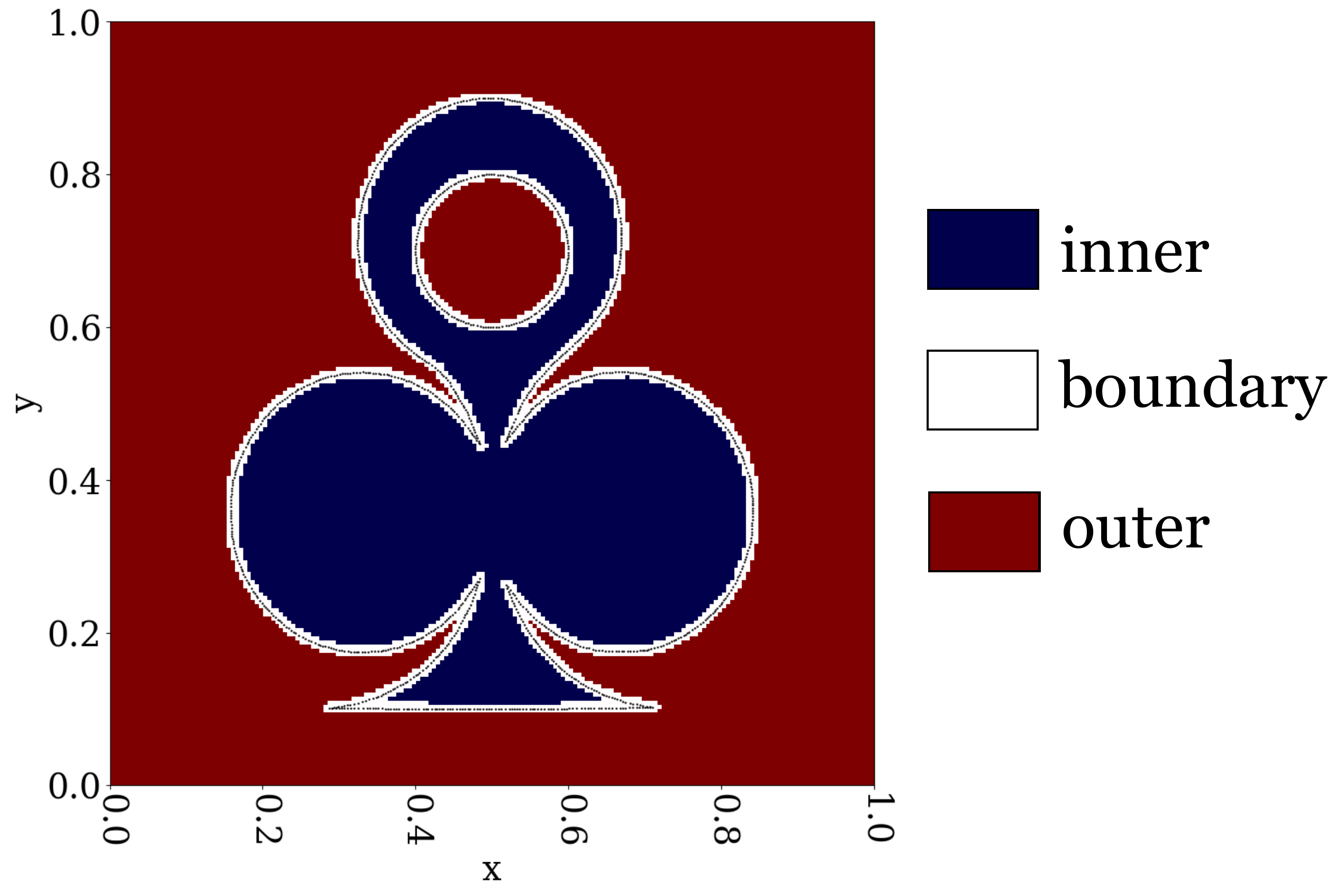}
  \caption{
  The underlying geometry grid generated for the custom shape, with $N_{\text{total}}=5,000$ points, and geometry grid dimensions $n_y^{\text{geo}}=n_x^{\text{geo}}=195$.
  ~\label{fig:geo_grid_ctgr}}
\end{figure}

Figure \ref{fig:geo_grid_ctgr} shows the custom shape and the classification of geometry grid cells into the three types. It is generated with $N_{\text{total}}=5,000$, and the geometry grid has dimensions $n_y^{\text{geo}}=n_x^{\text{geo}}=195$.

\subsection{Element sizing field}
\label{sec:element sizing field}
The element sizing field represents the desired relative edge lengths of the triangle elements. It can be a user-defined function, or automatically constructed based on the SDF, following the procedure described by Alliez et al.~\cite{alliez2005}. They use the local feature size, $\lfs:\R^2 \to \R$~\cite{amenta1998} of the geometry, which captures both local curvature and local thickness of the shape. Local curvature measures the amount the boundary curve deviates from a straight line. Local thickness is a measure of the distance of a boundary point to the opposite boundary; for example, a tree shape has a smaller local thickness for the branches and a larger local thickness for the tree trunk. Using $\lfs(\vec{x})$, the sizing field is aware of the shape's boundary curvature, and the approximation error at the boundary does not exceed the local thickness. The sizing field value at a point $\vec{x}$ in the domain, $\mu(\vec{x})$, is
\begin{equation}
\label{eqn:sizing field}
\mu(\vec{x})=\inf_{\vec{s}\in \delta \Omega}\left[K\cdot d(\vec{s},\vec{x})+\lfs(\vec{s}) \right],
\end{equation}
where $\Omega$ is the geometry domain, $\delta \Omega$ the geometry boundary, and $d(\vec{s},\vec{x})$ the distance from $\vec{x}$ to a boundary point $\vec{s}\in \delta \Omega$. $\lfs(\vec{s})$ is the local feature size at $\vec{s}$. The parameter $K$ controls mesh gradation, which determines how fast triangle elements vary in size locally. $K=0$ corresponds to a uniform sizing field. The sizing field is a maximal $K$-Lipschitz function that does not exceed $\lfs(\vec{x})$ on $\delta \Omega$~\cite{alliez2005}. It corresponds to a mesh that minimizes the number of triangles at the specified mesh gradation rate by forcing the interior triangles to be large.

To generate the sizing field defined in Eq.~\eqref{eqn:sizing field}, we first need a set of geometry boundary approximating points, $\mathcal{S}=\{\vec{s}_1,\vec{s}_2,\ldots\}$. Boundary points $\vec{s}\in \mathcal{S}$ are used in Eq.~\eqref{eqn:sizing field}; They are also used to generate the medial axis approximation points, which are then used to compute the local feature size, $\lfs(\vec{s})$ in Eq.~\eqref{eqn:sizing field}. In practice, we found the set of points $\mathcal{S}$ should be roughly evenly spaced, with some distance $d_{\vec{s}}$ proportional to the resolution of the mesh, which reduces approximation errors for the medial axis approximation points they generate. They should also capture detailed features of the shape (e.g.~sharp corners, concave edges, and narrow regions) well. The procedure for the generation of $\mathcal{S}$ is detailed in Appendix~\ref{appendix:bdry pt generation}.

Once we have the set of boundary points, $\mathcal{S}$, we follow the procedure described by Alliez et al.~\cite{alliez2005} to generate a set of medial axis approximating points. The medial axis~\cite{quadros2004} is the locus of all the centers of maximal balls included in either the shape or its complement. A simple procedure~\cite{amenta1998} can be used to obtain a discrete approximation of the medial axis. First, we compute the Voronoi diagram of the boundary points. Then, for each boundary point $\vec{s}$ and its Voronoi cell, we select the Voronoi vertex $\vec{v}_1$ farthest away from $\vec{s}$ as a medial axis approximating point. We then divide the Voronoi cell with a half space through $\vec{s}$ and normal to the vector $\vec{v}_1-\vec{s}$. We find the farthest away Voronoi vertex $\vec{v}_2$ in the half space not containing $\vec{v}_1$, and add $\vec{v}_2$ into the set of points approximating the medial axis.

\begin{figure}
  \centering
  \includegraphics[width=1\textwidth]{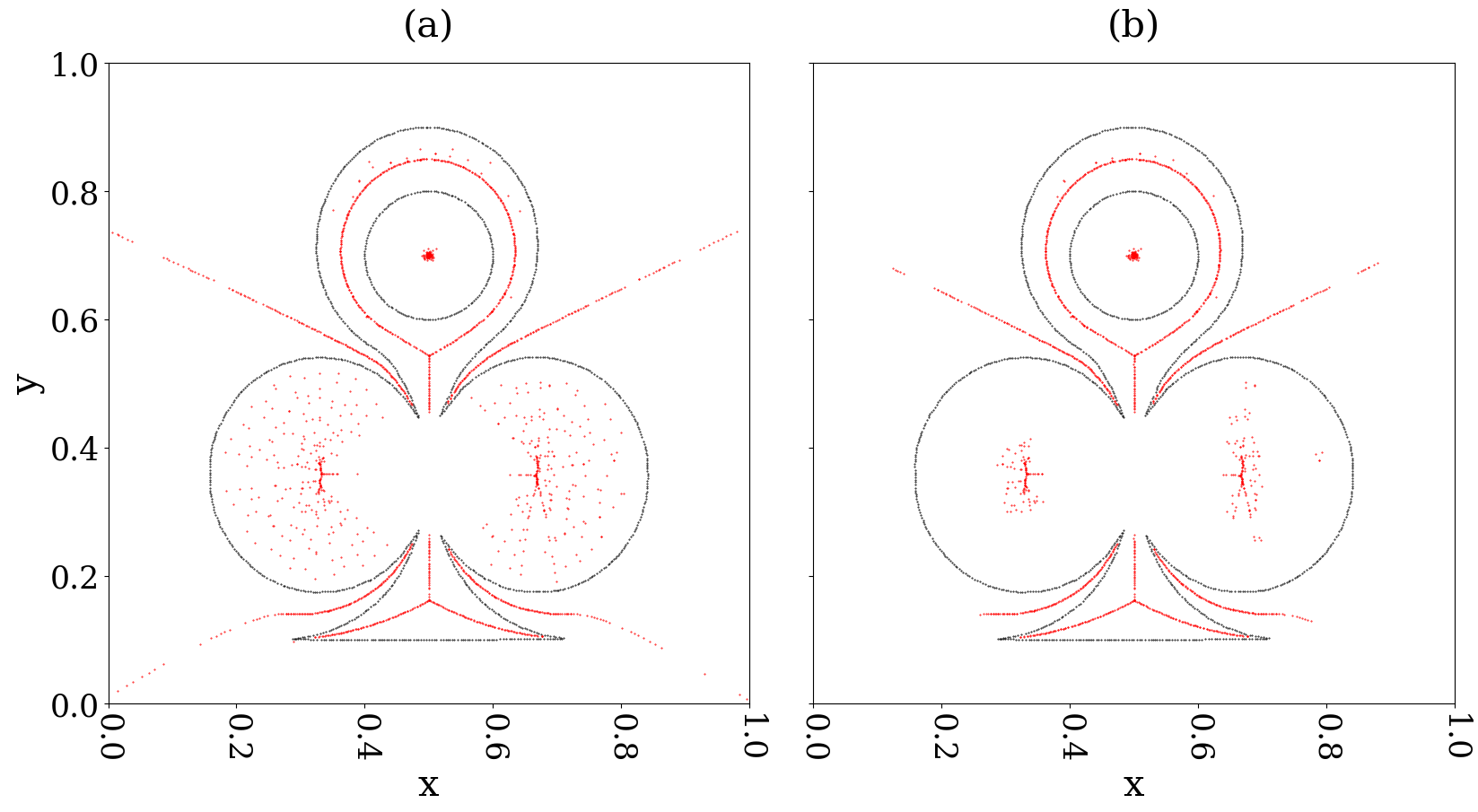}
  \caption{(a) The set of geometry boundary points, and the corresponding medial axis approximating points they generated. (b) The set of geometry boundary points, and the trimmed set of medial axis approximating points. These are the points we use to generate the element sizing field.
  ~\label{fig:bdry_medial_pts}}
\end{figure}

For the custom shape, the set of boundary points and the corresponding medial axis approximating points are shown in Fig.~\ref{fig:bdry_medial_pts}(a). Due to approximation errors, there are sparsely scattered points in the initial set of medial axis approximating points. Therefore, we further trim them by assuming that the valid medial axis approximating points should be near other points, since the medial axis should be continuous line segments in general, or a single dot like the circle case, where many approximating points overlap. Therefore, any single point in isolation to others is invalid. We sort the approximating points using the geometry grid. For a medial axis approximating point, we check its neighboring grid cells, and count the neighboring points within a certain radius, $r_{\text{nei}}$, in these grid cells. The radius $r_{\text{nei}}$ should be dependent on the spacing of the boundary points $\vec{s}\in\mathcal{S}$, since denser boundary points result in denser and closer located medial axis approximating points. If the number of neighbors within $r_{\text{nei}}$ is less than a threshold, $n_{\text{nei}}^{\text{thres}}$, we treat the point as invalid and discard it. Here $r_{\text{nei}}= \textit{fac}_{\text{nei}}\cdot n_{\text{nei}}^{\text{thres}} \cdot d_{\vec{s}}$. Figure~\ref{fig:bdry_medial_pts}(b) shows the resulting medial axis approximating points after trimming.

With the set of boundary points and trimmed medial axis approximating points, we can proceed to compute $\lfs(\vec{s})$ for each boundary point $\vec{s}\in\mathcal{S}$, as the shortest distance to a medial axis approximating point. Now we can calculate the sizing field on the geometry grid. The sizing field only needs to be defined on inner and boundary grids, since points never reside in outer grids. For a grid cell, we use its midpoint $\vec{m}$ and Eq.~\eqref{eqn:sizing field} to determine the sizing value of the cell. The values of $\lfs(\vec{s})$ are precomputed so that they can be accessed rapidly during the mesh construction. The sizing field for the custom shape is shown in Fig.~\ref{fig:sizing_density_fields}(a). It is generated with $N_{\text{total}}=5,000$ and $K=0.05$.

\subsection{Density field}
\label{sec:density_field}
\begin{figure}
  \centering
  \includegraphics[width=1\textwidth]{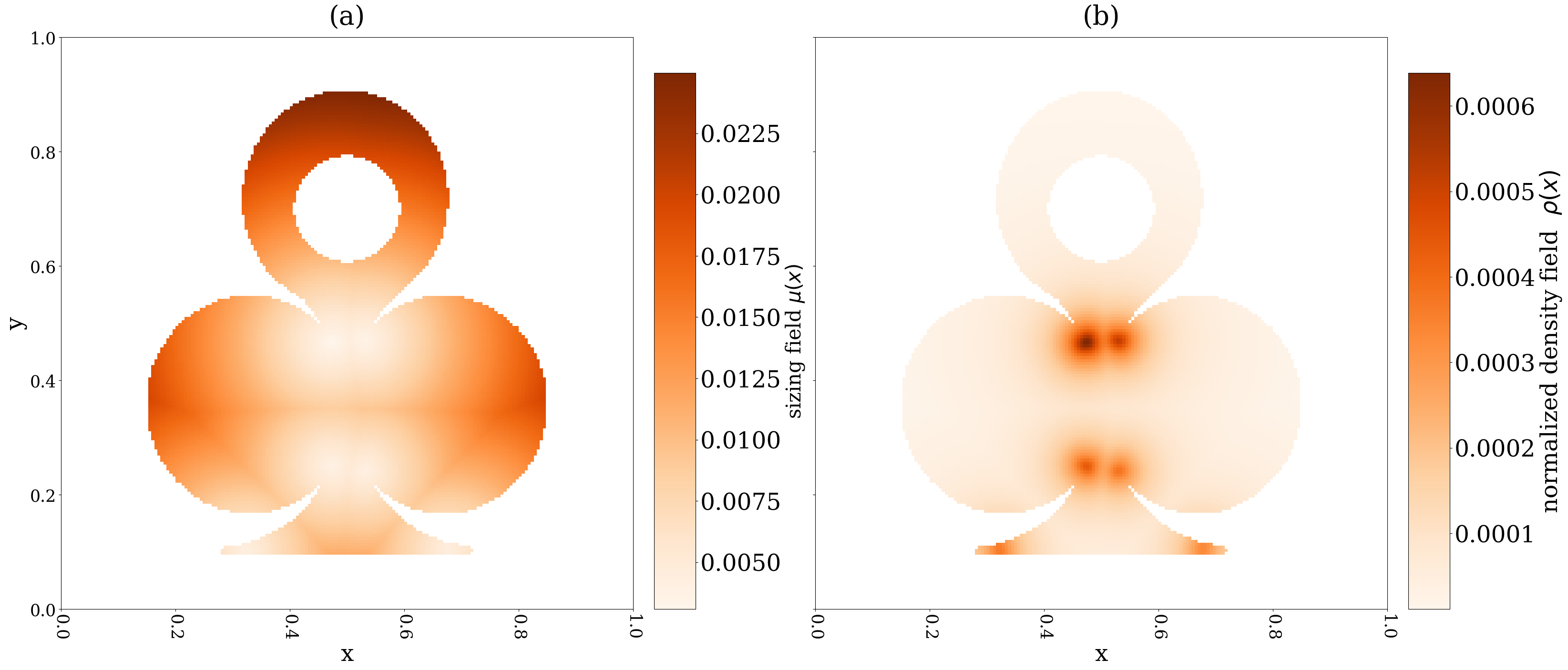}
  \caption{(a) The element sizing field for the custom shape, generated with $N_{\text{total}}=5,000$ points and $K=0.05$. (b) The corresponding density field.
  ~\label{fig:sizing_density_fields}}
\end{figure}

The density field value at a point $\vec{x}$, $\rho(\vec{x})$, can be calculated using the relationship with the sizing value~\cite{alliez2005,Persson04,persson2005implicitgeometries}, $\mu(\vec{x})$,
\begin{equation}
  \label{eq:density sizing relationship}
  \rho(\vec{x})=\frac{1}{\mu(\vec{x})^2}.
\end{equation}
Again, the geometry grid is used to precompute density field values for inner and boundary grid cells. The density field for the custom shape is shown in Fig.~\ref{fig:sizing_density_fields}(b).

Density values are not needed on outer grid cells for DistMesh, but for CVD meshing these values are required. This is because Voronoi cells for boundary points of the mesh may have some area in outer grid cells. Therefore, the outer grid cells' density information is needed to compute centroids for these Voronoi cells. In the code, upon initiation of the CVD meshing method, the same procedure described above to compute sizing and density field values is used to obtain outer grid cells' sizing and density values.

\subsection{Point initialization} \label{step:point init}
Once we have the density field, we can use it to initialize points for meshing, by ensuring that their distribution is approximately proportional to the density field. Similar to the strategy suggested by Alliez et al.~\cite{alliez2005}, we generate $N_\text{init}$ points by using the underlying geometry grid. We make sure that none of the points are outside of the geometry.

\begin{figure}
  \centering
  \includegraphics[width=0.3\textwidth]{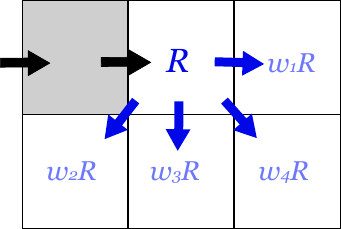}
  \caption{The Floyd--Steinberg dithering scheme to diffuse an residual amount, $R$, to nearby grid cells. We loop the grid cells in order, from top to bottom, from left to right, in serial. The residual $R$ in a cell is distributed with even weights to its valid neighboring grid cells that have not been looped through yet, in the directions indicated with the arrows in the plot. A neighboring grid cell is valid if it is either an inner cell or a selected boundary cell. Supposed there are $m_{\text{nei}}$ valid neighboring cells, then the weights are $w_i=m_{\text{nei}}^{-1}$ for $i=1,\ldots, m_{\text{nei}}$.\label{fig:residual_dithering}}
\end{figure}

We first loop through the boundary grid cells, and test whether their midpoints $\vec{m}$ satisfy $\sdf(\vec{m})\leq \eta\cdot \min(\Delta x^{\text{geo}},\Delta y^{\text{geo}})$, where $\eta$ is a user-defined scaling factor. If so, they are selected as a subset of the boundary cells that are in a narrower band along the boundary. We generate points only in the inner grid cells and the selected boundary cells. By including the selected boundary cells to generate points, as opposed to only using the inner cells as suggested by Alliez et al.~\cite{alliez2005}, we find in empirical testing that the initial points approximate the geometry boundary better. We then loop through the inner and selected boundary cells, $i$, and calculate their normalized density values, $\rho_i^{\text{norm}}$. We can use $\rho_i^{\text{norm}}$ to compute the expected number of points to generate in each of these cells, $n_i^{\text{E}}=\rho_i^{\text{norm}} \cdot N_{\text{init}}$. For a grid cell, we round the number to the nearest integer, compute the residual, $R$, and use the Floyd--Steinberg dithering scheme~\cite{floyd_steinberg1976} to diffuse $R$ to valid neighboring cells. The scheme is illustrated in Fig.~\ref{fig:residual_dithering}. The inner and selected boundary grid cells are looped in order from left to right, from top to bottom, serially. At a grid cell, we diffuse the residual $R$ to the valid neighbor cells (i.e.~inner or selected boundary grids) that have not been looped through yet. Suppose there are $m_{\text{nei}}$ valid neighbor grid cells, we use equal weighting, $w_i=m_{\text{nei}}^{-1}$ for $i=1, \ldots, m_{\text{nei}}$, in our implementation. The neighboring grids then account for the residual amount when calculating the number of points to generate.

Once we know the number of points to generate for each valid grid cell, we generate uniform random points in them, and overall, the number of points will conform to the density field. For the selected boundary grid cells, the generated points may lie outside of the geometry. To catch these cases, for each selected boundary cell $i$, we test its points' SDF against a local geometry boundary tolerance defined on the cell, $\geps_i$, whose calculation is detailed in Sec.~\ref{geometric adaptive quantities}. If a point's SDF is larger than $\geps_i$, the point lies outside of the geometry. We use the point projection procedure described in Sec.~\ref{sec: new points treatment}:~\ref{point projection method} to project outside points back onto the geometry boundary. The projected point is then kept as an initial meshing point.

The initial points are categorized as inner or boundary points. All points lying in inner grids are inner points. For a point $\vec{x}$ lying in a boundary grid $i$, if $\lvert \sdf(\vec{x})\rvert\leq\geps_i$, the point is categorized as a boundary point; otherwise, it is an inner point.

\subsection{Point addition scheme}
\label{set-up-step-point-addition}
Alliez et al.~\cite{alliez2005} suggested generating only a small fraction of the total number of points at the beginning of meshing, and adding points from time to time while meshing. The approach saves time, since Voronoi computation for a smaller number of points is cheaper. To add points, Alliez et al.~\cite{alliez2005} suggested to sort triangles in each iteration based on the ratio of circumradius divided by the local desired edge length, and add points to centroids of triangles with the larger ratios. The approach refines regions of the mesh that need points the most.
We follow this point addition scheme. Specifically, $N_{\text{current}}=N_{\text{init}}=\textit{fac}_{\text{init}}\cdot N_{\text{total}}$ points are initialized in the beginning. To monitor mesh quality, in each meshing iteration, we compute each triangle's aspect ratio, $\alpha$, and edge ratio, $\beta$,
\begin{equation}
  \label{eq:aspect, edge ratios}
  \alpha=\frac{R_{\text{circum}}}{2\cdot R_{\text{in}}}, \quad \beta=\frac{l_{\max}}{l_{\min}},
\end{equation}
where $R_{\text{circum}}$ and $R_{\text{in}}$ are the circumradius and the inradius of the triangle, respectively. $l_{\max}$ and $l_{\min}$ are the longest and shortest edge lengths of the triangle, respectively. Both $\alpha$ and $\beta$ have minimum values of $1$, corresponding to an equilateral triangle. $\alpha$ has large values for badly shaped triangles, including both needle-type and flat-type triangles shown in Fig.~\ref{fig:needle_flat_triangles}. $\beta$ has large values for needle-type triangles.

We want a measure of the overall mesh quality. Medians of the ratios are good choices, since they are not sensitive to outliers. However, computing the medians requires sorting the ratios, which is expensive and inefficient for parallelization. Therefore, we compute the generalized $\frac{1}{2}$-mean of both ratios,
\begin{equation}
  \label{eq:1/2 mean}
  M_{\frac{1}{2}}(x_1,\ldots,x_n)=\left( \frac{1}{n}\sum_{i=1}^n x_i^{\frac{1}{2}} \right)^2.
\end{equation}
We use the $\frac{1}{2}$-mean because it is less sensitive to large outliers than the arithmetic mean, and thus more suitable as an indicator for overall mesh quality.

When both ratios have relative changes in $\frac{1}{2}$-mean less than a threshold $T^{\text{add}}_{\text{quality}}$, indicating that limited improvements are being made, new points are added into the mesh. The number of new points is $N_{\text{add}}=\min(\textit{fac}_{\text{add}}N_{\text{current}}, N_{\text{total}}-N_{\text{current}})$, and $N_{\text{current}}$ is updated to $N_{\text{current}}+N_{\text{add}}$. Following the approach in Alliez et al.~\cite{alliez2005}, we sort triangles in the current iteration based on the ratio of circumradius divided by the local desired edge length, $h_i$, whose calculation is detailed in Sec.~\ref{geometric adaptive quantities}. Points are added to centroids of triangles with the larger ratios. This point addition scheme is repeated until $N_{\text{current}}=N_{\text{total}}$.

\begin{figure}
  \centering
  \includegraphics[width=1\textwidth]{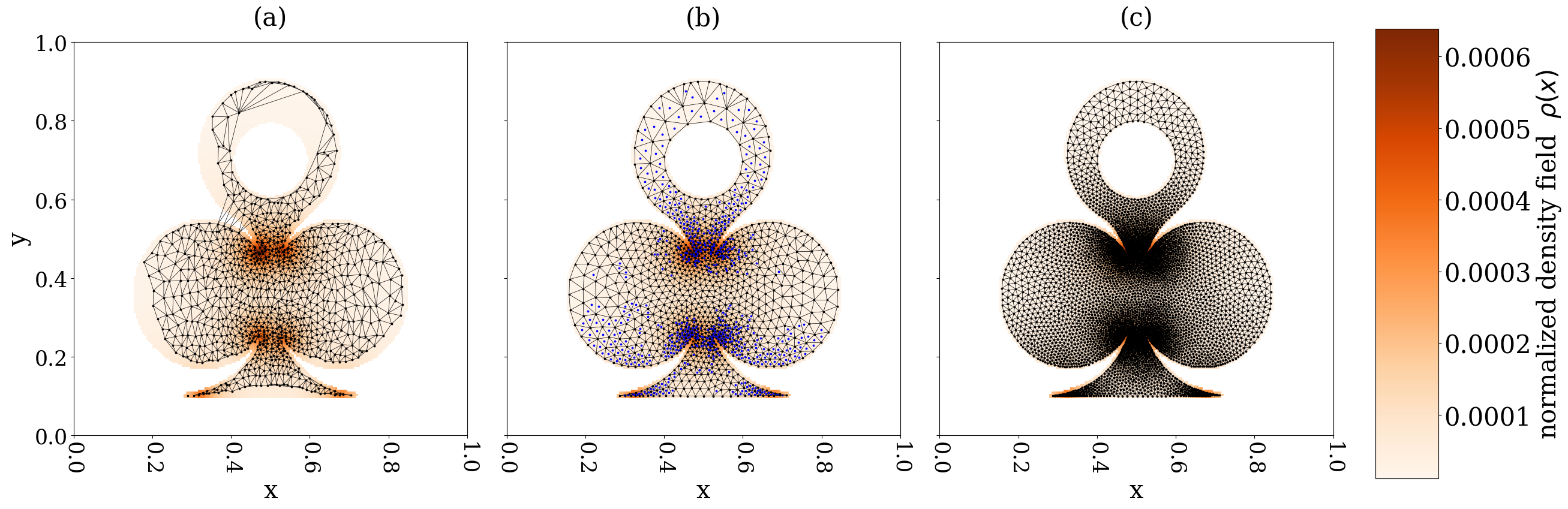}
  \caption{(a) An example point initialization according to the density field, with $N_{\text{init}}=\textit{fac}_{\text{init}}\times N_{\text{total}}=0.2\times 5,000=1,000$ initial points, and the corresponding triangle mesh. (b) New points are added into the mesh, when the current mesh reaches a good quality and not many quality improvements can be made any more. $N_{\text{add}}=\textit{fac}_{\text{add}}N_{\text{current}}=0.6\times 1,000=600$ points are added to the centroids of triangles with large ratios of circumradius divided by their local desired edge length, $h_i$. (c) The $N_{\text{total}}$ points at termination and the corresponding high quality mesh we obtain.\label{fig:pt_addition}}
\end{figure}

An example of the point initialization and addition procedure is shown in Fig.~\ref{fig:pt_addition} on the custom shape. In this example, we use CVD meshing. More details of the meshing process can be found in Sec.~\ref{sec:main meshing part}. With $N_{\text{total}}=5,000$, Fig.~\ref{fig:pt_addition}(a) shows the initial points generated according to the density field, and the corresponding initial mesh. Figure~\ref{fig:pt_addition}(b) shows that when the initial mesh reaches a good quality, new points are added to refine mesh regions that needed them most. The process is repeated, until $N_{\text{current}}=N_{\text{total}}$. Figure~\ref{fig:pt_addition}(c) shows the final set of points and a good quality mesh obtained at termination.

\subsection{Geometric adaptivity} \label{geometric adaptive quantities}
Geometric adaptive quantities are defined on the geometry grid. Some of these quantities were used in the original DistMesh implementation. However, they were defined as fixed quantities. We make improvements by defining local versions of the quantities on the geometry grid, and thus making them adaptive to the geometry shape, the density field, and the mesh resolution.
\begin{itemize}
    \item $h_i$: Upon point initialization in Sec.~\ref{step:point init}, each inner and boundary grid cell $i$ computes and stores a local desired edge length, $h_i$. $h_i$ is computed based on the normalized density $\rho^{\text{norm}}_i$ and the current number of points in the mesh, $N_{\text{current}}$. As discussed in Sec.~\ref{step:point init}, we obtain the expected number of points in an inner or selected boundary grid by $n^E_i=\rho^{\text{norm}}_i \cdot N_{\text{current}}$. For these grids, $h_i=\sqrt{(\Delta x^{\text{geo}}\cdot \Delta y^{\text{geo}})/ n^E_i}$, where $(\Delta x^{\text{geo}}\cdot \Delta y^{\text{geo}})/ n^E_i$ is an approximate of each triangle's area in the grid, and the square root approximates the edge length. For boundary grids that are not selected in Sec.~\ref{step:point init}, the density values are normalized by the same factor (i.e.~the sum of densities of inner and selected boundary grids), so that the relative densities on these grids are maintained locally with the surrounding inner and boundary grids. $h_i$ for these boundary grids are then calculated in the same way using their $\rho^{\text{norm}}_i$.
\end{itemize}
Once we have the local desired edge length, $h_i$, for each inner and boundary grid, we can use it to define other geometric adaptive quantities on the grid. On each inner and boundary grid cell, $i$:
\begin{itemize}
    \item $T^{\text{retria}}_{\text{mvmt},i}$: This is a threshold defined when using the DistMesh algorithm. In an iteration, after points' positions are updated via force balancing, each point's movement since the previous triangulation is tested against this local threshold. If any point's movement is larger than its local threshold (i.e.~the point has large point movement), re-triangulation is required in the next iteration. In the implementation, $T^{\text{retria}}_{\text{mvmt},i}= \textit{fac}^{\text{retria}}_{\text{mvmt}} \cdot h_i$. This adaptive movement threshold is smaller in finer regions of the mesh, and larger in coarser regions.
    \item $T^{\text{end}}_{\text{mvmt},i}$: This is a threshold to determine small movements of inner points for termination. In an iteration, after point positions are updated, we test inner points' movements in the current iteration against this threshold. If all inner points' movements are smaller than their local thresholds, meshing is terminated. In the implementation, $T^{\text{end}}_{\text{mvmt},i}= \textit{fac}^{\text{end}}_{\text{mvmt}} \cdot h_i$.
    \item $T^{\text{pt}}_{\text{mvmt},i}$: This is a threshold to bound point movements. In an iteration, when updating point positions, we require that points cannot move for a distance more than their local threshold. If a point's movement distance is more than the threshold, it is shortened to the threshold value. In the implementation, $T^{\text{pt}}_{\text{mvmt},i}=\textit{fac}^{\text{pt}}_{\text{mvmt}} \cdot h_i$.
\end{itemize}
On every boundary grid cell, $i$:
\begin{itemize}
    \item $\geps_i$: This is the local geometry boundary tolerance. If a point $\vec{x}$ on a boundary grid cell, $i$, has $\lvert \sdf(\vec{x})\rvert \leq \geps_i$, it is categorized as a boundary point. The quantity is only defined for boundary grids, since we only need to test if a point lies inside, outside, or on the boundary if it is in a boundary grid. In the implementation, $\geps_i=\textit{fac}_{\text{geps}} \cdot \min(h_i, h_{\text{avg}})$, where $h_{\text{avg}}$ is the average $h_i$ of inner and selected boundary grids. We use a minimum of $h_i$ and $h_{\text{avg}}$ here to conservatively define $\geps_i$, to avoid large values for grids that have low density values (i.e.~large element sizes).
    \item $\deps_i$: This is the step size used in finite difference when approximating derivatives in the point projection step, which projects an outside point back onto the geometry boundary and is detailed in Sec.~\ref{sec: new points treatment}:~\ref{point projection method}. The quantity is only defined for boundary grid cells, because we restrict points to only lie in inner or boundary cells, and a point can only lie outside of the geometry if it is in a boundary cell. Therefore, point projection only happens in boundary cells. In the implementation, $\deps_i= \sqrt{\epsilon} \cdot  \min(h_i,h_{\text{avg}})$, where $\epsilon$ is the machine precision.
\end{itemize}
All of the above quantities are updated accordingly based on resolution of the mesh. When $N_{\text{add}}$ new points are added into the mesh, we update $N_{\text{old}}=N_{\text{current}}$, $N_{\text{current}}=N_{\text{current}}+N_{\text{add}}$, the quantities are scaled by a factor $\sqrt{N_{\text{old}}/N_{\text{current}}}$.

\begin{figure}
  \centering
  \includegraphics[width=0.3\textwidth]{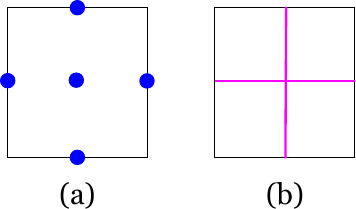}
  \caption{(a) For an ADF cell, we use bilinear interpolation to calculate the ADF values on the five points: the four midpoints of the edges, and the cell centroid. We compare their ADF values with their SDF values calculated from the geometry shape input. (b) If the maximum error of the ADF values of the five points in (a) exceeds $E_{\text{tol},i}^{\text{adf}}$, the ADF cell is subdivided into a quadtree with four children cells, as long as the maximum depth allowed, $T_{\text{depth}}^{\text{adf}}$, is not reached.
  ~\label{fig:quadtree_point_check}}
\end{figure}

Another quantity, the adaptive signed distance field (ADF)~\cite{frisken2000}, is also defined on boundary grid cells.
\begin{itemize}
  \item $\textit{adf}_i$: An adaptive signed distance field, $\textit{adf}_i$, is constructed using quadtree, for each boundary grid cell, $i$. We use the top-down approach~\cite{frisken2000} to construct the quadtree. In the top-down approach, we define an error tolerance, $E_{\text{tol},i}^{\text{adf}}$, and the maximum depth allowed for the quadtree, $T_{\text{depth}}^{\text{adf}}$. We start from the coarsest cell, which is the whole grid cell. We subdivide it into a quadtree with four children cells, and we keep subdividing cells in the quadtree recursively until either the error tolerance or the maximum depth of the tree is reached. The subdivision criteria for a quadtree cell is as follows: For each cell, the corner points have ADF values exactly equal to their SDF computed from the shape input. We then compute the ADF values of the four mid-edge points and the centroid point, as shown in Fig.~\ref{fig:quadtree_point_check}(a), using bilinear interpolation on the element. We also compute the SDF values for these five points using the shape input. We calculate the errors of the points' ADF against their SDF, and if the maximum error exceeds the error tolerance, the cell is subdivided into four children cells, as shown in Fig.~\ref{fig:quadtree_point_check}(b), as long as the maximum depth of the quadtree is not reached.

\end{itemize}

In the implementation, we use a fixed maximum depth threshold $T_{\text{depth}}^{\text{adf}}$ for all ADF cells. Furthermore, we define adaptive local error tolerance for an ADF cell, $E_{\text{tol},i}^{\text{adf}}$, to be proportional to the local geometry boundary tolerance, $\geps_i$. It is calculated as $E_{\text{tol},i}^{\text{adf}}=\textit{fac}_{E_{\text{tol}}}^{\text{adf}} \cdot \geps_i \cdot \textit{fac}_{\text{pt}}^{\text{adf}}$, where $\textit{fac}_{\text{pt}}^{\text{adf}}=\sqrt{N_{\text{current}}/N_{\text{total}}}$. Here, a small factor $\textit{fac}_{E_{\text{tol}}}^{\text{adf}}$ is used to ensure that the ADF error is much smaller than the geometry boundary tolerance, so it can be used effectively to distinguish whether a point is inside/outside/on the boundary. A $\textit{fac}_{\text{pt}}^{\text{adf}}$ is used, so that upon point initialization with $N_{\text{current}}=N_{\text{init}}\leq N_{\text{total}}$, the ADF cells are already constructed with small error tolerances, as if the mesh is of full resolution with $N_{\text{total}}$ points. In this way, we only need to construct the ADF cells once at the beginning of meshing, without updating them when new points are added, and we can just use them throughout the meshing procedure.

\begin{figure}
  \centering
  \includegraphics[width=0.6\textwidth]{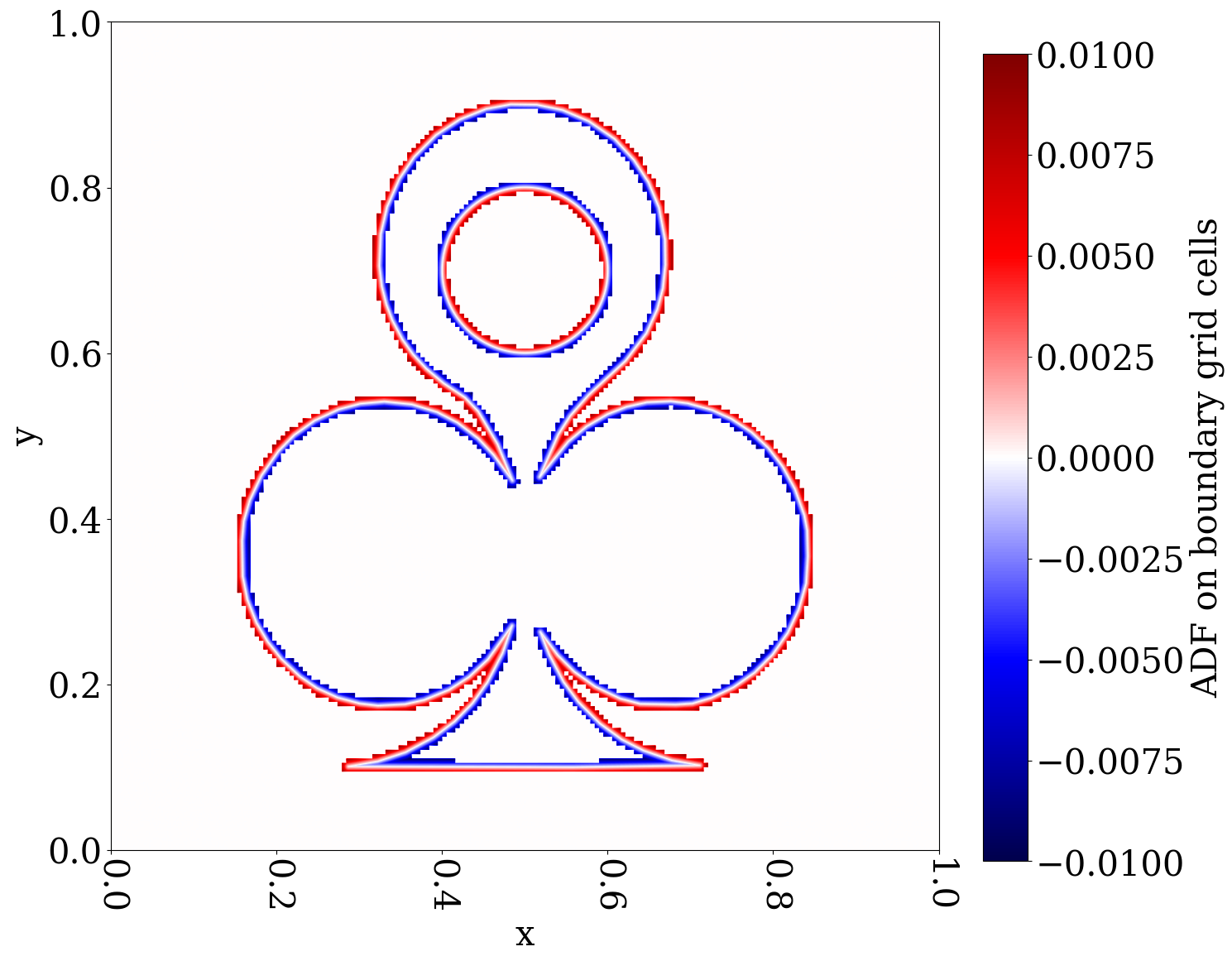}
  \caption{The ADF cells defined on the boundary grids for the custom shape.
  ~\label{fig:adf_bdry_cells}}
\end{figure}

The construction of ADF cells is an important feature that is computationally efficient. Firstly, they are only constructed on the boundary grids, which are much less in number compared to the size of the full geometry grid. They are used for the purpose of deciding whether a point lying in the boundary grid is inside/outside/on the geometry boundary. Since all points in inner grids are inner points, and no points ever lie in outer grids, there is no need to construct ADF cells on inner and outer grids. Secondly, finding the signed distance value using bilinear interpolation in an ADF cell is computationally cheap. Therefore, the ADF cells can be used to quickly look up accurate signed distance values for points in the boundary grids. This avoids potentially expensive computation of the SDF from the geometry shape input or SDF input function. For the custom shape, Fig.~\ref{fig:adf_bdry_cells} shows the ADF cells generated on boundary grids.

\section{Meshing}
\label{sec:main meshing part}
After the set-up part, we enter the main meshing part, where point positions of the mesh are iteratively improved until some termination criteria are met. One set of termination criteria we use is with respect to mesh qualities. For meshing to terminate, the relative changes in $\frac{1}{2}$-means of $\alpha$ and $\beta$ in Eq.~\eqref{eq:aspect, edge ratios} are both required to be smaller than a threshold $T^{\text{end}}_{\text{quality}}$. This indicates that not many improvements can be made on the overall mesh quality any more. We also require the relative change in maximum aspect ratio to be smaller than a threshold $T^{\text{end}}_{\alpha_{\text{max}}}$, indicating the maximum aspect ratio is stabilized, and not many more improvements can be made to the worst triangle anymore. If these criteria are satisfied, meshing terminates.

Another set of termination criteria we use is with respect to small point movements. In an iteration, if all inner points' movements are smaller than their local $T^{\text{end}}_{\text{mvmt},i}$, meshing terminates. This is a strict termination criterion, and usually the mesh quality termination criterio are reached before it.

\subsection{Procedure~\ref{procedure: key meshing step: Voro, Delaunay}: Voronoi tessellation and Delaunay triangulation on current points}
\label{sec: Procedure I: Voronoi tessellation and Delaunay triangulation on current points}

\begin{figure}
  \centering
  \includegraphics[width=1\textwidth]{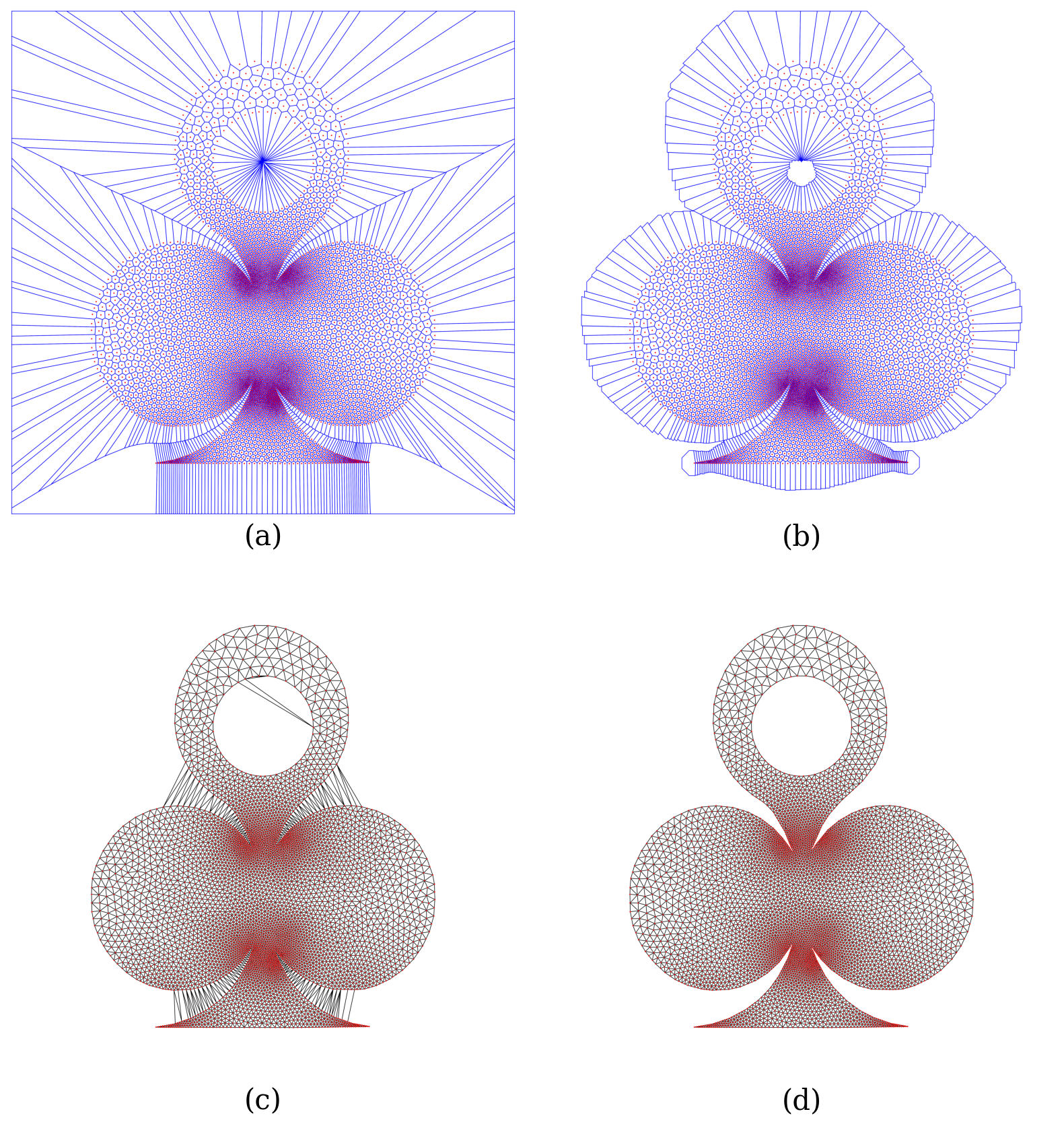}
  \caption{(a) Points and their full Voronoi diagram in the computation domain. (b) The adaptively bounded Voronoi diagram for the points. (c) The Delaunay triangulation obtained from the bounded Voronoi diagram. With the bounded Voronoi diagram, points far away are not neighbors. Therefore, the Delaunay triangulation is already quite clean. (d) After extracting valid triangles, we obtain a Delaunay triangulation on the shape.\label{fig:full_clipped_voro_tria}}
\end{figure}

First, we compute the Voronoi diagram for $\vec{p}_n$ using the multi-threaded \vpp{}. Voronoi cells on the geometry boundary can be much larger in size compared to the interior ones, as shown in Fig.~\ref{fig:full_clipped_voro_tria}(a). In \vpp{}, large Voronoi cells are more expensive to compute than smaller ones~\cite{lu2022}. Therefore, to achieve computational efficiency, we bound the sizes of the Voronoi cells adaptively. Each Voronoi cell is initialized as an octagon with span $S=\textit{fac}_{\text{voro}}^{\text{bound}}\cdot h_i$, which is adaptive to the local element size. The repeated cell cutting procedure described in Sec.~\ref{sec: intro, multi-threaded voro++} is applied to the octagon. Using this approach, as shown in Fig.~\ref{fig:full_clipped_voro_tria}(b), the boundary Voronoi cells are much smaller in size, and a significant speed up is achieved.

\begin{figure}
  \centering
  \includegraphics[width=1\textwidth]{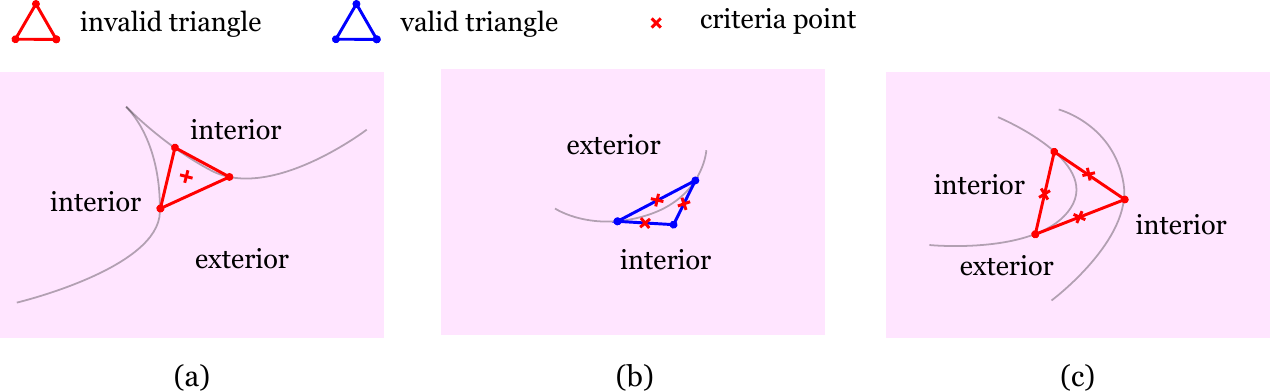}
  \caption{(a) An example invalid triangle whose centroid is outside of the geometry, and can be detected and discarded in the centroid test. (b) An example valid triangle on a concave geometry boundary. It approximates the geometry well and should be kept. Since it has two edge midpoints that are inside/on the boundary, it is kept as valid in the edge midpoints test. (c) An example invalid triangle that connects two separate boundaries, one convex and the other concave. Since it only has one edge midpoint inside/on the boundary, it can be detected and discarded in the edge midpoints test.
  ~\label{fig:tria_extract}}
\end{figure}

Once we have the Voronoi diagram, the Delaunay triangulation is obtained by connecting points sharing the same Voronoi cell edges, as shown in Fig.~\ref{fig:full_clipped_voro_tria}(c). An added benefit of bounding Voronoi cell sizes is that points that are far away will not be neighbors to each other, and the Delaunay triangulation is already quite clean. We then loop through the triangles, discard the ones outside of the geometry domain, and keep the valid ones that approximate the geometry shape well, as shown in Fig.~\ref{fig:full_clipped_voro_tria}(d). A series of tests are done on each triangle to determine its validity:
\begin{itemize}
  \item \textbf{Test 1: Centroid.} The triangle's vertices are all inside/on geometry boundary. Here, we test the centroid $\vec{c}$ of the triangle. We first look up the geometry grid the centroid is in.There are three possible cases:
     \begin{itemize}
         \item Case 1: If the geometry grid is an inner grid, the triangle is a valid triangle.
         \item Case 2: If it is an outer grid, discard the triangle.
     \end{itemize}

     \begin{itemize}
         \item Case 3: If $\vec{c}$ is in a boundary grid, we use the grid's ADF cell to compute its signed distance, and compared with the local boundary tolerance, $\geps_i$. If $\textit{adf}(\vec{c})>\geps_i$, meaning the centroid is outside of the geometry, we discard the triangle.
     \end{itemize}
     Case 1 and 2 take care of the majority of triangles. In case 3, we use the ADF cell defined on the boundary grid for a fast computation of the signed distance. Test 1 is effective at eliminating most outside triangles, such as the one shown in Fig.~\ref{fig:tria_extract}(a).
   \item \textbf{Test 2: Edges' midpoints.} After test 1, the remaining triangles all have vertices and centroids inside/on the geometry boundary. In test 2, we compute the triangle edges' midpoints. For each midpoint, $\vec{m}$, we look up the geometry grid it lies in. We also keep a count of edge midpoints inside/on the geometry:
    \begin{itemize}
        \item If $\vec{m}$ is in an inner grid, increment the count by one.
        \item If $\vec{m}$ is in a boundary grid, we use the ADF cell to do a fast computation of the signed distance. If $\textit{adf}(\vec{m})\leq\geps_i$, increment the count by one.
    \end{itemize}
	After checking all three edges' midpoints, if the count is at least two, the triangle is valid. Test 2 is effective at keeping triangles that approximate the shape well, as shown in Fig.~\ref{fig:tria_extract}(b). It can eliminate invalid triangles as shown in Fig.~\ref{fig:tria_extract}(c).
   \item \textbf{Test 3: Circumcenter.} Test 1 and 2 should account for most triangles. If there are any triangles whose validity is not determined after test 1 and 2, an additional test using the circumcenter of the triangle is performed, to extract triangles within a small distance away from the geometry boundary~\cite{alliez2005}.

    We compute the circumcenter, $\vec{c}_{\text{circum}}$, of the triangle, and its signed distance, $\sdf(\vec{c}_{\text{circum}})$. We also compute the circumradius, $R_{\text{circum}}$. If the ratio $\lvert \sdf(\vec{c}_{\text{circum}})\rvert/R_{\text{circum}}$ is larger than a threshold, $T^{\text{tria}}_{\vec{c}_{\text{circum}}}$, the triangle is discarded; otherwise, the triangle is kept as valid.

    Test 3 may be slightly more expensive than test 1 and 2, due to the computation of $\vec{c}_{\text{circum}}$ and $R_{\text{circum}}$. In addition, $\vec{c}_{\text{circum}}$ may not lie in a geometry boundary grid. In that case, its signed distance is computed by using the shape input information (user-defined signed distance function, or automatic calculation of SDF from input geometry contour line segments), which can be potentially expensive. However, test 3 should happen rarely.
\end{itemize}

\subsection{Algorithm to compute new point positions}
\label{sec: Algorithm to compute new point positions}
We now describe some details of the meshing algorithms. A selected meshing algorithm is used in an iteration to generate an initial set of new point positions.

\begin{enumerate}[wide, labelwidth=!, labelindent=0pt]
  \item \textbf{DistMesh meshing.} To obtain the set of initial new point positions in an iteration, we follow Persson's force balancing steps to resolve inner spring forces acting on points. Compared to the original implementation by Persson, we use the geometry grid structure for improved computational efficiency, and we use geometric adaptive quantities for improved geometric adaptivity both in space and across different resolutions of meshing.
  \item \textbf{CVD meshing.} An important step in CVD meshing is the computation of Voronoi cell centroids. As mentioned in Sec.~\ref{sec:intro-CVD}, we can use a quadrature rule, which divides a Voronoi cell into triangles, to compute its centroid. Here, we use an adaptive scheme for triangle division on each cell, to bound the error of centroid approximation. The centroid approximation error threshold for a Voronoi cell $V$ of a point $j$ in an iteration is, $E_{\text{thres},V}^{\vec{c}}=l_V \cdot \epsilon_{\vec{c}}$, where $l_V$ is the local characteristic length-scale of $V$. It is approximated by $l_V=\sqrt{A_V}$, where $A_V$ is the area of $V$. $\epsilon_{\vec{c}}$ is a small dimensionless number. We relate $\epsilon_{\vec{c}}$ with the previous point movement distance, $d_{\text{prev}}^{j}$, and the local desired edge length, $h_i$. We define $\epsilon_{\vec{c}}$ as $\epsilon_{\vec{c}}=\max(d_{\text{prev}}^{j}/h_i, \textit{fac}^{\text{end}}_{\text{mvmt}})$. Recall that our termination threshold for small point movement is
    $T^{\text{end}}_{\text{mvmt},i}= \textit{fac}^{\text{end}}_{\text{mvmt}} \cdot h_i$. We bound $\epsilon_{\vec{c}}$ with a minimum of $\textit{fac}^{\text{end}}_{\text{mvmt}}$, consistent with the small point movement threshold factor, which represents the smallest scaling of the error we need.

Figure~\ref{fig:Voro_tria_division} shows an example Voronoi cell division and the corresponding quadrature points at each level. Level $0$ represents the Voronoi cell itself, and there is no subdivision. The centroid approximation uses the point itself as a quadrature point. In level $1$, the Voronoi cell is divided into triangles by connecting the point to its cell vertices. In each triangle, the centroid is used as a quadrature point. If we want to advance from level $k$ to level $k+1$, for each triangle in level $k$, we subdivide it into two, by connecting the midpoint of its longest edge to the opposite triangle vertex. Level $2$ and $3$ are shown as examples.

\begin{figure}
  \centering
  \includegraphics[width=1\textwidth]{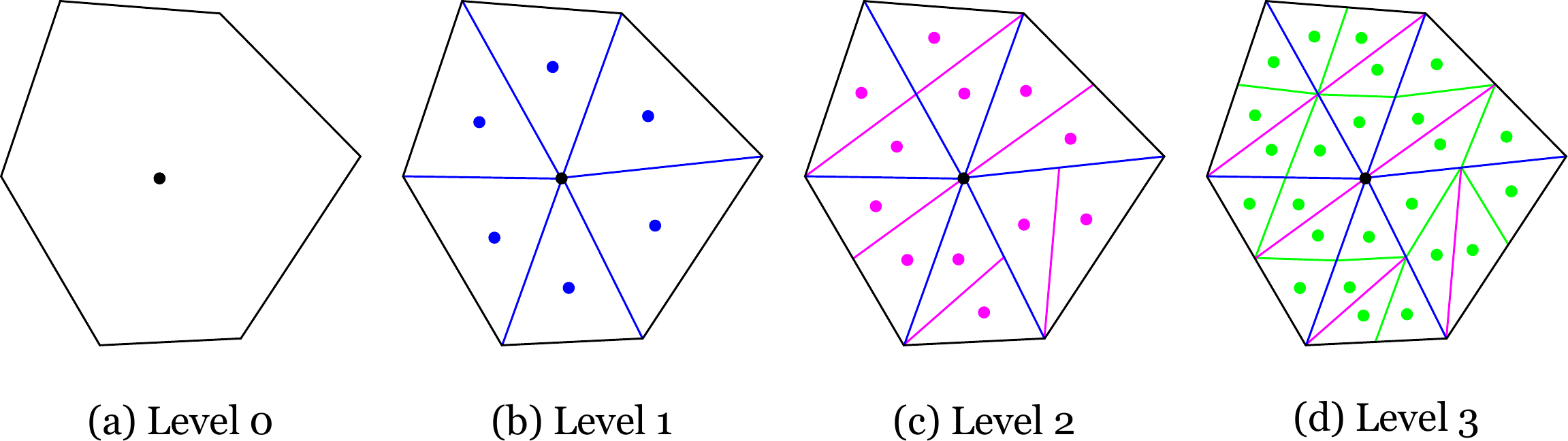}
  \caption{Triangle division of a Voronoi cell for the use of quadrature rule to calculate the Voronoi cell centroid. (a) At level $0$, the Voronoi cell is not divided, and the generator point itself is used as quadrature point. (b) At level $1$, we divide the Voronoi cell by connecting the point to its Voronoi cell vertices. The triangle centroids are used as quadrature points. For each of the following levels, we divide each triangle at the current level into two, by connecting the longest edge's midpoint to the opposite triangle vertex. (c)(d) The triangle divisions at level $2$ and $3$, respectively.
  ~\label{fig:Voro_tria_division}}
\end{figure}

At level $k$, we use $\hat{\vec{c}}_{k+1}$ as an approximation of the true centroid position, and test if $\lVert \hat{\vec{c}}_{k}-\hat{\vec{c}}_{k+1} \rVert < E_{\text{thres},V}^{\vec{c}}$. If not, we proceed to level $k+1,$ and test its error against level $k+2$. Otherwise, we have bounded the error of $\hat{\vec{c}}_{k}$, and the division stops. Since $\hat{\vec{c}}_{k+1}$ is already computed, we use $\hat{\vec{c}}_{k+1}$ as our centroid approximation, because it is more accurate than $\hat{\vec{c}}_{k}$.

The adaptive triangle division scheme bounds the Voronoi centroid approximation error. The error scales with the size of the Voronoi cell, conforming with local density. When the density is large, the Voronoi cell is small, and the error bound is small, and vice versa. The error bound also considers previous point movement. In the beginning meshing iterations, point movements are large, and the centroid approximations can be less accurate. As meshing progresses, point movements become small. To improve point position, accurate centroid approximation is needed. We obtain a dimensionless relative factor of the local movement, calculated as the previous point movement scaled by the local desired edge length. The error bound is proportional to this factor.

The Voronoi cell centroid calculation is more expensive when there is a larger number of dividing triangles. Since the adaptive triangle division scheme is implemented on every Voronoi cell, it achieves spatial adaptivity. We save computation time in regions that have larger error bounds and use more computation resources in regions that require more accurate centroid approximations.

  \item \textbf{Hybrid meshing.} We also combine the two methods, by performing DistMesh iterations to begin with. We switch to CVD meshing iterations only when $N_{\text{current}}=N_{\text{total}}$ and the relative change in $\frac{1}{2}$-mean of aspect ratio is smaller than a threshold, $T^{\text{switch}}_{\text{quality}}$, which is close to the mesh quality termination threshold discussed in Sec.~\ref{sec:main meshing part}, $T^{\text{end}}_{\text{quality}}$. Therefore, DistMesh iterations are used most of the time, and CVD meshing is only used in the last few iterations leading to termination. In the hybrid method, DistMesh is used to quickly obtain an overall high-quality mesh, and the more expensive CVD meshing is used for mesh quality refinement steps in the end.
\end{enumerate}

\subsection{Treatment for the updated points}
\label{sec: new points treatment}
In an iteration, by using one of the meshing algorithms above, we obtain an initial set of updated point positions.

\begin{figure}
  \centering
  \includegraphics[width=1\textwidth]{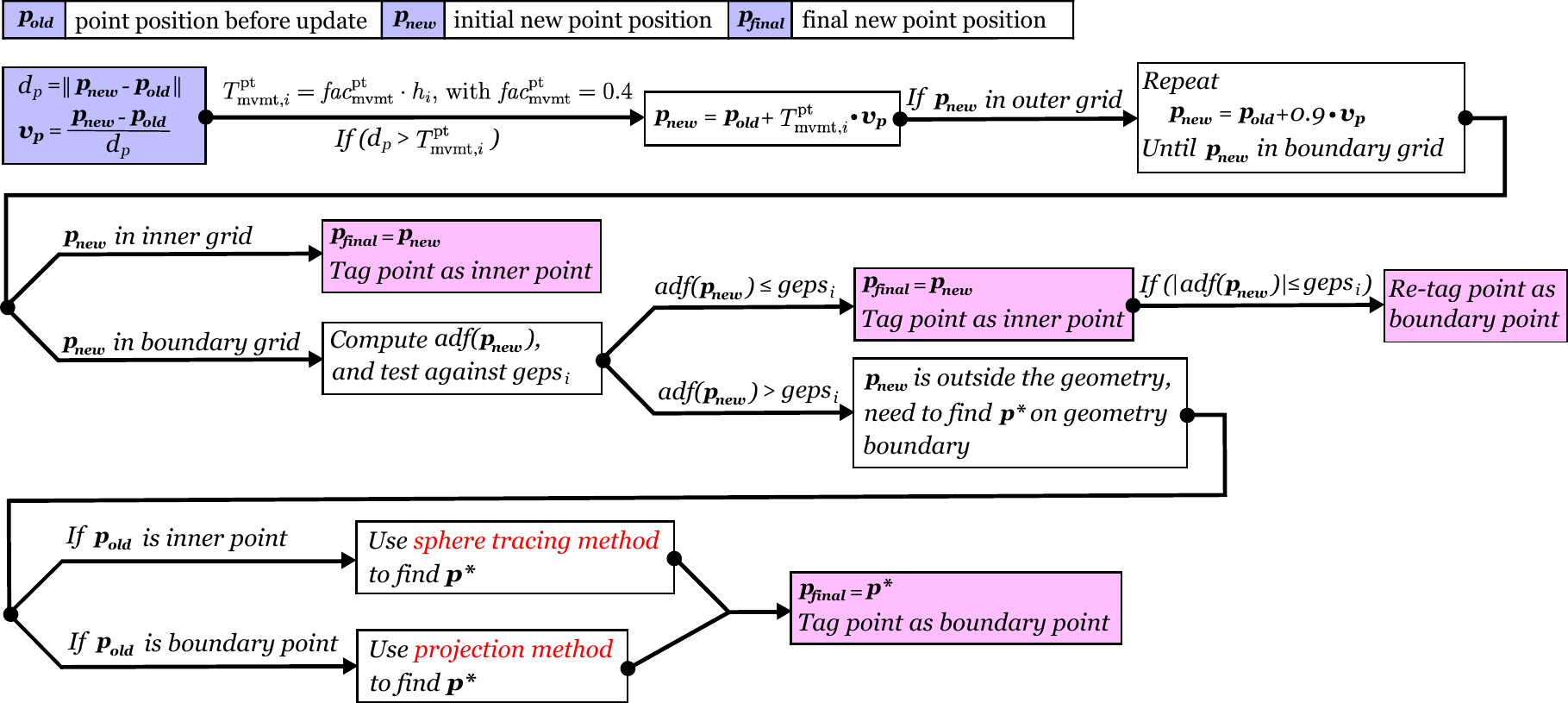}
  \caption{Procedure to check and correct $\vec{p}_{\text{new}}$, the initial set of updated point positions obtained from one of the meshing algorithms used. After the procedure, we obtain the final set of point positions, $\vec{p}_{\text{final}}$, to move the points $\vec{p}_{\text{old}}$ to.
  ~\label{fig:new_pt_treatment}}
\end{figure}

We now describe a procedure, as shown in Fig.~\ref{fig:new_pt_treatment}, that we use to check and correct each of the updated points, $\vec{p}_{\text{new}}$, to obtain the final set of point positions, $\vec{p}_{\text{final}}$, to move the points to.

The procedure above is optimized for computational efficiency by using the geometry grid, geometric adaptive quantities, and point category information. For a point, we start by restricting its point movement to not be too large by using the point movement threshold introduced in Sec.~\ref{sec: set up part}, $T^{\text{pt}}_{\text{mvmt},i}=\textit{fac}^{\text{pt}}_{\text{mvmt}} \cdot h_i$. We further restrict $\vec{p}_{\text{new}}$ to lie in inner or boundary grid only, assuming outer grids are too far away from the geometry for $\vec{p}_{\text{new}}$ to be an accurate updated point position. Now $\vec{p}_{\text{new}}$ lies in either an inner or a boundary grid. If $\vec{p}_{\text{new}}$ is in an inner grid, we set $\vec{p}_{\text{final}}=\vec{p}_{\text{new}}$, and tag it as an inner point. If $\vec{p}_{\text{new}}$ is in a boundary grid, we further test its signed distance by using the ADF cell on the boundary grid, $\textit{adf}(\vec{p}_{\text{new}})$, against the local geometry boundary tolerance, $\geps_i$. If $\vec{p}_{\text{new}}$ is inside/on the geometry boundary, we set $\vec{p}_{\text{final}}=\vec{p}_{\text{new}}$, and tag the point as an inner or boundary point accordingly. Otherwise, if $\vec{p}_{\text{new}}$ is outside of the geometry, we need to correct it to be a point on the geometry boundary, $\vec{p}^*$. Here, depending on the point category of $\vec{p}_{\text{old}}$, the point position before the current iteration, two methods are used. If $\vec{p}_{\text{old}}$ is an inner point, the sphere tracing method~\cite{hart1995} is used to find $\vec{p}^*$. If $\vec{p}_{\text{old}}$ is a boundary point, the point projection method using the damped Newton method~\cite{Persson04,persson2005implicitgeometries} is used. Lastly, we update $\vec{p}_{\text{final}}=\vec{p}^*$, and tag it as a boundary point. Next, we describe the sphere tracing and the point projection method for finding $\vec{p}^*$.

\begin{enumerate}[wide, labelwidth=!, labelindent=0pt]
  \item \textbf{Sphere tracing method.}\label{sphere tracing method} As seen in Fig.~\ref{fig:sphere_tracing_newton_projection}(a), when $\vec{p}_{\text{old}}$ is an inner point, and $\vec{p}_{\text{new}}$ is outside of the geometry, a reasonable $\vec{p}^*$ is the intersection of the line segment defined by $\vec{p}_{\text{old}}\vec{p}_{\text{new}}$ with the geometry boundary. To find the intersection point, an efficient algorithm, sphere tracing~\cite{hart1995}, is used. We start from $\vec{p}_{\text{old}}$, and define a circle with radius $R=\lvert \sdf(\vec{p}_{\text{old}})\rvert$. We find the intersection point of the circle with the line segment $\vec{p}_{\text{old}}\vec{p}_{\text{new}}$. We then apply the same process on the intersection point. This is repeated, until we find an intersection point within the geometry boundary tolerance. We set the final intersection point as $\vec{p}^*$. The sphere tracing algorithm is more efficient than a binary search for $\vec{p}^*$, since we are able to take adaptive step sizes towards $\vec{p}^*$, that are larger to begin with, and smaller when $\vec{p}^*$ is close.

  \item \textbf{Point projection method.}\label{point projection method} As seen in Fig.~\ref{fig:sphere_tracing_newton_projection}(b), when $\vec{p}_{\text{old}}$ is on the geometry boundary, and $\vec{p}_{\text{new}}$ is outside of the geometry, $\vec{p}^*$ is the closest point of $\vec{p}_{\text{new}}$ on the geometry boundary. A point projection method is used to find $\vec{p}^*$. Assuming that $\vec{p}_{\text{new}}$ is close to the geometry boundary, $\nabla\sdf(\vec{p}_{\text{new}})$ points in the direction of the closest point, normal to the boundary tangent. We use the damped Newton method, to find $\vec{p}^*$~\cite{Persson04,persson2005implicitgeometries}. Since $\vec{p}^*$ is on geometry boundary, it should have $\sdf(\vec{p}^*)=0$. Moreover, $\vec{p}_{\text{new}}-\vec{p}^*$ should be parallel to $\nabla \sdf(\vec{p}_{\text{new}})$. Therefore, we want $\vec{p}^*$ to satisfy,
    \begin{equation}
      \label{eq:projection_point_requirement}
      \vec{L}(\vec{p}^*)=\left[ \sdf(\vec{p}^*), \quad (\vec{p}_{\text{new}}-\vec{p}^*)\times\nabla \sdf(\vec{p}_{\text{new}}) \right] = \vec{0}.
    \end{equation}
Equation~\eqref{eq:projection_point_requirement} can be solved via the damped Newton method, with $\vec{p}_{\text{new}}=(x_{\text{new}},y_{\text{new}})$ as initial guess. The Jacobian of $\vec{L}$ is
\begin{equation}
  \label{eq:projection_Newton_Jacobian}
  \vec{J}(\vec{p})=\frac{\partial \vec{L}}{\partial \vec{p}}=\left[ \begin{matrix}
    f_x &
    f_y+(x-x_{\text{new}})f_{xy}-(y-y_{\text{new}})f_{xx}\\
    f_y  &
    -f_x-(y-y_{\text{new}})f_{xy}+(x-x_{\text{new}})f_{yy}\\
  \end{matrix}
  \right]^\trans.
\end{equation}
where $f$ represents the SDF. The derivatives in the above equations are approximated with the finite
difference method of step size, $\deps_i$. We update the solution by
\begin{equation}
  \label{eq:projection_pt_update}
  \vec{p}_{k+1}=\vec{p}_{k}-\hat{\alpha} \vec{J}^{-1}(\vec{p}_k)\vec{L}(\vec{p}_k),
\end{equation}
with damping parameter $\hat{\alpha}$. We iterate until
$\vec{p}_{k+1}$ is within its local geometry boundary tolerance, $\geps_i$, or
when a maximum Newton's step threshold $T^{\text{newton}}_{\text{ct}}$ is reached.

Lastly, we check the resulting point, $\vec{p}_{k+1}$. If it is a boundary
point, $\vec{p}^*=\vec{p}_{k+1}$. Otherwise, the projection step diverges and
fails to find the boundary point. In that case, we set
$\vec{p}^*=\vec{p}_{\text{old}}$, and the point remains in its position without
updating.

\begin{figure}
  \centering
  \includegraphics[width=0.6\textwidth]{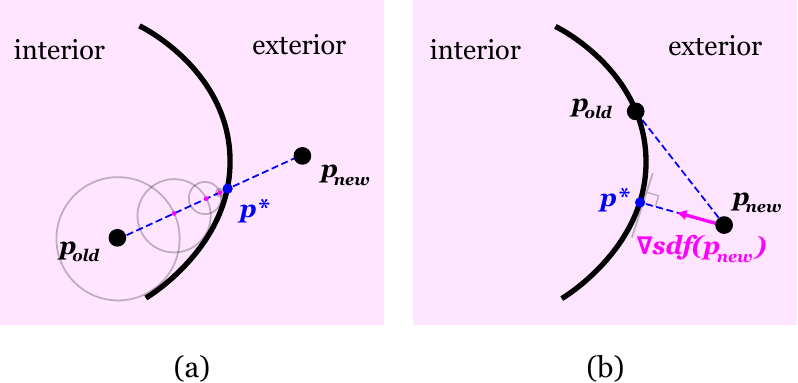}
  \caption{(a) When $\vec{p}_{\text{old}}$ is an inner point and $\vec{p}_{\text{new}}$ is outside of the geometry, $\vec{p}^*$ is the intersection point of $\vec{p}_{\text{old}}\vec{p}_{\text{new}}$ and the geometry boundary, and we can use the sphere tracing algorithm to find $\vec{p}^*$ efficiently. (b) When $\vec{p}_{\text{old}}$ is a boundary point and $\vec{p}_{\text{new}}$ is outside of the geometry, $\vec{p}^*$ is the closest point of $\vec{p}_{\text{new}}$ on the geometry boundary. We can use a point projection scheme with damped Newton's method to find $\vec{p}^*$. ~\label{fig:sphere_tracing_newton_projection}}
\end{figure}

\end{enumerate}

\section{Multi-threaded parallel implementation}
\label{sec: parallel implementation}
We use the multi-threaded version of \vpp{} for computing the Voronoi diagram and Delaunay triangulation in Procedure~\ref{procedure: key meshing step: Voro, Delaunay}. Points are assigned to different parallel threads, and threads compute their assigned points' Voronoi cells simultaneously. We use the OpenMP dynamic load balancing strategy (enabled with the OpenMP directive \co{schedule(dynamic)}), which was shown by Lu et al.~\cite{lu2022} to provide good parallel performance across a wide range of cases.

For a DistMesh iteration shown in Algorithm~\ref{algorithm:DistMesh meshing}, as discussed in Sec.~\ref{sec:intro-DistMesh}, the internal force balancing procedure to compute new point positions is less suitable for parallelization. For example, the calculation of $\textit{fac}_{\mu}$ in Eq.~\eqref{eq:DistMesh-size scaling} requires calculating the sums $\sum l_i^2$ and $\sum \mu(\vec{x}_i)^2$ from all edges in the triangulation. Therefore, the internal force balancing for a point is dependent on its neighboring points as well as the connecting edges, and requires information on the whole spring system. It cannot be done for each point independently in a single parallel loop over the points. Rather, we need to first loop through the edges, and compute some quantities of the edges, such as edge lengths, and the sum of the squared edge length. Here, calculating the sum in a parallel loop is not ideal, since it requires threads to update the same quantity in serial, to avoid a race condition. To resolve this, OpenMP has a built-in directive, \texttt{\#pragma omp parallel for reduction(...)}, which can ensure summation quantities in a parallel for loop are updated by one thread at a time. We use the directive to calculate $\sum l_i^2$ and $\sum \mu(\vec{x}_i)^2$ in a parallel loop over the edges.

After the loop, we use the obtained sums to calculate information of the entire spring system, $\textit{fac}_{\mu}$. We then loop through the edges again, and for each edge, we compute the desired edge length using Eq.~\eqref{eq:DistMesh-size scaling}, and the corresponding repulsive force generated using Eq.~\eqref{eq:DistMesh-spring force}. We also look up the two connecting points of the edge and apply the edge repulsive force on them. When applying edge forces on connecting points to update their positions, the same point's position can only be updated by one thread at a time, to avoid a race condition. To resolve this, in a parallel loop over the edges, we use another OpenMP built-in directive, \texttt{\#pragma omp atomic}, when updating point positions. This directive ensures that the same point position is only updated by one thread at a time, although it incurs a small performance penalty.

In contrast, a CVD meshing iteration shown in Algorithm~\ref{algorithm:CVD meshing} is ideal for parallelization. As discussed in Sec.~\ref{sec:intro-CVD}, for a point, the computations of its Voronoi cell and cell centroid are independent from other points. Therefore, they can be done in parallel in one single loop of the points. This distributes the most expensive computations in the algorithm to parallel threads efficiently.

\section{Example code}
\label{sec:example code}
An example meshing code on a 2D circle is shown in Listing~\ref{code:example code}. We first set up parameters and create mesh output directory. We create a \co{container\_2d} class object to use the multi-threaded \vpp{} software to compute Voronoi diagrams. We then create the circle shape, and generate a sizing field on the shape, with the specified mesh gradation parameter $K$. The density field is automatically calculated in the sizing field object. Next, we create a \co{parallel\_meshing\_2d} class object, \co{pm2d}, which gathers all information from the set up part, and initializes points in the mesh. Lastly, we choose a meshing method (DistMesh, CVD meshing, or the hybrid method) and initialize a corresponding class object of the method. We use the chosen meshing method to do meshing with \co{pm2d}.

\begin{lstlisting}[caption={Example code of multi-threaded meshing on a 2D circle}, label={code:example code}]
int main() {

//-----Parameter set up-----
    //Number of parallel threads
    int num_t=4;
    //Total number of points in the mesh
    int Ntotal=5000;
    //Mesh gradation parameter. K=0 corresponds to a uniform sizing field.
    double K=0.005;
    //If output_interval=0, no output;
    //If output_interval=-1, only output mesh at the beginning and at termination;
    //If output_interval=m, where m is a positive integer, output mesh every m iterations.
    //Here, output_interval=1, output mesh in every iteration.
    int output_interval=1;

//-----Create output directory-----
    char output_directory[256];
    sprintf(output_directory,"meshing_outputs");
    mkdir(output_directory,S_IRWXU|S_IRWXG|S_IROTH|S_IXOTH);

//-----Create container_2d class object to use multi-threaded VORO++-----
    //Construct a 2D container as a non-periodic unit square defined on [0,1]x[0,1], divided into a cnx by cny grid of blocks. The grid is constructed so that if Ntotal points are homogeneously distributed in the domain, on average 3.3 points are in each grid.
    //Each block initially holds up to 16 points.
    //num_t threads are used for the parallel computing of Voronoi diagrams.
    int cnx=sqrt(Ntotal/3.3); int cny=cnx;
    container_2d con(0.0,1.0,0.0,1.0,cnx,cny,false,false,16,num_t);

//-----Create shape-----
    //Define a circle shape centered at (0.5,0.5) with radius 0.1
    shape_2d_circle shp(con,num_t,0.1,0.5,0.5);

//-----Create sizing field-----
    //Automatic generation of the sizing and density fields on the shape, with mesh gradation parameter K
    sizing_2d_automatic size_field(&shp,K);

//-----Create parallel_meshing_2d object, pm2d-----
    //The parallel_meshing_2d class gathers all geometry set up information, and initialize points in the mesh
    parallel_meshing_2d pm2d(&con, &shp, &size_field, num_t, output_interval, output_directory);
    pm2d.pt_init(Ntotal);

//-----Parallel meshing-----
    //Choose the meshing algorithm to use and do meshing
    //DistMesh: mesh_alg_2d_dm;
    //CVD meshing: mesh_alg_2d_cvd;
    //Hybrid method: mesh_alg_2d_hybrid.
    //Here, we use DistMesh
    mesh_alg_2d_dm mesh_method(&pm2d);
    pm2d.meshing(&mesh_method);
}
\end{lstlisting}

\section{Performance}
\label{sec:performance}
We test the performance of \textsc{TriMe++} on two meshing cases, both with $N_\text{total} = 10^6$ meshing points: (I) a square shape with a uniform density field, and (II) the custom shape with an adaptive density field. For each case, we test three different meshing methods (DistMesh, CVD, and the hybrid method), and analyze the following aspects of performance:
\begin{enumerate}[wide, labelwidth=!]
    \item \textbf{Parallel performance.} We measure the wall clock time $t_p$ using a variable number of $p$ threads, and calculate the parallel efficiency using
    \begin{equation}
      T_e(p) =\frac{t_1}{p \cdot t_p},
      \label{eq:parallel_eff}
    \end{equation}
    which measures the effective slowdown from the hypothetically perfect parallel
    scaling.
    \item \textbf{Optimal point scheme.} The above tests are repeated using two different point schemes:
    \begin{itemize}
      \item All $N_\text{total}$ points are initialized prior to meshing.
      \item As described in Sec.~\ref{set-up-step-point-addition}, a small fraction of points are first initialized, and during the meshing iterations, additional points are inserted, until $N_\text{total}$ points are reached.
    \end{itemize}
    We compare the timing performance and mesh quality of the point schemes.
  \item \textbf{Mesh quality.} We provide mesh quality statistics from a typical run using the optimal point scheme, including the medians, means, maximum values and standard deviations of the triangles' $\alpha$ and $\beta$. 
    \item \textbf{System size.} Using the preferred point scheme, we vary the number of meshing points $N_\text{total}$, and examine the parallel performance of the code as system size varies.
\end{enumerate}
Our tests are done on a Ubuntu Linux computer with 256\,GB of memory and dual Intel Xeon E5-2650L v4 processors with 14 low-power cores using a 1.7\,GHz base clock speed. Since the computer has 28 physical cores, we focus our attention to the first $28$ threads. The Intel processors use a technology called Turbo Boost that alters the clock speed depending on the total active cores, which complicates the measurement of parallel efficiency. We turn off the Turbo Boost feature for the tests. 

\subsection{Uniform meshing of a square}
\begin{figure}[h!]
  \centering
  \includegraphics[width=0.4\textwidth]{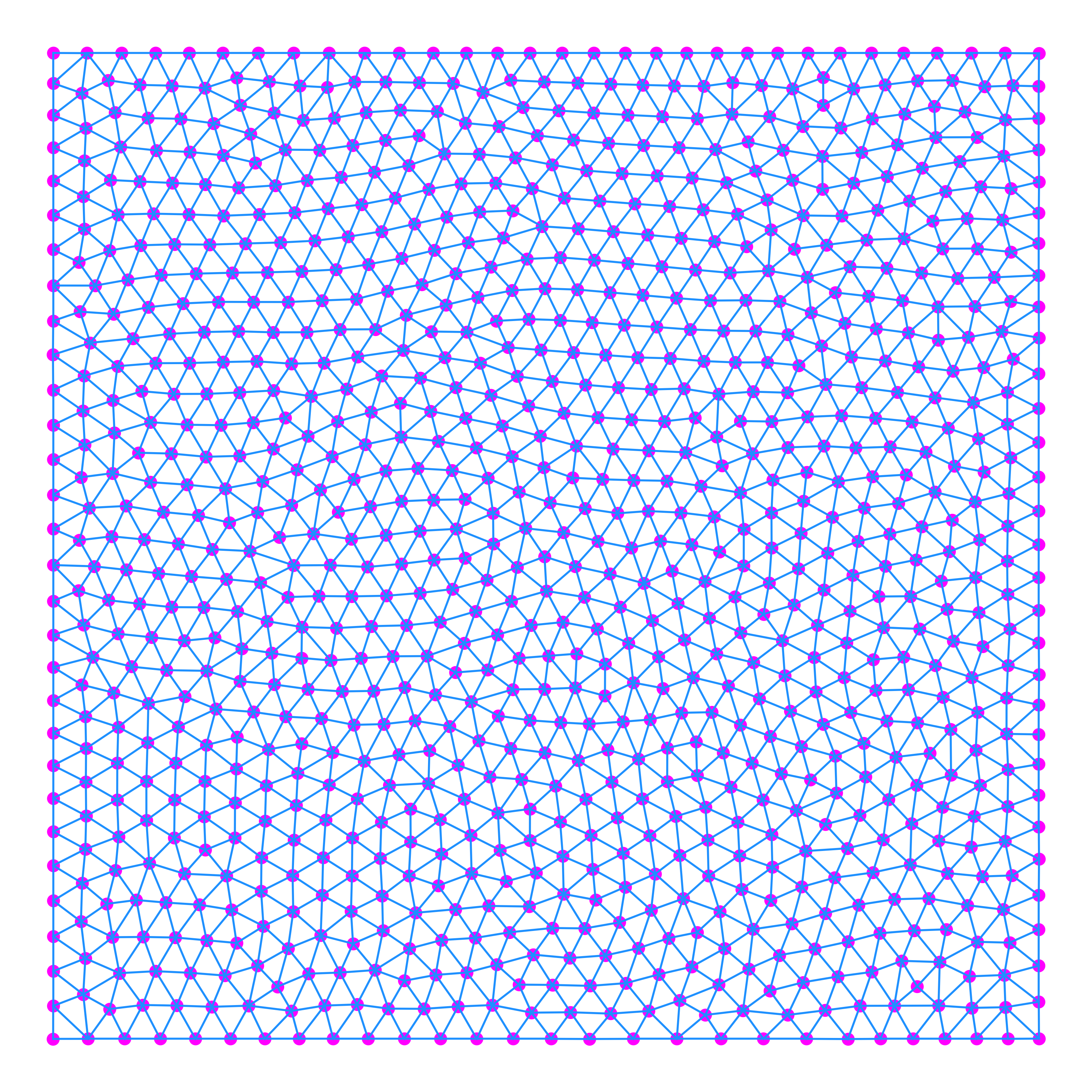}
  \caption{Schematic illustration of an example mesh from uniform meshing on a sqaure, using $10^3$ points. \label{fig:square_mesh_1000}}
\end{figure}

Uniform meshing of a square represents a situation where both the geometry and the sizing field of the mesh are simple. An example mesh with $10^3$ points is shown in Fig.~\ref{fig:square_mesh_1000}. In general, we found that for simple and convex shapes (e.g.\@ rectangles, circles) where points can easily move around without obstruction, and when the element sizing field varies gradually, there is no need to use the point addition scheme. Instead, initializing all points to start meshing gives better performance.

\begin{figure}[h!]
  \centering
  \includegraphics[width=1\textwidth]{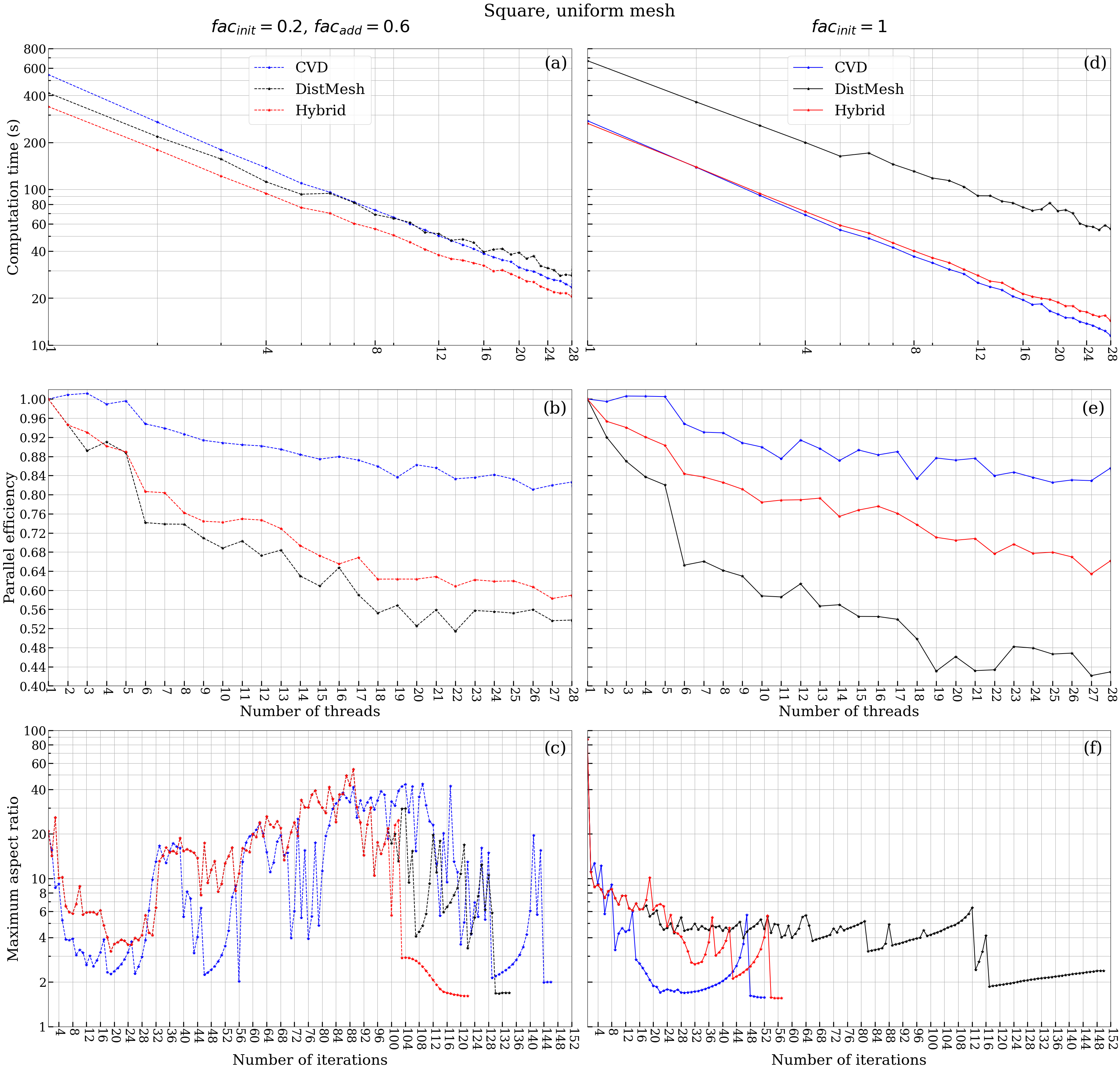}
  \caption{Performance data for uniform meshing of a square, with $N_{\text{total}}=10^6$ points. Comparison of the point addition scheme (a,b,c) and the scheme where we initialize all point to start meshing (d,e,f). (a,d) Computation time of meshing using the three meshing methods against the number of threads. (b,e) The corresponding parallel efficiency against the number of threads for the three meshing methods. (c,f) The worst triangle qualities of the three meshing methods measured by the maximum aspect ratios. \label{fig:Square_uniform_parallel_quality}}
\end{figure}

Figure~\ref{fig:Square_uniform_parallel_quality} compares the parallel performance and the worst triangle qualities of the two point schemes. Figures~\ref{fig:Square_uniform_parallel_quality}(c,f) show the worst triangle qualities measured by the maximum aspect ratios, from typical runs of the two point schemes, respectively. For this simple meshing case, the point addition scheme has no significant advantage. All meshing methods in both schemes achieve low maximum aspect ratios at termination, meaning that the worst triangles in the meshes are already of very high qualities.

Figures \ref{fig:Square_uniform_parallel_quality}(a,d) show the computation time of the three meshing methods against the number of parallel threads, using the different point schemes, respectively. Table~\ref{tbl:uniform square meshing computation time} lists the corresponding wall clock time using the serial code, and the parallel code with $28$ threads. Since the number of iterations at termination vary across methods, and can vary across different number of threads for the same method, we also provide the time per triangulation iteration in Table~\ref{tbl:uniform square meshing computation time}. For the serial code, using the point addition scheme, both CVD meshing and hybrid meshing take a significantly longer time. Since the shape and the sizing field are both simple, CVD meshing can effectively and steadily improve mesh quality and stably lead to termination. Using the point addition scheme disrupts the mesh from time to time, rather than refining it; this deters termination, requiring more meshing iterations. Therefore, the optimal point scheme is to initialize all points prior to meshing.

DistMesh takes significantly longer computation time, as shown in Fig.~\ref{fig:Square_uniform_parallel_quality}(d). This can sometimes happen when a specific set of meshing points results in problematic triangles during the meshing process, causing more iterations to terminate. This situation rarely happens using CVD meshing. Since hybrid meshing switches to CVD meshing once the overall mesh achieves a good quality, it also stably leads to termination.

\begin{table}[h!]
\centering
\begin{tabular}{|p{30mm}|c|c|c|c|}
\hline
Point scheme &  \multicolumn{2}{c|}{Point addition scheme} & \multicolumn{2}{c|}{Initialize all points}
\\ \hline
Computation Time (s)&1 thread&28 threads&1 thread&28 threads
\\ \hline
  DistMesh&415.842&28.034&669.497&55.6992
\\ &(3.0784)&(0.204491)&(4.46162)&(0.37121)
\\ \hline
  CVD&544.973&23.5537&275.623&11.5011
\\ meshing&(3.73193)&(0.161254)&(5.2983)&(0.220976)
\\ \hline
  Hybrid&339.598&20.5707&265.257&14.3139
\\ method&(2.7814)&(0.168459)&(4.64912)&(0.250808)
\\ \hline
\end{tabular}
\caption{Comparison of the point schemes, using uniform meshing on a square: computation time of meshing for the three methods, including both the total time, and in parentheses, the time per triangulation iteration, using 1-thread serial code and 28-thread parallel code.\label{tbl:uniform square meshing computation time}}
\end{table}

Figures \ref{fig:Square_uniform_parallel_quality}(b,e) show the corresponding parallel efficiency of the different meshing methods, using the two point schemes, respectively. We see similar performance in both point schemes. As expected, CVD meshing has high parallel efficiency, since in each meshing iteration, the centroid of the Voronoi cells can be computed simultaneously among parallel threads. DistMesh is less suitable for parallelization and has lower parallel efficiency, due to the force balancing step explained in Sec.~\ref{sec: parallel implementation}. Hybrid meshing has parallel efficiency in between DistMesh and CVD meshing; more CVD meshing iterations in the end increases the parallel efficiency of hybrid meshing.

\subsubsection{Mesh quality}
\begin{figure}[h!]
  \centering
  \includegraphics[width=1\textwidth]{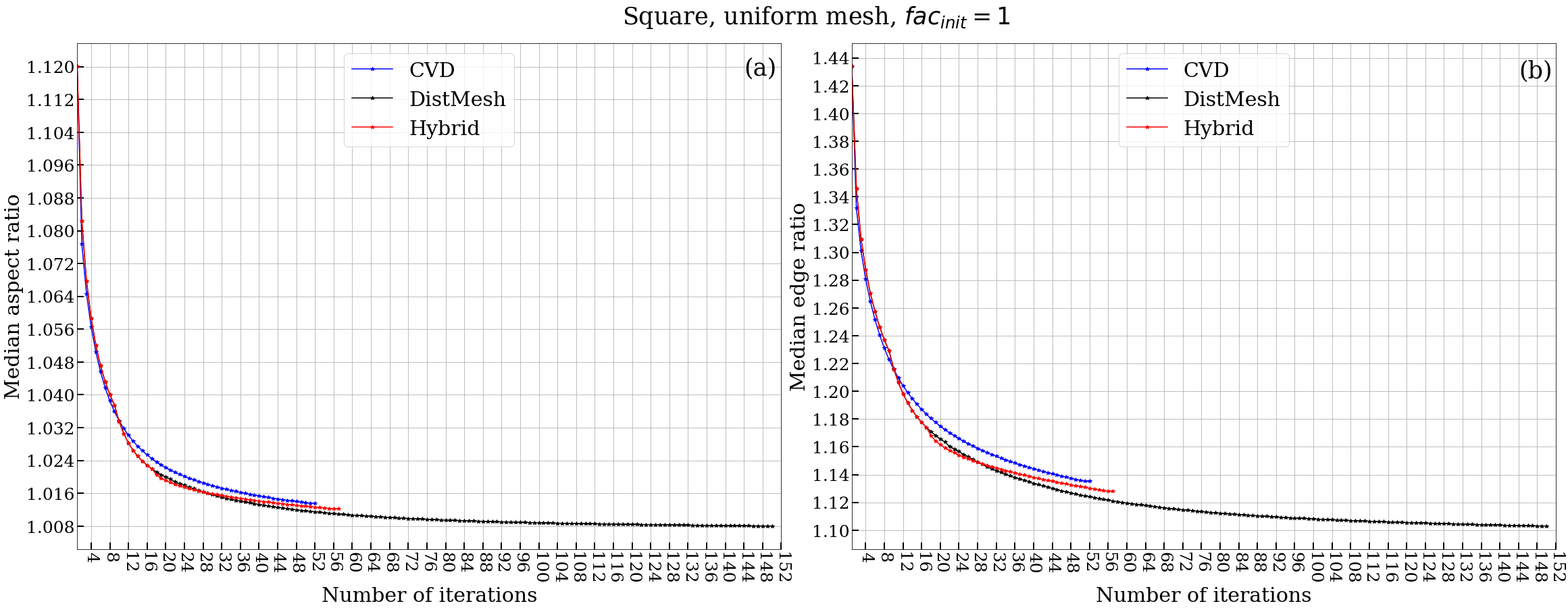}
  \caption{Uniform meshing on a square, where we initialize all points to start meshing:  (a) Median aspect ratios for the three meshing methods against the number of iterations. (b) Median edge ratios for the three meshing methods against the number of iterations. \label{fig:uniform_square_performance_median_ratios}}
\end{figure}

Using the optimal point scheme where we initialize all points to start meshing, Fig.~\ref{fig:uniform_square_performance_median_ratios} shows the median aspect and edge ratios against the number of iterations for the three meshing methods, respectively. Table~\ref{tbl:Uniform square meshing termination quality stats} gives mesh quality statistics at termination. All three methods achieve very good overall mesh quality at termination, with almost all triangles having $\alpha <2$. Furthermore, the maximum aspect and edge ratios are all small, meaning the worst triangle qualities are all very good.

\begin{table}[h!]
\centering
\footnotesize
\begin{tabular}{|p{22mm}||p{18mm}|p{18mm}|p{18mm}|p{18mm}|p{19mm}|p{19mm}|}
\hline
Mesh quality statistics at termination&Number of triangles&Median $\alpha$, $\beta$&Mean $\alpha$, $\beta$&Maximum $\alpha$, $\beta$&St.\@ dev.\@ from mean for $\alpha$, $\beta$& \% triangles with $\alpha<1.2$, $\alpha<2$
\\ \hline
    DistMesh&1,996,088&1.00802,&1.01778,&2.38954,&0.0242561,&99.96\%,\\
    & &1.10301&1.12803&1.77153&0.0912871&100.00\%
\\ \hline
  CVD&1,996,112&1.01356,&1.02764,&1.57996,&0.0383476,&99.16\%, \\
  meshing& &1.13518&1.15913&1.88335&0.10467&100.00\%
\\ \hline
  Hybrid&1,996,183&1.01221,&1.02545,&1.55909,&0.0361209,&99.30\%,\\
  method&&1.12804&1.15226&1.77469&0.101996&100.00\%
\\ \hline
\end{tabular}
\caption{Uniform meshing on a square, where we initialize all points to start meshing: Mesh quality statistics at termination.\label{tbl:Uniform square meshing termination quality stats}}
\end{table}

\subsubsection{Parallel performance for different system sizes}
\begin{figure}[h!]
  \centering
  \includegraphics[width=0.6\textwidth]{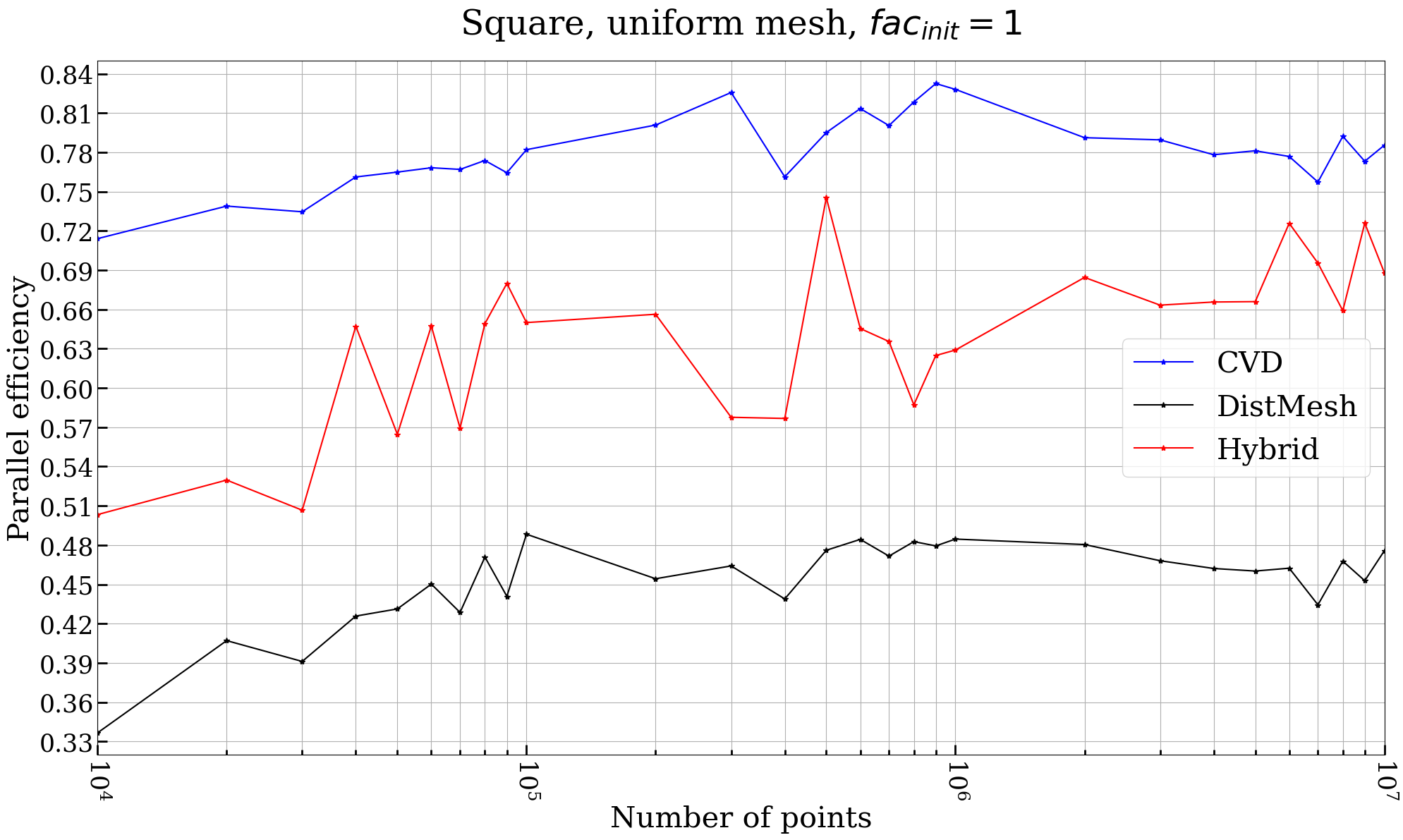}
  \caption{Parallel efficiency as a function of the number of points, for uniform meshing on a square where all points are initialized prior to meshing.\label{fig:Square_uniform_varyN}}
\end{figure}

Using the optimal point scheme for uniform meshing on a square, Fig.~\ref{fig:Square_uniform_varyN} shows the parallel efficiency of the three meshing methods as the number of points increases. As expected, CVD meshing has the best efficiency, and the efficiency of DistMesh is the lowest. Furthermore, we see a gradually increasing trend of parallel efficiency for the three meshing methods from $10^4$ to $10^6$ particles. After this, parallel efficiencies remain roughly at the same level as the system size continues to increase. The higher parallel efficiency for large systems is desirable for large-scale meshing.

\subsection{Adaptive meshing of the custom shape}
\begin{figure}[h!]
  \centering
  \includegraphics[width=0.5\textwidth]{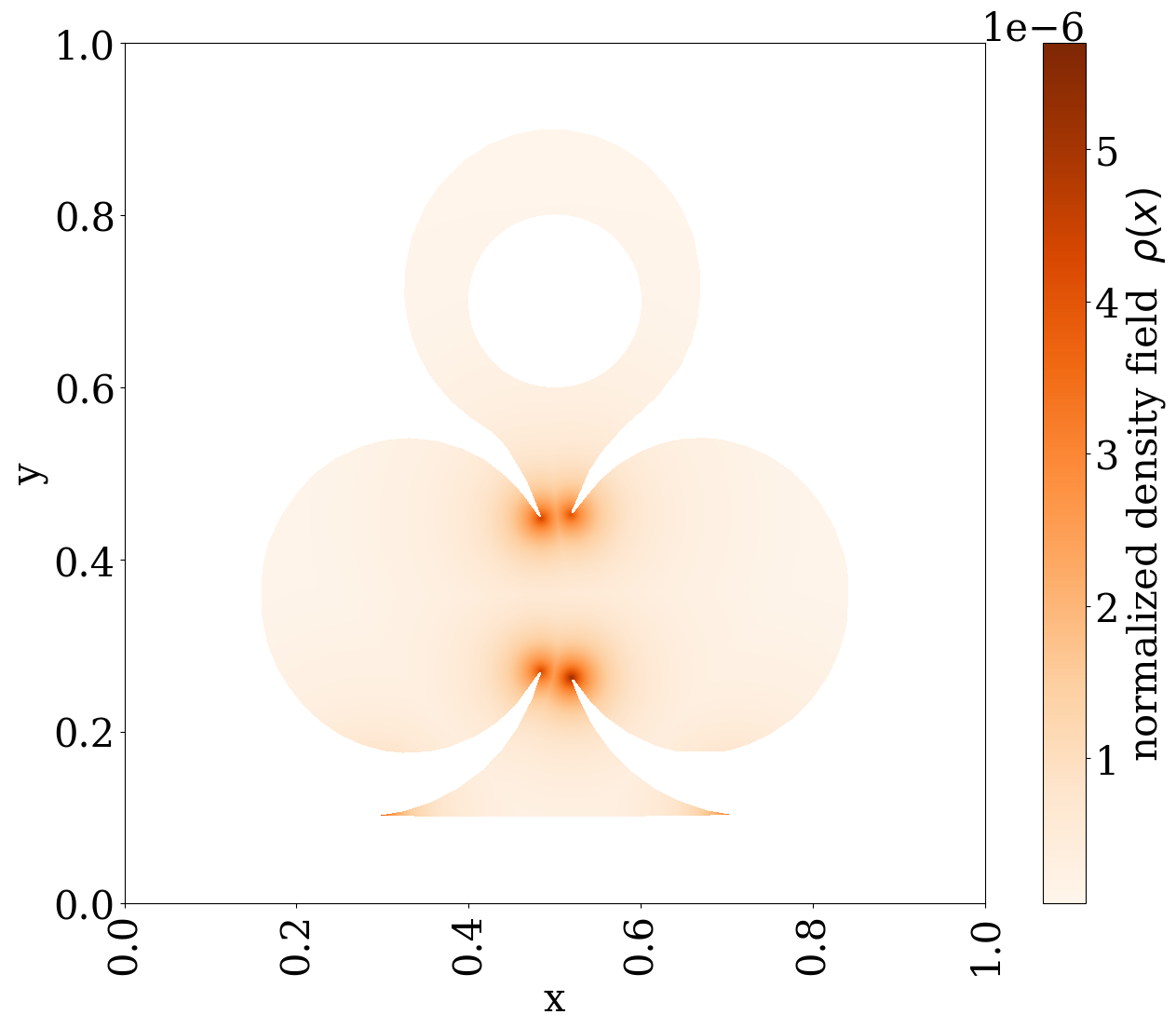}
  \caption{The normalized density field for the custom shape, generated with $N_{\text{total}}=10^6$ points.
  ~\label{fig:density_large_system}}
\end{figure}

Adaptive meshing on the custom shape represents a case where the geometry and the sizing field are complicated. With $K=0.005$ for $N_\text{total} = 10^6$ points, the corresponding density field is shown in Fig.~\ref{fig:density_large_system}.

\begin{figure}[h!]
  \centering
  \includegraphics[width=1\textwidth]{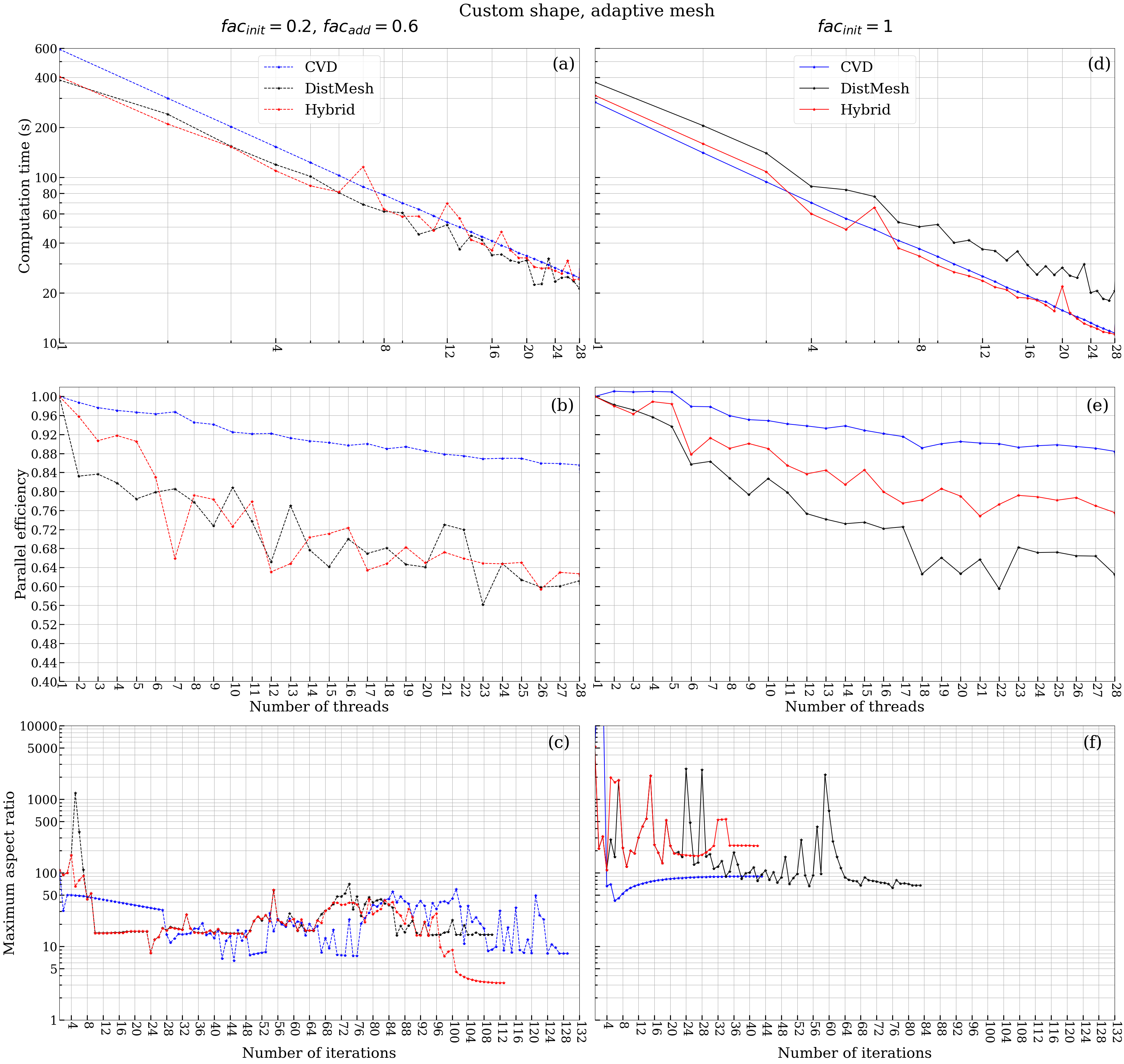}
  \caption{Performance data for adaptive meshing of the custom shape, with $N_{\text{total}}=10^6$ points. Comparison of the point addition scheme (a,b,c) and the scheme where we initialize all points to start meshing (d,e,f). (a,d) Computation time of meshing using the three meshing methods against the number of threads. (b,e) The corresponding parallel efficiency against the number of threads for the three meshing methods. (c,f) The worst triangle qualities of the three meshing methods measured by the maximum aspect ratios. \label{fig:Poker_adaptive_parallel_quality}}
\end{figure}

Figure \ref{fig:Poker_adaptive_parallel_quality} shows the comparison of the two point schemes. Figures~\ref{fig:Poker_adaptive_parallel_quality}(c,f) show the maximum aspect ratio (i.e.~worst triangle quality) of the mesh versus the meshing iterations for the three meshing methods. We see that using the point addition scheme, at termination, we obtain significantly lower maximum aspect ratios for all meshing methods. Although for all meshing methods, initializing all points to start meshing tends to terminate earlier with fewer iterations, the worst triangle qualities are much worse. This is expected: according to the highly non-uniform density field, points in the mesh require more movements and adjustments. However, since the geometry is complicated, points cannot move around freely without obstruction. This leads to some badly shaped triangles. On the other hand, using the point addition scheme effectively refines regions of the mesh that need points most during the meshing iterations, and leads to much better shaped triangles conforming to the density field.

Figures~\ref{fig:Poker_adaptive_parallel_quality}(a,d) show the computation time against the number of parallel threads for the two point schemes, and Figs.~\ref{fig:Poker_adaptive_parallel_quality}(b,e) show the corresponding parallel efficiencies. We see that the parallel performance is similar to the case of uniform meshing on a square. CVD meshing has the highest parallel efficiency, DistMesh the lowest, and hybrid meshing in between.

\begin{table}[h!]
\centering
\begin{tabular}{|p{30mm}|c|c|c|c|}
\hline
Point scheme &  \multicolumn{2}{c|}{Point addition scheme} & \multicolumn{2}{c|}{Initialize all points}
\\ \hline
Computation Time (s)&1 thread&28 threads&1 thread&28 threads
\\ \hline
  DistMesh&385.997&21.3186&374.578&20.5796
\\ &(3.47505)&(0.202846)&(4.67893)&(0.267014)
\\ \hline
  CVD meshing&592.178&24.7195&283.954&11.4668
\\ &(4.58964)&(0.191537)&(6.60092)&0.266414)
\\ \hline
  Hybrid method&404.872&24.2899&311.869&11.262
\\ &(3.54899)&(0.202259)&(5.66519)&(0.267662)
\\ \hline
\end{tabular}
\caption{Comparison of the point schemes, using adaptive meshing of the custom shape: computation time of meshing for the three methods, including both the total time, and in parentheses, the time per triangulation iteration, using 1-thread serial code and 28-thread parallel code.\label{tbl:adaptive poker meshing computation time}}
\end{table}

Table~\ref{tbl:adaptive poker meshing computation time} lists the total computation time for the three meshing methods using the serial code, and the parallel code with 28 threads, of the two point schemes. For the point addition scheme with the serial code, CVD meshing is much more expensive than DistMesh and the hybrid method. Since the hybrid method uses DistMesh iterations most of the time, and only uses CVD meshing at the end as refinement steps, DistMesh and the hybrid method are similar in their computation time. All three methods achieve significant speedup in time using the parallel code. With 28 threads, the three methods take a similar amount of time.

\subsubsection{Mesh quality}
\begin{figure}
  \centering
  \includegraphics[width=1\textwidth]{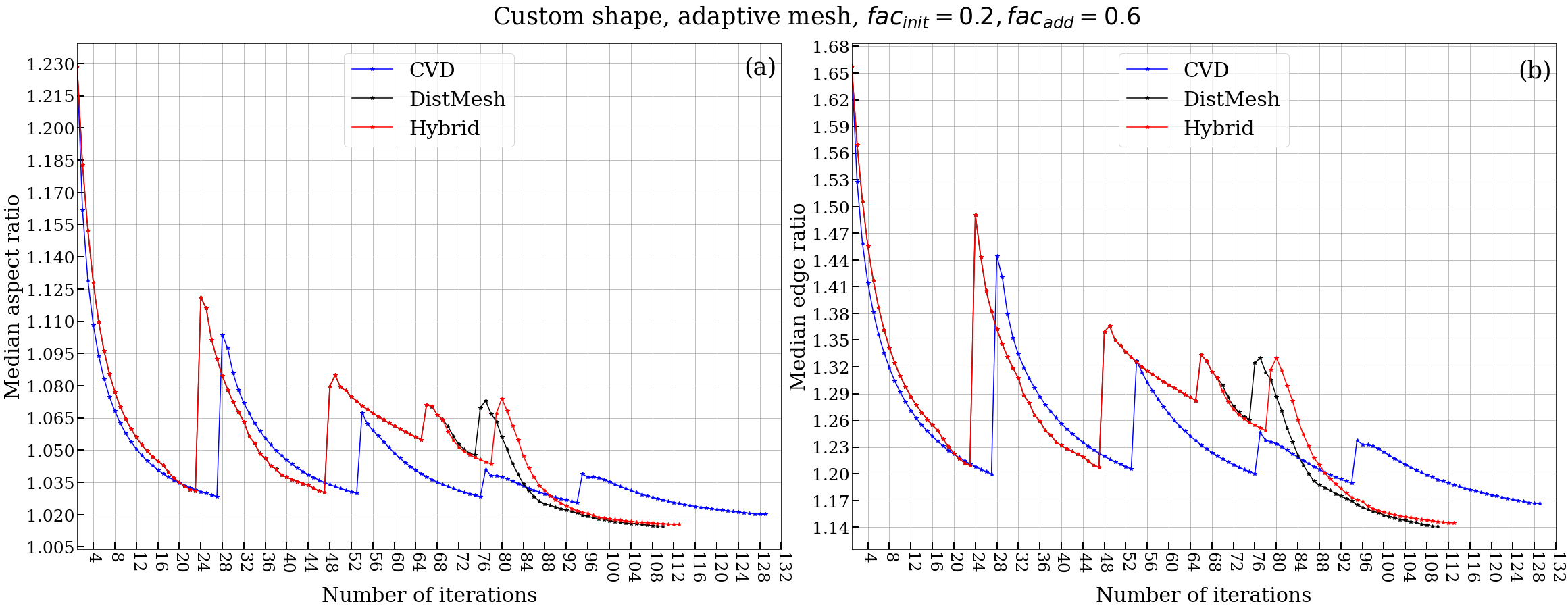}
  \caption{Adaptive meshing on the custom shape, where we use the point addition scheme. (a) Median aspect ratios for the three meshing methods against the number of iterations. (b) Median edge ratios for the three meshing methods against the number of iterations. \label{fig:Poker_adaptive_median_quality}}
\end{figure}

Figures \ref{fig:Poker_adaptive_median_quality}(a,b) show the median aspect and edge ratio against the number of iterations using the point addition scheme for the three meshing methods, respectively. The peaks in the plots correspond to when points are added into the mesh according to the point addition scheme. All three meshing methods achieve very good median ratios at termination.

Table~\ref{tbl:Adaptive poker meshing termination quality stats} shows the mesh quality statistics at termination using the point addition scheme. All three methods achieve good overall mesh quality at termination, with most triangles having $\alpha<2$. The maximum aspect ratio for the mesh generated with DistMesh is much larger than the other two. Generally, through our empirical testing, the needle-type triangle is not a concern for either DistMesh or CVD meshing, as their meshes all have maximum edge ratios at reasonably low values. However, DistMesh may result in extremely flat triangles with very large aspect ratios, while CVD meshing does not have this problem. In all our testing, the hybrid method generates meshes with reasonable maximum aspect ratios similar to that of CVD meshing.

The hybrid method can improve upon the triangulation quality from DistMesh, and like CVD meshing, it tends to avoid extremely flat or needle type triangles. The hybrid method also has a lower computation time similar to that of DistMesh. Therefore, it is especially advantageous in serial meshing, or in parallel meshing with a low number of threads.

\begin{table}
\centering
\footnotesize
\begin{tabular}{|p{22mm}||p{18mm}|p{18mm}|p{18mm}|p{18mm}|p{19mm}|p{19mm}|}
\hline
Mesh quality statistics at termination&Number of triangles&Median $\alpha$, $\beta$&Mean $\alpha$, $\beta$&Maximum $\alpha$, $\beta$&St.\@ dev.\@ from mean for $\alpha$, $\beta$& \% triangles with $\alpha<1.2$, $\alpha<2$
\\ \hline
    DistMesh&1,991,758&1.0147,&1.02693,&14.4941,&0.0376411,&99.41\%,\\
    & &1.14128&1.16165&2.83109&0.101256&100.00\%
\\ \hline
  CVD&1,992,449&1.02015,&1.03668,&8.03026,&0.0464582,&98.48\%,\\
  meshing  & &1.1667&1.18905&3.63211&0.116198&100.00\%
\\ \hline
  Hybrid&1,991,820&1.01547,&1.02944,&3.20321,&0.0390257,&91.75\%,\\
  method  & &1.145&1.16682&5.80708&0.105397&99.93\%
\\ \hline
\end{tabular}
\caption{Adaptive meshing on the custom shape, where we use the point addition scheme: Mesh quality statistics at termination.\label{tbl:Adaptive poker meshing termination quality stats}}
\end{table}

\subsubsection{Parallel performance for different system sizes}
\begin{figure}[h!]
  \centering
  \includegraphics[width=0.6\textwidth]{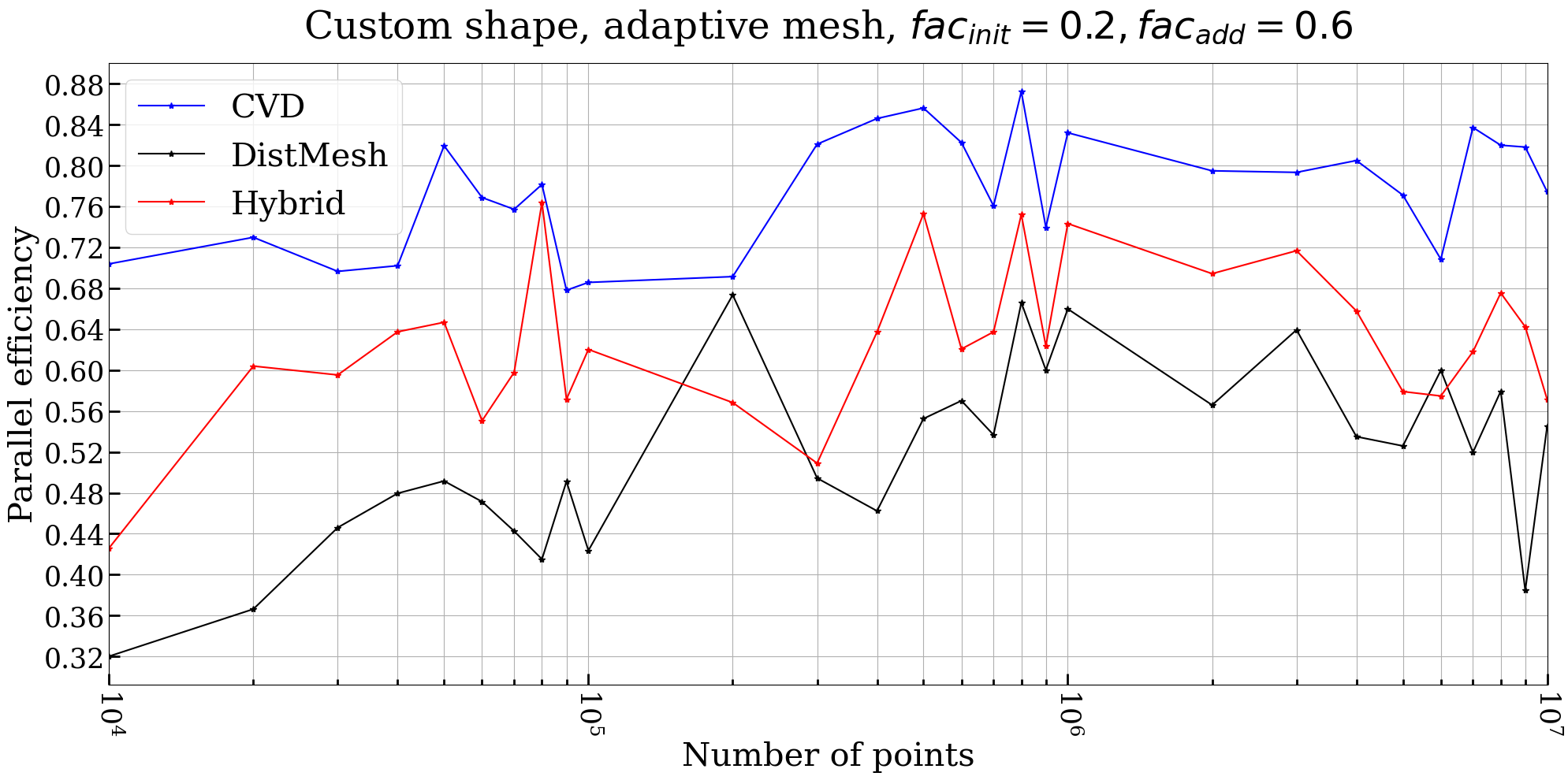}
  \caption{Adaptive meshing on the custom shape, where we use the point addition scheme: Parallel efficiency of the three meshing methods as the number of points increases.\label{fig:Poker_adaptive_varyN}}
\end{figure}

As shown in Fig.~\ref{fig:Poker_adaptive_varyN}, similar to the uniform meshing of square case, we observe similar trend of a gradual increase in parallel efficiency as the system size increases from $10^4$ to $10^6$ particles. Then the efficiency remains at roughly the same level as the system size continues to increase. Furthermore, CVD meshing has the highest efficiency, DistMesh the lowest, and hybrid meshing in between.

\subsection{North America continental map}
\label{sec:performance-map}

\begin{figure}[h!]
  \centering
  \includegraphics[width=1\textwidth]{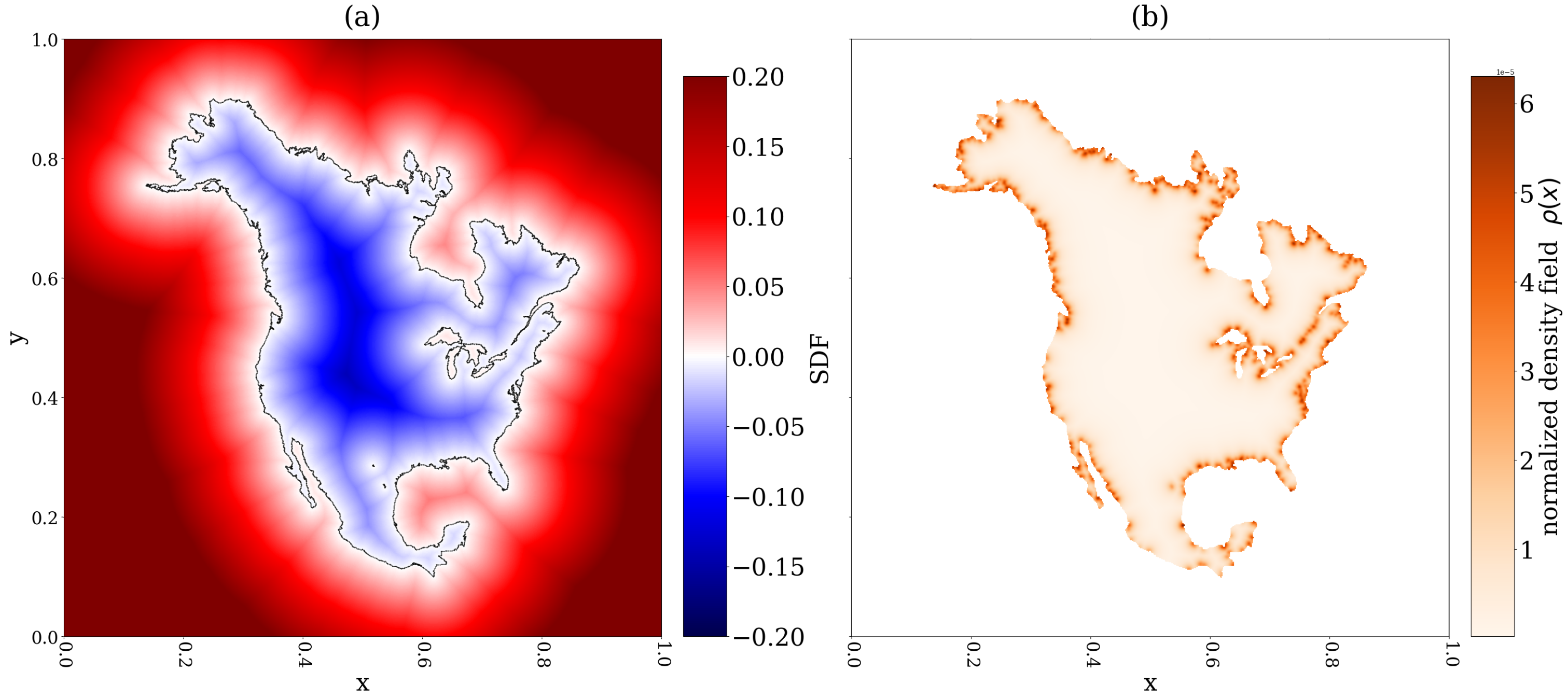}
  \caption{(a) SDF of the North America continent map, with shape boundaries overlaid. (b) Density field for $N=70,000$ points with $K=0.1$.\label{fig:NA_bdry_SDF_density}}
\end{figure}

To demonstrate the capability of \textsc{TriMe++} to handle complicated shapes, we present an example meshing of a map taken from the North American continent. The shape input is taken from an SVG file~\cite{NA_map_SVG_source}. It has very fine resolution and consists of $4,833$ boundary line segments. We use $N=70,000$ and $K=0.1$. Fig.~\ref{fig:NA_bdry_SDF_density} shows the shape boundary, the SDF and the corresponding density field.

With hybrid meshing and $28$ parallel threads, it takes $4.67348$ s to terminate in $127$ triangulation iterations. Table~\ref{tbl:NA meshing termination quality stats} shows the triangle quality statistics at termination. We see that the triangle quality is very good, where $97.59\%$ triangles have $\alpha<1.2$, and $99.99\%$ triangles have $\alpha<2$.

\begin{table}[H]
\centering
\footnotesize
\begin{tabular}{|p{22mm}||p{18mm}|p{18mm}|p{18mm}|p{18mm}|p{19mm}|p{19mm}|}
\hline
Mesh quality statistics at termination&Number of triangles&Median $\alpha$, $\beta$&Mean $\alpha$, $\beta$&Maximum $\alpha$, $\beta$&St.\@ dev.\@ from mean for $\alpha$, $\beta$& \% triangles with $\alpha<1.2$, $\alpha<2$
\\ \hline
  Hybrid&131,710&1.02166,&1.04084,&2.48391,&0.0569425,&97.59\%,\\
  method& &1.17334&1.20061&3.45416&0.132576 &99.99\%
\\ \hline
\end{tabular}
\caption{North America continent map: Mesh quality statistics at termination. \label{tbl:NA meshing termination quality stats}}
\end{table}

\begin{figure}[h!]
  \centering
  \includegraphics[width=1\textwidth]{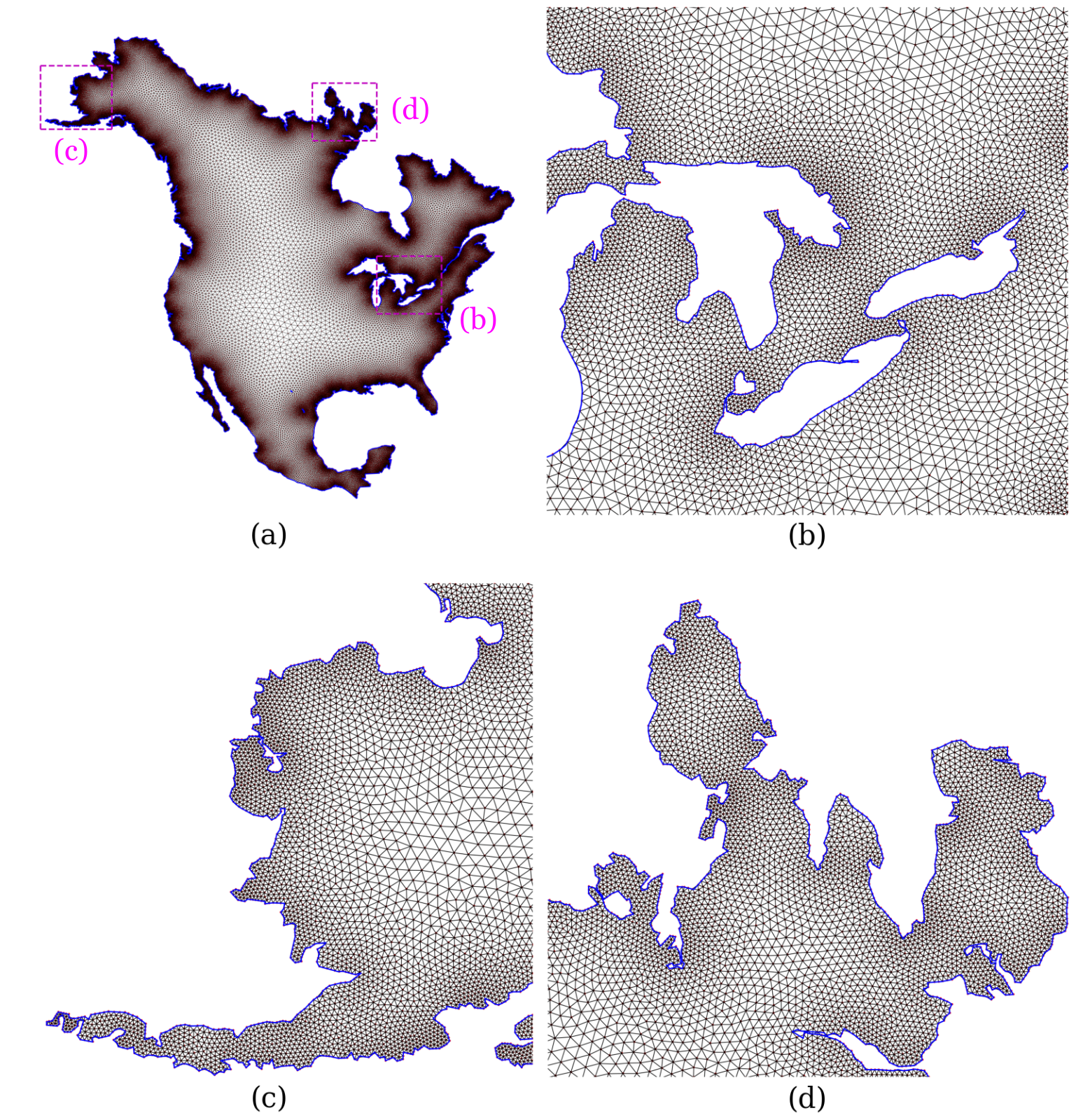}
\caption{(a) Overview of the whole mesh for the North American continent, with shape boundaries overlaid using blue lines. Zoomed-in views of (b) Michigan and Ontario, (c) western Alaska, (d) and Nunavut are also shown, with their locations highlighted on the whole map usings dotted pink lines.\label{fig:NA_triangulation_overview_zoom}}
\end{figure}

Figure \ref{fig:NA_triangulation_overview_zoom} shows the final mesh. The blue lines overlaying on the mesh are the input shape boundaries. Panel (a) shows an overview of the adaptive mesh. Panels (b) to (d) are zoomed-in views at various boundary regions with detailed features. The mesh is able to resolve these small features well, with an error tolerance proportional to the mesh resolution/element sizes in the region.

\section{Comparison with existing software}
\label{sec:comparison with existing software}
Here, we compare the performance of \textsc{TriMe++} with two existing software packages: (I) PyDistMesh~\cite{pydistmesh}, a popular Python implementation of the DistMesh algorithm, and (II) CGAL~\cite{cgal:eb-23b, cgal_website}, a widely used C++ library with efficient geometric algorithms.

\subsection{Comparison with PyDistMesh}
PyDistMesh uses SciPy~\cite{2020SciPy-NMeth} to perform the Delaunay triangulation, which implements the QuickHull algorithm with the Qhull~\cite{qhull_lib} library. PyDistMesh is implemented in serial. 

\begin{figure}[h!]
  \centering
  \includegraphics[width=0.5\textwidth]{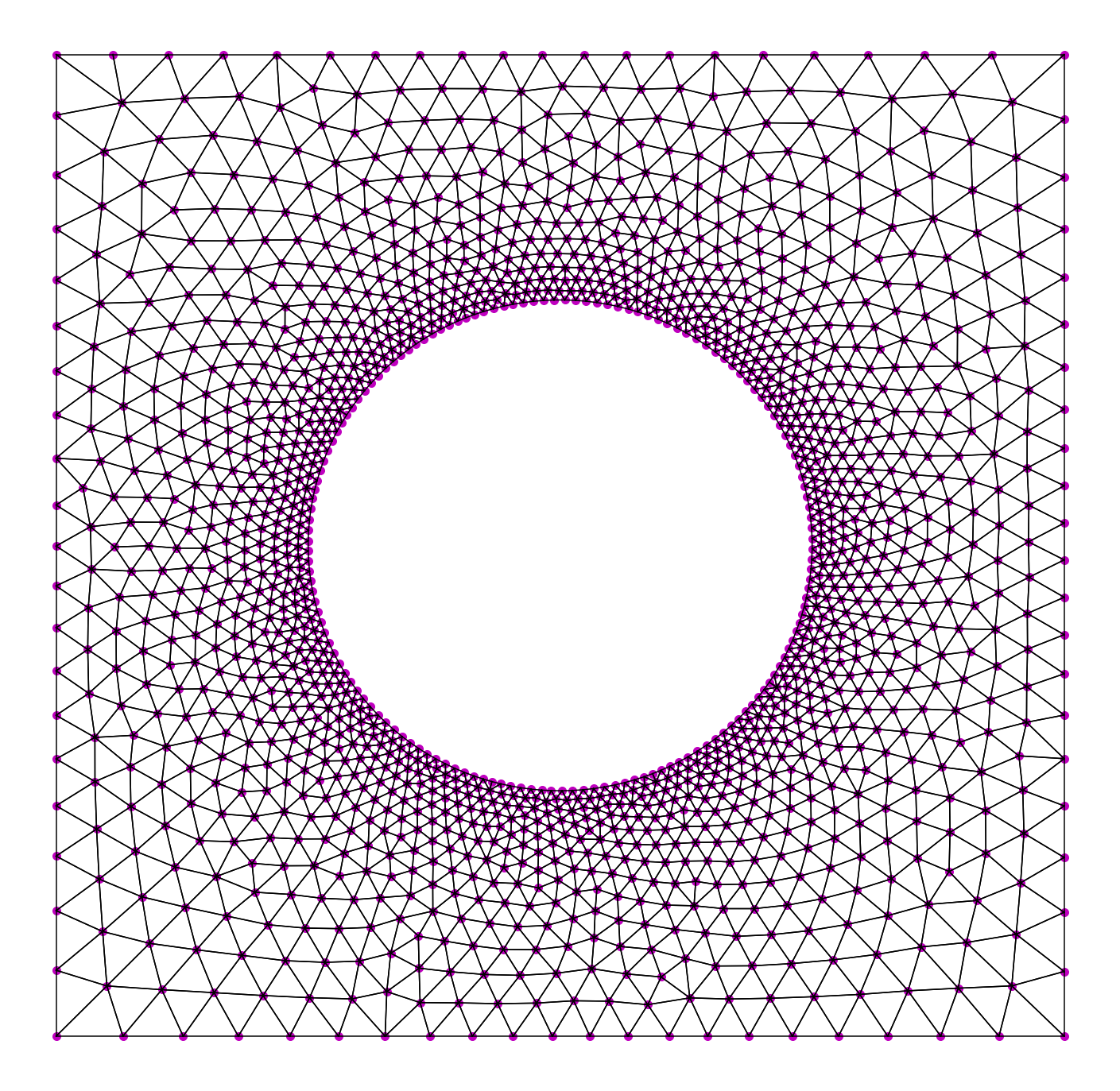}
\caption{Schematic illustration for adaptive meshing of a square with a hole, using $1,603$ points. \label{fig:square_with_hole}}
\end{figure}

Here, we look at a case of adaptive meshing on a square with a hole in the middle. To construct the shape, we first define a square $x,y\in [-1,1]$, and a circle with center $(0,0)$ and radius $0.5$. We then take the difference of the two shapes, using the set operations introduced in Sec.~\ref{sec: Geometry input and shape representation}. The adaptive sizing field is given by the function, $\texttt{fh}=0.05+0.3*\texttt{dcircle}(x,y)$, where $\texttt{dcircle}(x,y)$ is the signed distance field of the middle circle shape. An example mesh of the case using $1,603$ meshing points is shown in Fig.~\ref{fig:square_with_hole}. 

In our tests, PyDistMesh failed to terminate if we do not specify the corners of the square as fixed points. On the other hand, \textsc{TriMe++} can converge without this specification. For consistency of comparison, we specify shape corners as fixed points for both PyDistMesh and \textsc{TriMe++}.
We specify an average target edge length of $\texttt{h0}=0.002$ for PyDistMesh. It generates a mesh with $N_{\text{total}}=\text{156,506}$ points. We then perform comparable meshing with the same number of points using the three meshing methods on \textsc{TriMe++}.

Table~\ref{tbl:uniform square meshing, compare pydistmesh} shows some timing and mesh statistics at termination. For mesh quality, all methods produce very high-quality final meshes, with median $\alpha$ and $\beta$ close to $1$. Their maximum $\alpha$ and $\beta$ are all very reasonable, meaning that the worst triangles in the meshes are already of good shape, close to equilateral triangles. We see PyDistMesh produces slightly better mesh quality. The reason is that it uses significantly more iterations to terminate. In fact, PyDistMesh terminates with $8,105$ total iterations, where $1,191$ are triangulation iterations. In contrast, the DistMesh implementation in \textsc{TriMe++} terminates with $371$ iterations, where $112$ are triangulation iterations. PyDistMesh only uses small point movement as its termination criteria. If we care most about mesh quality, this termination criteria could cause the program to run a lot more iterations than needed, and only get a marginal improvement on mesh quality.

In terms of computation time, for \textsc{TriMe++}, as expected, CVD meshing is more expensive than the other two methods in serial code. Overall, the serial computation time of the three methods is in the range of $48$--$130$\,s. In comparison, parallel \textsc{TriMe++} significantly speeds up all three methods. Their $28$-thread parallel performance are close in computation time, in the range of $3$--$5$\ s. On the other hand, PyDistMesh only runs in serial. PyDistMesh also takes a prohibitively long time to terminate, partly due to the large number of iterations needed to terminate. This results in PyDistMesh takes almost $100$-fold and $2500$-fold longer computation time than serial \textsc{TriMe++} and $28$-thread parallel \textsc{TriMe++}, respectively. 

\begin{table}[h!]
\centering
\begin{tabular}{|p{14mm}|c|c|c|p{20mm}|c|c|}
\hline
 \multicolumn{2}{|c|}{\multirow{2}{*}{Method}} & \multicolumn{2}{c|}{Computation time (s)} & Number of & Median& Maximum
\\
\multicolumn{2}{|c|}{}&1 thread&28 threads & triangles  & $\alpha,\beta$ & $\alpha,\beta$

\\ \hline
\multicolumn{1}{|c|}{\multirow{6}{*}{\textsc{TriMe++}}}& DistMesh&56.063&4.644 & 310,345 & \centering\arraybackslash 1.01, 1.13&1.54, 2.14\\

    \multicolumn{1}{|c|}{}& &  (0.5004) & (0.0380) &  &  &\\

    \multicolumn{1}{|c|}{}&CVD&  126.909 & 4.237 & 310,530 & \centering\arraybackslash 1.02, 1.17 &2.31, 2.93\\

    \multicolumn{1}{|c|}{}&&  (0.6972) & (0.0286) &  &  &\\
    
    \multicolumn{1}{|c|}{}&Hybrid& 48.899 & 3.163 &310,312 & \centering\arraybackslash 1.02, 1.16&1.58, 2.07\\

    \multicolumn{1}{|c|}{}&&  (0.4362) & (0.0326) &  &  &\\
    
    \hline

\multicolumn{2}{|c|}{\multirow{2}{*}{PyDistMesh}}&12,418.732& -- & 310,658 & \centering\arraybackslash  1.01, 1.12& 1.41, 1.58\\ 
  
  \multicolumn{2}{|c|}{}& (10.4271)&  & &  &  \\
  \hline
\end{tabular}
\caption{Comparison of \textsc{TriMe++} and PyDistMesh. Adaptive meshing on a square with a hole in the middle, using $156,506$ points. We present wall clock computation time, and in parenthesis, computation time per triangulation iteration. We also present mesh statistics of a typical final mesh. \label{tbl:uniform square meshing, compare pydistmesh}}
\end{table}

\subsection{Comparison with CGAL}
In 3D, CGAL offers many advanced features for tetrahedral meshing, including geometry adaptivity and parallelization. However, in 2D, CGAL only offers serial meshing. Here, we look at a test case of uniform meshing on a square, defined on $x,y\in [0.1,0.9]$.

\begin{figure}[h!]
  \centering
  \includegraphics[width=0.8\textwidth]{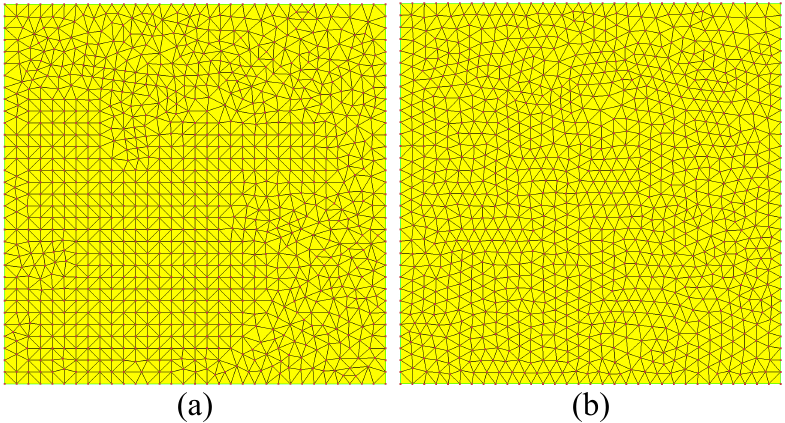}
\caption{CGAL meshing schematic illustration: (a) A uniform mesh on the square $x,y\in [0.1,0.9]$ generated by the mesh refinement algorithm, using constraints $B=\sqrt{2}$ and $L=0.04$.  (b) We perform $64$ iterations of the Lloyd's algorithm for CVD meshing on the mesh in (a), and obtain a mesh whose triangle angles are optimized, and the triangles are much closer to equilateral triangles. \label{fig:refinement_VS_lloyd_mesh}}
\end{figure}

CGAL implements functionalities to construct a quality mesh using the mesh refinement algorithm by Schewchuk~\cite{shewchuk2000, shwechuk2002_delaunay_refinement}. The mesh refinement can take in two constraints. The first constraint is a bound $B$, such that
\begin{equation}
    \frac{R_{\text{circum}}}{l_{\text{min}}}\leq B.
\end{equation}
$B$ is related to the minimum angle of the triangle $\theta_{\text{min}}$ by
\begin{equation}
    \sin{\theta_{\text{min}}}=\frac{1}{2B}.
\end{equation}
Setting $B=\sqrt{2}$ corresponds to bounding $\theta_{\text{min}}\geq \ang{20.7}$, and this is the best bound to use that guarantees the termination of the refinement algorithm. Another constraint is the maximum edge length, $L$, such that $l_{\text{max}}<L$. Figure~\ref{fig:refinement_VS_lloyd_mesh}(a) shows an example uniform mesh on the square, generated from the mesh refinement algorithm, using $B=\sqrt{2}$ and $L=0.04$. The mesh contains $1,074$ points. The triangles in the mesh have good qualities in terms of sizes and minimum angles. However, their angles can be further optimized for the triangles to be as close to equilateral triangles as possible.

CGAL offers a global optimization method that runs Lloyd's algorithm~\cite{du1999centroidal,lloyd82} for CVD meshing to further improve the quality of the generated refined mesh. In their implementation, the shape boundary points are fixed, and only the interior points move. Furthermore, there is no way to input our own mesh sizing field. Instead, the mesh sizing field is automatically constructed based on the point density of the refined mesh before optimization. As shown in Fig.~\ref{fig:refinement_VS_lloyd_mesh}(b), after performing $64$ iterations of the Lloyd's algorithm on the original mesh in Fig.~\ref{fig:refinement_VS_lloyd_mesh}(a), we obtain a mesh whose triangles are much closer to equilateral triangles.

Next, we compare CGAL's two-step meshing procedure (refinement, optimization) with the three meshing methods of \textsc{TriMe++}. Setting CGAL's refinement algorithm constraints to be $B=\sqrt{2}$ and $L=0.002$, we obtain an initial mesh with $443,357$ points. We then perform Lloyd's iterations on the initial mesh. We set the termination criteria of the Lloyd's optimization to be comparable to the mesh quality termination criteria of \textsc{TriMe++}, requiring the relative changes in the $\frac{1}{2}$-mean of $\alpha$ and $\beta$ to be smaller than threshold $T^{\text{end}}_{\text{quality}}$, and the relative change in the maximum $\alpha$ to be smaller than $T^{\text{end}}_{\alpha_{\text{max}}}$.

Table~\ref{tbl:uniform square meshing, compare cgal} shows the comparison of CGAL with \textsc{TriMe++} using the same number of points on the case of uniform meshing of a square. We see that all methods generate high quality mesh, with median $\alpha$, $\beta$ very close to $1$, and very reasonale and small maximum $\alpha$ and $\beta$. The biggest difference is in computation time. We observe that the Lloyd's optimization procedure in CGAL takes a prohibitively long time for large-scale mesh. The per-iteration optimization time, shown in parentheses, is significantly higher in CGAL than in \textsc{TriMe++}'s CVD meshing. Therefore, although both methods use a similar number of Lloyd's iterations, \textsc{TriMe++}'s CVD meshing terminates much faster. Furthermore, CGAL only offers serial meshing, which takes hours to terminate. In comparison, using $28$-thread parallel code, all three meshing methods of \textsc{TriMe++} terminate in seconds.

\begin{table}[h!]
\centering
\small
\begin{tabular}{|p{14mm}|c|c|c|p{20mm}|c|c|}
\hline
 \multicolumn{2}{|c|}{\multirow{2}{*}{Method}} & \multicolumn{2}{c|}{Computation Time (s)} & Number of & Median& Maximum
\\
\multicolumn{2}{|c|}{}&1 thread&28 threads & triangles  &$\alpha,\beta$ & $\alpha,\beta$

\\ \hline
\multicolumn{1}{|c|}{\multirow{6}{*}{\textsc{TriMe++}}}& DistMesh&63.6192&5.07552 & 884,219 & 1.01, 1.13 &1.48, 1.72\\
\multicolumn{1}{|c|}{}&& (1.7166) & (0.1370) &  &  &\\

    \multicolumn{1}{|c|}{}&CVD& 79.5313 & 3.75685 & 884,178 & 1.02, 1.15 &1.82, 1.82\\
\multicolumn{1}{|c|}{}&& (2.2710) & (0.1072) &  &  &\\

    \multicolumn{1}{|c|}{}&Hybrid& 69.1126 & 4.25905 &884,239 & 1.02, 1.15&1.64, 1.87 \\
    \multicolumn{1}{|c|}{}&& (1.9166) & (0.1181) &  &  &\\

    \hline

  \multicolumn{1}{|c|}{\multirow{4}{*}{CGAL}}& \textit{I. Refinement}&56.963&\centering\arraybackslash -- &  & &\\
    \multicolumn{1}{|c|}{}&\textit{II. Optimiza-}& 22,861.8 & \centering\arraybackslash -- &  & & \\
    \multicolumn{1}{|c|}{}&\textit{tion}  & (557.605) &\centering\arraybackslash --  &  &  &\\

    \multicolumn{1}{|c|}{}&\textbf{Total}& \textbf{22,918.7} &\centering\arraybackslash -- &\textbf{884,516} &\textbf{1.02, 1.17} &\textbf{1.80, 2.20}\\
    \hline
\end{tabular}
\caption{Comparison of \textsc{TriMe++} and CGAL. Uniform meshing on a square with $443,357$ points: computation time and mesh statistics of a typical final mesh. Here we list both total computation time and, in parenthesis, computation time per triangulation iteration. \label{tbl:uniform square meshing, compare cgal}}
\end{table}

\section{Conclusion and future work}
\label{sec:conclusion_future_work}
We have developed an easy-to-use 2D meshing software, \textsc{TriMe++}, that exploits parallelization via the multi-threaded version of \vpp{}. The parallelization is efficient and results in significant speedup in time for large-scale mesh generation.

We also compared different meshing algorithms implemented in the software, DistMesh, CVD meshing, and a hybrid method of the two. DistMesh is cheaper in serial time, but has lower parallel efficiency, and may result in extremely flat triangles. On the other hand, CVD meshing is computationally more expensive, but has higher parallel efficiency. Moreover, CVD meshing has theoretical support on mesh quality, and tends to result in high-quality meshes free of extremely bad-shaped triangles. We then developed a hybrid of the two methods, which uses DistMesh most of the time, and uses CVD meshing as a refinement method for the last few iterations before termination. The hybrid method tends to have higher mesh quality like CVD meshing and has a cheaper serial computation time like DistMesh.

For future work, we may improve on the current construction of the geometry grid, described in Sec.~\ref{sec: Underlying geometry grid}. In general, the time it takes is small. However, if the shape inputs are a large number of very fine boundary line segments, and the meshing system is large (eg. millions of points), then the geometry grid is very fine and is expensive to compute. 
One potential improvement is to take a multi-layered approach. We begin with a coarse geometry grid, and determine its inner/boundary/outer cells. We then refine the boundary cells, and categorize the refined cells. Repeating the process, we can obtain the desired fine geometry grid, without calculating SDF for every one of its grid cells.

We can also extend the CVD meshing algorithm for distributed parallel computation on computer clusters via the Message Passing Interface (MPI) library~\cite{gropp2014,pacheco1997}. As each point's Voronoi cell centroid can be computed independently from other points, it is well-suited for distributed computing. Suppose the computation domain is divided to different computers. In a meshing iteration, suppose the processes hold information of the points in their domains, and information of the corresponding Voronoi cells, then each process can update their points' positions independently from other computers.

Another potential extension is parallel meshing using GPU. Bernaschi et al.~\cite{bernaschi17} proposed and implemented a GPU-parallelized procedure to construct approximated Voronoi diagrams and the corresponding dual Delaunay triangulation for a set of moving points. Although the work was originally developed to study plastic events in flow of confined emulsions, the methodology is general. We could implement CVD meshing using the procedure and perform GPU-parallelized meshing.

We can extend \textsc{TriMe++} to 3D meshing. We can implement DistMesh for its time saving advantage. For mesh quality control, instead of CVD meshing, we can implement an alternative variational tetrahedral meshing algorithm proposed by Alliez et al.~\cite{alliez2005}, that minimizes an energy $E_{\text{ODT}}$. This is because in 3D, there is no theoretical support that a CVD corresponds to a dual Delaunay mesh that has nearly regular tetrahedrons~\cite{eppstein2001,alliez2005}. In particular, slivers~\cite{Cheng2000Sliver} can appear in 3D CVD meshing. Alliez et al.~\cite{alliez2005} showed that the $E_{\text{ODT}}$-variational meshing algorithm leads to the elimination of slivers in the interior of the geometry domain. After the iterations terminate, they use a boundary vertex jittering method to eliminate slivers that may exist for geometry boundary vertices. Like CVD meshing, the $E_{\text{ODT}}$-variational meshing is ideal for parallel computation, where each point can be updated independently from others. Therefore, we expect this method to achieve high parallel efficiency.

As in 2D, we can develop a hybrid method in 3D. We can use DistMesh to obtain an overall high-quality mesh in a short time, and then use the $E_{\text{ODT}}$-variational meshing near termination as refinement steps to eliminate slivers in the interior of the mesh. Lastly, we can implement boundary vertex jittering method on the final mesh to eliminate slivers near geometry boundary. We can design a hybrid method that is fast and has theorectical support on tetrahedron qualities.

\section*{Acknowledgements}
This research was supported by a grant from the United States--Israel Binational
Science Foundation (BSF), Jerusalem, Israel through grant number 2018/170.
C.~H.~Rycroft was partially supported by the Applied Mathematics
Program of the U.S. DOE Office of Advanced Scientific Computing Research under
contract number DE-AC02-05CH11231.

\begin{appendices}
\counterwithin{figure}{section}

\section{Generation of geometry boundary approximation points}
\label{appendix:bdry pt generation}
As discussed in Sec.~\ref{sec:element sizing field}, the first step in generating the element sizing field is to generate a set of boundary approximation points, $\mathcal{S}=\{\vec{s}_1,\vec{s}_2,\ldots\}$. We want $\vec{s}\in \mathcal{S}$ to be roughly evenly spaced according to the mesh resolution, and to capture detailed features of the shape well. Our approach is to first generate a dense set of boundary points, to make sure that all detailed features of the shape are captured. Then we go through a trimming process of this initial set of boundary points, until the remaining points are at least $d_{\vec{s}}=\textit{fac}_{\vec{s}}\cdot\min(\Delta x^{\text{geo}},\Delta y^{\text{geo}})$ away from its closest neighbors on the same boundary, where $\Delta x^{\text{geo}}$ and $\Delta y^{\text{geo}}$ are edge lengths of the geometry grid cells.

\begin{figure}
  \centering
  \includegraphics[width=0.7\textwidth]{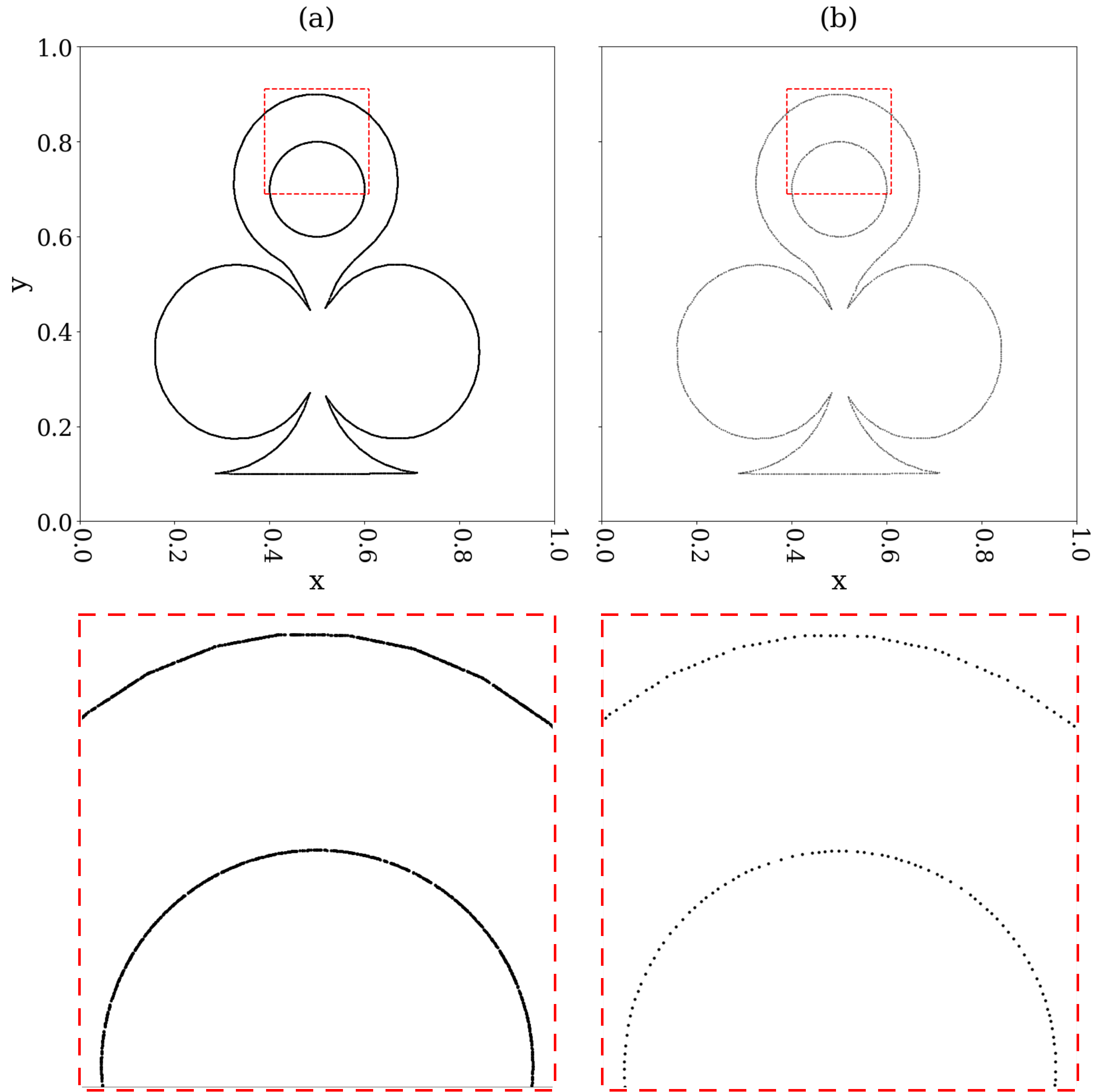}
  \caption{The overview of boundary points and the enlarged view of points in region $x\in[0.39, 0.61]$, $y\in[0.69,0.91]$ for (a) the initial set of dense geometry boundary points, generated with $N_{\text{total}}=5,000$; (b) the trimmed set of geometry boundary points from (a). ~\label{fig:bdry_pt_trim}}
\end{figure}

\begin{figure}
  \centering
  \includegraphics[width=0.3\textwidth]{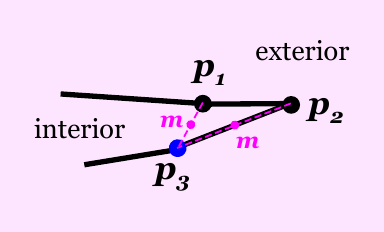}
  \caption{When finding the valid closest point for a boundary point, we check their midpoint, $\vec{m}$, and require that $\lvert \sdf(\vec{m})\rvert\leq \hat{\geps}$ for the closest point to be valid. That is, we require $\vec{m}$ to be on the boundary as well. Otherwise, the closest point may not be on the same geometry boundary feature and is invalid. Here, $\vp_1$ is an invalid closest point to $\vp_3$, as their midpoint is not on the geometry boundary, and they are not on the same geometry feature. Instead, $\vp_2$ is the valid closest point to $\vp_3$. ~\label{fig:bdry_pt_closest_nei_midpoint_check}}
\end{figure}

If the geometry input are the shape contour line segments, we can generate the initial set of boundary points by using all the line segment end points, and adding evenly spaced points along line segments that have length larger than $d_{\vec{s}}$, so that the distance between neighboring added points is smaller than $d_{\vec{s}}$.
If we only have the geometry SDF, and no boundary points are readily available, we can utilize the underlying geometry grid to generate the dense set of initial boundary points efficiently. We first generate $n_{\text{grid}}$ points uniformly in each boundary grid cell. These points are close to the geometry boundary, and they may be inside/outside/on the boundary. We test whether each point is on the boundary by comparing its SDF with a geometry boundary tolerance dependent on the mesh resolution, $\hat{\geps}=\textit{fac}_{\geps}\cdot\min(\Delta x^{\text{geo}},\Delta y^{\text{geo}})$. If a point is not on the boundary, we project it onto the boundary, using the point projection procedure detailed in Sec.~\ref{sec: new points treatment}~\ref{point projection method}, with a fixed geometry boundary tolerance, $\hat{\geps}$, and a fixed finite difference step size for approximating derivatives, $\hat{\deps}=\sqrt{\epsilon}\cdot\min(\Delta x^{\text{geo}},\Delta y^{\text{geo}})$, where $\epsilon$ is the machine precision. As a result, we obtain a dense set of initial boundary points. For the custom shape, an example initial set of boundary points is shown in Fig.~\ref{fig:bdry_pt_trim}(a), generated with $N_{\text{total}}=5,000$.

Next, we trim the initial set of boundary points. The points are sorted in the geometry grid. We loop through the boundary points. For the current boundary point $\vp_i$, we first find its closest point $\vp^*$. We use an efficient outward grid searching procedure, so we only need to check points close to $\vp_i$. We then find the midpoint $\vec{m}$ of $\vp_i$ and $\vp^*$. The closest point is valid only if $\lvert \sdf(\vec{m})\rvert\leq \hat{\geps}$. Otherwise, we treat $\vp^*$ as if it is not on the same geometry boundary feature as $\vp_i$, and therefore invalid. In that case, we find the next closest point, and continue until a valid $\vp^*$ is found. This is illustrated in Fig.~\ref{fig:bdry_pt_closest_nei_midpoint_check}, where $\vp_1$ would be an invalid closest point to $\vp_3$, and $\vp_2$ would be the valid one. Once we determine the valid closest point $\vp^*$, we can test whether it is less than $d_{\vec{s}}$ distance away from $\vp_i$. If so, we delete $\vp^*$, and move $\vp_i$ to the midpoint position $\vec{m}$. If $\vp^*$ is already at least $d_{\vec{s}}$ away, we are done with $\vp_i$, and there is no need to check $\vp_i$ in future iterations. We continue on to the next boundary point, $\vp_{i+1}$, with this procedure of finding and checking its closest point, until all points are checked in the loop. The loop is repeated in iterations until all boundary points have valid closest points at least $d_{\vec{s}}$ distance away. For the custom shape, an example trimmed set of boundary points is shown in Fig.~\ref{fig:bdry_pt_trim}(b), and we can see that compared to Fig.~\ref{fig:bdry_pt_trim}(a), it is more evenly spaced according to the resolution of the mesh desired. We use the trimmed set of boundary points in Sec~\ref{sec:element sizing field} to compute the medial axis approximation points and then the element sizing field.

\section{Symbols and control parameters}
\label{appendix: control parameters}

We provide a summary list (Table~\ref{tbl:meshing_parameters}) of the symbols used in the paper, as well as the control parameters used in the meshing pipeline. For the control parameters, their recommended default values are listed in the last column.

\begin{longtable}{|p{20mm}|p{25mm}|p{90mm}|p{13mm}|}
    \hline
    Sec.~\ref{sec:intro-DistMesh} & $N$ & Number of points & --
    \\ \hline
    Sec.~\ref{sec:intro-DistMesh} & $\vp$ & $\vp=(\vx,\vy)$ where $\vp\in \R^{N\times 2}$ and $\vx,\vy\in \R^N$. Point positions in matrix form in the DistMesh algorithm.  & --
    \\ \hline
    Sec.~\ref{sec:intro-DistMesh} & $\vF_\text{int}(\vp)$ & Net internal repulsive forces from the springs & --
    \\ \hline
    Sec.~\ref{sec:intro-DistMesh} & $\vF_\text{ext}(\vp)$ & Net external normal forces from the geometry boundaries & --
    \\ \hline
    Sec.~\ref{sec:intro-DistMesh} & $\vF(\vp)$ & Net force experienced by the points & --
    \\ \hline
    Sec.~\ref{sec:intro-DistMesh} &$f(l,l_0)$& Internal repulsive spring force for a triangle edge with length $l$ and desired edge length $l_0$& --
    \\ \hline
    Sec.~\ref{sec:intro-DistMesh} &$k$ & The spring stiffness parameter in calculating $f(l,l_0)$ & --
    \\ \hline
    Sec.~\ref{sec:intro-DistMesh} &$l_0$& The desired edge length & --
    \\ \hline
    Sec.~\ref{sec:intro-DistMesh} &$l_i$& The length of edge $i$ & --
    \\ \hline
    Sec.~\ref{sec:intro-DistMesh} &$\textit{fac}_F$ & A factor slightly larger than $1$ that scales the desired edge length in the DistMesh algorithm, so that points experience repulsive forces and spread out across the domain & 1.2
    \\ \hline
    Sec.~\ref{sec:intro-DistMesh} & $\mu(\vec{x})$ & Element size function. It gives the relative edge length distribution over the domain. & --
    \\ \hline
    Sec.~\ref{sec:intro-DistMesh} & $M$& Number of edges & --
    \\ \hline
    Sec.~\ref{sec:intro-DistMesh} & $\vec{m}_i$ & Edge midpoint for edge $i$, where $i=1,\dots , M$& --
    \\ \hline
    Sec.~\ref{sec:intro-DistMesh} & $\textit{fac}_{\mu}$ & $\textit{fac}_{\mu}=\left(\sum_{i=1}^{M} l_i^2/\sum_{i=1}^{M}{\mu(\vec{m}_i)^2}\right)^\frac{1}{2}$, a scaling factor in calculating the desired edge length $l_{0,i}=\mu(\vec{m}_i)\cdot \textit{fac}_F\cdot\textit{fac}_{\mu}$ for edge $i$. This factor takes into account the actual element sizes of the mesh as well as the relative desired edge length distribution. It is a scaling factor to compute the actual desired edge length distribution. & --
    \\ \hline
    Sec.~\ref{sec:intro-DistMesh} &$t$ & Time variable in the ODE system constructed for the DistMesh algorithm &--
    \\ \hline
    Sec.~\ref{sec:intro-DistMesh} &$\Delta t$ & Time step size in the forward Euler method to solve the ODE in the DistMesh algorithm &--
    
    \\ \hline\hline
    Sec.~\ref{sec:intro-CVD} & $V_i$ & The Voronoi cell corresponding to point $i$ &--
    \\ \hline
    Sec.~\ref{sec:intro-CVD} & $\rho(\vx)$ & Particle density field &--
    \\ \hline
    Sec.~\ref{sec:intro-CVD} & $\vec{c}_i$ & The weighted Voronoi cell centroid of point $i$ &--
    \\ \hline
    Sec.~\ref{sec:intro-CVD} & $E_{\text{CVD}}$ & The quadratic energy of the Voronoi diagram that is minimized by a CVD, which optimizes the compactness of the Voronoi cells&--
    
    \\ \hline\hline
    Sec.~\ref{sec: Geometry input and shape representation} & SDF & Signed distance field & --
    \\ \hline
    Sec.~\ref{sec: Geometry input and shape representation} & $\vec{n}$ & Inward-pointing normal of shape & --
    \\ \hline
    Sec.~\ref{sec: Geometry input and shape representation} & $L_{12}$ & Line segment from $\vp_1=(x_1,y_1)$ to $\vp_2=(x_2,y_2)$ & --
    \\ \hline
    Sec.~\ref{sec: Geometry input and shape representation} & $\vp^*$ & Closest point of $\vp^*$ on line segment $L_{12}$ & --
    \\ \hline
    Sec.~\ref{sec: Geometry input and shape representation} & $\Delta x , \Delta y$ & The horizontal and vertical side lengths of each grid cell &--
    \\ \hline
    Sec.~\ref{sec: Geometry input and shape representation} & $\mathcal{B}$ & Used in the efficient scheme to find the closest line segment for a point. $\mathcal{B}$ is the square grid region looping outwards from the point, where a line segment is first encountered. & --
    \\ \hline
    Sec.~\ref{sec: Geometry input and shape representation} & $\mathcal{C}$ & A bounding circle of $\mathcal{B}$ & --
    \\ \hline
    Sec.~\ref{sec: Geometry input and shape representation} & $l$ & The number of grid layers of $\mathcal{B}$ & --
    \\ \hline
    Sec.~\ref{sec: Geometry input and shape representation} & $R$ & Radius of $\mathcal{C}$, $R=\sqrt{(l\Delta x)^2+(l\Delta y)^2}$ & --
    \\ \hline
    Sec.~\ref{sec: Geometry input and shape representation} & $l_x, l_y$ & Length and height of the domain & --
    \\ \hline
    Sec.~\ref{sec: Geometry input and shape representation} & $\bar{l}_{\text{in}}$ &  The average length of the input line segments & --
    \\ \hline
    Sec.~\ref{sec: Geometry input and shape representation} & $n_x, n_y$ & Horizontal and vertical dimension of the grid used in the efficient scheme to find the closest line segment for a point, $n_x=\lceil l_x/\bar{l}_{\text{in}}\rceil$ and $n_y=\lceil l_y/\bar{l}_{\text{in}}\rceil$ & --

    \\ \hline\hline
    Sec.~\ref{sec: Underlying geometry grid} & $\geps_i$ & A geometry grid is generated based on the SDF. This quantity is a local geometry boundary tolerance defined on each boundary cell $i$. & --
    \\ \hline
    Sec.~\ref{sec: Underlying geometry grid}& $N_{\text{total}}$ & Total number of meshing points & --
    \\ \hline
    Sec.~\ref{sec: Underlying geometry grid}& $a_x, b_x$, $a_y, b_y$& Bounds of the rectangular domain for meshing, $[a_x,b_x]\times[a_y,b_y]$ & --
    \\ \hline
    Sec.~\ref{sec: Underlying geometry grid}&$\textit{fac}_{\text{grid}}$&Factor used to scale the resolution of the base gird to construct the underlying geometry grid &5
    \\ \hline
    Sec.~\ref{sec: Underlying geometry grid}& $\lambda$ & A parameter used in calculating the size of the geometry grid, $\lambda=\sqrt{N_{\text{total}}/(N_{\text{opt}}(b_x-a_x)(b_y-a_y))}$& --
    \\ \hline
    Sec.~\ref{sec: Underlying geometry grid}& $n_x, n_y$ & $n_x=\lceil \lambda (b_x-a_x) \rceil$ and $n_y=\lceil \lambda (b_y-a_y) \rceil$ & --
    \\ \hline
    Sec.~\ref{sec: Underlying geometry grid}&$N_{\text{opt}}$&If $N_{\text{total}}$ points are homogeneously distributed in the domain, this is the average number of points in a grid cell, for a grid with dimension $n_x\times n_y$ &3.3
    \\ \hline
    Sec.~\ref{sec: Underlying geometry grid}& $n_x^{\text{geo}}, n_y^{\text{geo}}$ & Size of the geometry grid in the horizontal and vertical dimension, $n_x^{\text{geo}}=\textit{fac}_{\text{grid}}\cdot n_x$ and $n_y^{\text{geo}}=\textit{fac}_{\text{grid}}\cdot n_y$ & --
    \\ \hline
    Sec.~\ref{sec: Underlying geometry grid}& $\vec{m}$ & Midpoint of the geometry grid cell & --
    \\ \hline
    Sec.~\ref{sec: Underlying geometry grid}&  $\sdf(\vec{m})$ & SDF of $\vec{m}$, used to categorize whether the grid cell is inner, boundary or outer grid cell & --
    \\ \hline
    Sec.~\ref{sec: Underlying geometry grid}& $\Delta x^{\text{geo}}$, $\Delta y^{\text{geo}}$ & A geometry grid cell's dimensions in the horizontal and vertical directions & --
    \\ \hline
    Sec.~\ref{sec: Underlying geometry grid}& $l_\text{diag}$ &  A criterion used to define the boundary cells, $l_\text{diag}=\sqrt{(\Delta x^{\text{geo}})^2+(\Delta y^{\text{geo}})^2}$ & --
    
    \\ \hline \hline
    Sec.~\ref{sec:element sizing field}&$\lfs(\vec{x})$&Local feature size at a point $\vec{x}$ in the domain, capturing both local curvature and local thickness of the shape ($\lfs:\R^2 \to \R$) &--
    \\ \hline
    Sec.~\ref{sec:element sizing field}&$\mu(\vec{x})$&The sizing field value at a point $\vec{x}$ in the domain &--
    \\ \hline
    Sec.~\ref{sec:element sizing field}&$\Omega$&The geometry domain&--
    \\ \hline
    Sec.~\ref{sec:element sizing field}&$\delta \Omega$&The geometry boundary &--
    \\ \hline
    Sec.~\ref{sec:element sizing field}&$\vec{s}$&A boundary point $\vec{s}\in \delta \Omega$&--
    \\ \hline
    Sec.~\ref{sec:element sizing field}&$d(\vec{s},\vec{x})$&The distance from $\vec{x}$ to a boundary point $\vec{s}$&--
    \\ \hline
    Sec.~\ref{sec:element sizing field}&$K$& A parameter controlling mesh gradation, which determines how fast triangle elements vary in size locally &--
    \\ \hline
    Sec.~\ref{sec:element sizing field}&$\vec{s}_i$& A boundary approximation point &--
    \\ \hline
    Sec.~\ref{sec:element sizing field}&$\mathcal{S}$& The set of boundary approximation points, $\mathcal{S}=\{\vec{s}_1,\vec{s}_2,\ldots\}$ &--
    \\ \hline
    Sec.~\ref{sec:element sizing field}& $d_{\vec{s}}$ & The minimum distance between adjacent boundary points on the same boundary. The boundary approximating points should be roughly evenly spaced according to this distance. &--
    \\ \hline
    Sec.~\ref{sec:element sizing field}&$\vec{v}_1$&For each boundary point $\vec{s}$ and its Voronoi cell, $\vec{v}_1$ is the Voronoi vertex farthest away from $\vec{s}$. Used as a medial axis approximating point. &--
    \\ \hline
    Sec.~\ref{sec:element sizing field}&$\vec{v}_2$&This is the farthest away Voronoi vertex in the half space not containing $\vec{v}_1$. Used as a medial axis approximating point.&--
    \\ \hline
    Sec.~\ref{sec:element sizing field}&$r_{\text{nei}}$&For a medial axis approximating point, this is an adaptive distance to search for neighboring medial axis approximating points. Used in the trimming procedure of the medial axis approximating points. &--
    \\ \hline
    Sec.~\ref{sec:element sizing field}&$n_{\text{nei}}^{\text{thres}}$&Number of neighbors of a valid medial axis approximation point should be larger than this threshold&3
    \\ \hline
    Sec.~\ref{sec:element sizing field}&$\textit{fac}_{\text{nei}}$&Factor used to calculate the distance to search for neighboring points, $r_{\text{nei}}=\textit{fac}_{\text{nei}}\cdot n_{\text{nei}}^{\text{thres}}\cdot d_{\vec{s}}$ &2

    \\ \hline
    Sec.~\ref{sec:element sizing field}&$\vec{m}$&The midpoint of a grid cell of the geometry grid. It is used to calculate the local sizing value of the grid cell. &--
    
\\ \hline \hline
    Sec.~\ref{sec:density_field}&$\rho(\vec{x})$&The density field value at a point $\vec{x}$&--

\\ \hline \hline
    Sec.~\ref{step:point init}&$N_\text{init}$&The number of meshing points at initialization&--
    \\ \hline
    Sec.~\ref{step:point init}&$\vec{m}$&Midpoint of a grid cell of the geometry grid&--
    \\ \hline
    Sec.~\ref{step:point init}&$\eta$&Factor used during point initialization to obtain a subset of boundary grid cells that are in a narrower band along the boundary &0.5
    \\ \hline
    Sec.~\ref{step:point init}&$\rho_i^{\text{norm}}$&The normalized density value at a grid cell $i$ &--
    \\ \hline
    Sec.~\ref{step:point init}&$n_i^{\text{E}}$& Used in meshing point initialization. The expected number of points to generate in a grid cell $i$; $n_i^{\text{E}}=\rho_i^{\text{norm}} \cdot N_{\text{init}}$.&--
    \\ \hline
    Sec.~\ref{step:point init}&$R$&The residual amount of the expected number of points to generate in a grid cell and the actual number of points generated &--
    \\ \hline
     Sec.~\ref{step:point init}&$m_{\text{nei}}$&The number of valid neighbor grid cells (i.e.\@ inner or selected boundary grid cells) of a grid cell. Used in the Floyd--Steinberg dithering scheme for the generation of initial mesh points. &--
    \\ \hline
    Sec.~\ref{step:point init}&$w_i$&The weighting used for the $i^{\text{th}}$ neighboring grid cell in the Floyd--Steinberg dithering scheme. Equal weighting is used and $w_i=m_{\text{nei}}^{-1}$ for $i=1, \ldots, m_{\text{nei}}$.&--
    \\ \hline
     Sec.~\ref{step:point init}&$\geps_i$& The local geometry boundary tolerance defined on the geometry grid cell $i$; its calculation is detailed in Sec.~\ref{geometric adaptive quantities}&--

\\ \hline \hline
    Sec.~\ref{set-up-step-point-addition}&$N_{\text{current}}$&The current number of meshing points &--
    \\ \hline
    Sec.~\ref{set-up-step-point-addition}&$\textit{fac}_{\text{init}}$&Factor to calculate the initial number of points in the mesh, $N_{\text{init}}=\textit{fac}_{\text{init}}\cdot N_{\text{total}}$. Default values are listed, for meshing on complicated shapes or complicated density fields, and in parenthesis, for meshing on simple shapes with gradually varying density field.&0.2 (1.0)
    \\ \hline
    Sec.~\ref{set-up-step-point-addition}&$\alpha$&The aspect ratio of a triangle &--
    \\ \hline
    Sec.~\ref{set-up-step-point-addition}&$\beta$&The edge ratio of a triangle &--
    \\ \hline
    Sec.~\ref{set-up-step-point-addition}&$R_{\text{circum}}$&The circumradius of a triangle &--
    \\ \hline
    Sec.~\ref{set-up-step-point-addition}&$R_{\text{in}}$&The inradius of a triangle &--
    \\ \hline
    Sec.~\ref{set-up-step-point-addition}&$l_{\max}$& The longest edge length of a triangle &--
     \\ \hline
    Sec.~\ref{set-up-step-point-addition}&$l_{\min}$&The shortest edge length of a triangle &--
     \\ \hline
    Sec.~\ref{set-up-step-point-addition}&$M_{\frac{1}{2}}(x_1,\ldots,x_n)$&The generalized $\frac{1}{2}$-mean, $M_{\frac{1}{2}}(x_1,\ldots,x_n)=\left( \frac{1}{n}\sum_{i=1}^n x_i^{1/2} \right)^2$ &--
    \\ \hline
    Sec.~\ref{set-up-step-point-addition}&$T^{\text{add}}_{\text{quality}}$&When $N_{\text{current}}<N_{\text{total}}$, add new points when relative changes in the $\frac{1}{2}$-mean of the triangle aspect and edge ratios, $\alpha$ and $\beta$, are smaller than this threshold&0.002
    \\ \hline
    Sec.~\ref{set-up-step-point-addition}&$N_{\text{add}}$&Number of new meshing points to be added&--
    \\ \hline
    Sec.~\ref{set-up-step-point-addition}&$\textit{fac}_{\text{add}}$&Factor used to calculate the number of new points to add to the mesh, $N_{\text{add}}=\min(\textit{fac}_{\text{add}} N_{\text{current}}, N_{\text{total}}-N_{\text{current}})$ &0.6

\\ \hline \hline
    Sec.~\ref{geometric adaptive quantities}&$h_i$&A local desired edge length in each inner and boundary grid cell of the geometry grid &--
    \\ \hline
    Sec.~\ref{geometric adaptive quantities}&$T^{\text{retria}}_{\text{mvmt},i}$&The threshold to detect large point movements for retriangulation in the DistMesh algorithm &--
    \\ \hline
    Sec.~\ref{geometric adaptive quantities}&$\textit{fac}^{\text{retria}}_{\text{mvmt}}$&Factor used to calculate the threshold of large point movements for retriangulation, $T_{\text{mvmt},i}^{\text{retria}}=\textit{fac}^{\text{retria}}_{\text{mvmt}}\cdot h_i$, where $h_i$ is the local desired edge length of grid $i$ &0.1
    \\ \hline
    Sec.~\ref{geometric adaptive quantities}&$T^{\text{end}}_{\text{mvmt},i}$&A threshold to determine small movements of inner points for termination&--
    \\ \hline
    Sec.~\ref{geometric adaptive quantities}&$\textit{fac}^{\text{end}}_{\text{mvmt}}$&Factor used to calculate the threshold of small inner point movements for termination, $T_{\text{mvmt},i}^{\text{end}}=\textit{fac}^{\text{end}}_{\text{mvmt}}\cdot h_i$&0.001
    \\ \hline
    Sec.~\ref{geometric adaptive quantities}&$T^{\text{pt}}_{\text{mvmt},i}$&A threshold to bound point movements&--
    \\ \hline
    Sec.~\ref{geometric adaptive quantities}&$\textit{fac}^{\text{pt}}_{\text{mvmt}}$&Factor to calculate the threshold that bounds point movement distance in successive meshing iterations, $T^{\text{pt}}_{\text{mvmt},i}=\textit{fac}^{\text{pt}}_{\text{mvmt}}\cdot h_i$&0.4
    \\ \hline
    Sec.~\ref{geometric adaptive quantities}&$\geps_i$&The local geometry boundary tolerance of a boundary grid cell $i$ of the geometry grid &--
    \\ \hline
    Sec.~\ref{geometric adaptive quantities}&$\textit{fac}_{\text{geps}}$&Factor to calculate the geometry boundary tolerance, $\geps{}_i=\textit{fac}_{\text{geps}}\cdot \min(h_i,h_{\text{avg}})$, where $h_{\text{avg}}$ is the average $h_i$ of inner and selected boundary grids &0.01
    \\ \hline
    Sec.~\ref{geometric adaptive quantities}&$h_{\text{avg}}$&The average $h_i$ of inner and selected boundary grids&--
    \\ \hline
    Sec.~\ref{geometric adaptive quantities}&$\deps_i$&The step size used in finite difference when approximating derivatives in the point projection step, which projects an outside point back onto the geometry boundary&--
    \\ \hline
    Sec.~\ref{geometric adaptive quantities}&$\epsilon$&Machine precision&--
    \\ \hline
    Sec.~\ref{geometric adaptive quantities}&$N_{\text{old}}$&The previous number of meshing points before the addition of new points &--
    \\ \hline
    Sec.~\ref{geometric adaptive quantities}&$\textit{adf}_i$&The adaptive signed distance field, $\textit{adf}_i$, for a boundary grid cell, $i$, constructed using quadtree&--
    \\ \hline
    Sec.~\ref{geometric adaptive quantities}&$E_{\text{tol},i}^{\text{adf}}$&The error tolerance of an ADF cell at grid $i$&--
    \\ \hline
    Sec.~\ref{geometric adaptive quantities}&$T_{\text{depth}}^{\text{adf}}$&Maximum depth allowed for quadtrees in the construction of AFD on boundary cells&10
    \\ \hline
    Sec.~\ref{geometric adaptive quantities}&$\textit{fac}_{\text{pt}}^{\text{adf}}$&Factor to calculate the error tolerance of an ADF cell at grid $i$; $\textit{fac}_{\text{pt}}^{\text{adf}} = \sqrt{N_{\text{current}}/N_{\text{total}}}$ &--
    \\ \hline
    Sec.~\ref{geometric adaptive quantities}&$\textit{fac}_{E_{\text{tol}}}^{\text{adf}}$&Factor to calculate the error tolerance of an ADF cell at grid $i$, $E^{\text{adf}}_{\text{tol},i}=\textit{fac}_{E_{\text{tol}}}^{\text{adf}}\cdot \geps{}_i \cdot\textit{fac}_{\text{pt}}^{\text{adf}}$&0.1
\\ \hline \hline

    Sec.~\ref{sec:main meshing part}&$T^{\text{end}}_{\text{quality}}$&One of the termination thresholds: when relative changes in the $\frac{1}{2}$-mean of $\alpha$ and $\beta$ are smaller than the threshold&0.001
    \\ \hline
    Sec.~\ref{sec:main meshing part}&$T^{\text{end}}_{\alpha_{\text{max}}}$&One of the termination thresholds: when relative change in the maximum $\alpha$ is smaller than the threshold&0.005

\\ \hline \hline
    Sec.~\ref{sec: Procedure I: Voronoi tessellation and Delaunay triangulation on current points}&$S$&The span of an octagon used to initialize the Voronoi cells &--
    \\ \hline
    Sec.~\ref{sec: Procedure I: Voronoi tessellation and Delaunay triangulation on current points}&$\textit{fac}_{\text{voro}}^{\text{bound}}$&Factor to calculate the spans of the octagons used to initialize the Voronoi cells, $S=\textit{fac}_{\text{voro}}^{\text{bound}}\cdot h_i$&5
    \\ \hline
    Sec.~\ref{sec: Procedure I: Voronoi tessellation and Delaunay triangulation on current points}&$\vec{c}$&The centroid of a triangle &--
    \\ \hline
    Sec.~\ref{sec: Procedure I: Voronoi tessellation and Delaunay triangulation on current points}&$\vec{m}$&The midpoint of a triangle edge&--
    \\ \hline
    Sec.~\ref{sec: Procedure I: Voronoi tessellation and Delaunay triangulation on current points}&$\vec{c}_{\text{circum}}$&The circumcenter of a triangle&--
    \\ \hline
    Sec.~\ref{sec: Procedure I: Voronoi tessellation and Delaunay triangulation on current points}&$T^{\text{tria}}_{\vec{c}_{\text{circum}}}$&Threshold used in circumcenter test for valid triangles in the mesh&0.4

\\ \hline \hline
    Sec.~\ref{sec: Algorithm to compute new point positions}&$V$&A Voronoi cell&--
    \\ \hline
    Sec.~\ref{sec: Algorithm to compute new point positions}&$A_V$&The area of Voronoi cell $V$&--
    \\ \hline
    Sec.~\ref{sec: Algorithm to compute new point positions}&$l_V$&The local characteristic length-scale of a Voronoi cell $V$; approximated by $l_V=\sqrt{A_V}$&--
    \\ \hline
    
    Sec.~\ref{sec: Algorithm to compute new point positions}&$d_{\text{prev}}^{j}$&The previous point movement distance&--
    \\ \hline
    Sec.~\ref{sec: Algorithm to compute new point positions}&$\epsilon_{\vec{c}}$&A small dimensionless number; $\epsilon_{\vec{c}}=\max(d_{\text{prev}}^{j}/h_i, \textit{fac}^{\text{end}}_{\text{mvmt}})$ &--
    \\ \hline
    Sec.~\ref{sec: Algorithm to compute new point positions}&$E_{\text{thres},V}^{\vec{c}}$&The centroid approximation error threshold for a Voronoi cell $V$ of a point $j$; $E_{\text{thres},V}^{\vec{c}}=l_V \cdot \epsilon_{\vec{c}}$ &--
    \\ \hline
    Sec.~\ref{sec: Algorithm to compute new point positions}&$k$&The level of triangle subdivision of the Voronoi cell in calculating the cell centroid using numerical quadrature &--
    \\ \hline
    Sec.~\ref{sec: Algorithm to compute new point positions}&$T^{\text{switch}}_{\text{quality}}$&Hybrid method: switch from DistMesh to CVD meshing when the relative change in the $\frac{1}{2}$-mean of $\alpha$ is smaller than the threshold&0.0015
    
\\ \hline \hline
    Sec.~\ref{sec: new points treatment}&$\vec{p}_{\text{new}}$&The initial set of updated point positions in a meshing iteration &--
    \\ \hline
    Sec.~\ref{sec: new points treatment}&$\vec{p}_{\text{final}}$&The final set of updated point positions in a meshing iteration &--
    \\ \hline
    Sec.~\ref{sec: new points treatment}&$\vec{p}^*$&If a point $\vec{p}_{\text{new}}$ is outside of the geometry, it is replaced by $\vec{p}^*$, a point on the geometry boundary. It is found by sphere tracing or the point projection method. &--
    \\ \hline
    Sec.~\ref{sec: new points treatment}&$\vec{p}_{\text{old}}$&The point position before the current iteration &--
    
\\ \hline \hline
    Sec.~\ref{sec: new points treatment}:~\ref{sphere tracing method}&$R$&The radius of the circle used in the sphere tracing method &--

\\ \hline \hline
    Sec.~\ref{sec: new points treatment}:~\ref{point projection method}&$\vec{L}(\vec{p}^*)$&Equation to be solved in order to find $\vec{p}^*$, the closest point to $\vec{p}_{\text{new}}$ on the geometry boundary &--
    \\ \hline
    Sec.~\ref{sec: new points treatment}:~\ref{point projection method}&$\vec{J}(\vec{p})$&The Jacobian of $\vec{L}$ &--
    \\ \hline
    Sec.~\ref{sec: new points treatment}:~\ref{point projection method}&$f$& Represents the SDF in the calculation of $\vec{J}(\vec{p})$ &--
    \\ \hline
    Sec.~\ref{sec: new points treatment}:~\ref{point projection method}&$\hat{\alpha}$&Damping parameter for the damped Newton method used to project outside points onto geometry boundary&1
    \\ \hline
    Sec.~\ref{sec: new points treatment}:~\ref{point projection method}&$T^{\text{newton}}_{\text{ct}}$&Maximum Newton steps allowed&10
    
\\ \hline \hline
    Sec.~\ref{sec:performance}&$p$&The number of parallel threads&--
    \\ \hline
    Sec.~\ref{sec:performance}&$t_p$&The wall clock time of the computation using $p$ threads&--
    \\ \hline
    Sec.~\ref{sec:performance}&$T_e(p)$&The parallel efficiency using $p$ threads &--
    
\\ \hline \hline
    Sec.~\ref{sec:comparison with existing software}&$B$&A bounding constraint used in the mesh refinement scheme by Shewchuk; It is related to the minimum angle of the triangles.&$\sqrt{2}$
    \\ \hline
    Sec.~\ref{sec:comparison with existing software}&$\theta_{\text{min}}$&A minimum angle constraint used in the mesh refinement scheme by Shewchuk; $\sin{\theta_{\text{min}}}=\frac{1}{2B}$. Setting $B=\sqrt{2}$ corresponds to bounding $\theta_{\text{min}}\geq \ang{20.7}$.&--
    \\ \hline
    Sec.~\ref{sec:comparison with existing software}&$L$&The maximum edge length constraint used in the  mesh refinement scheme by Shewchuk&0.04
    \\ \hline
    Sec.~\ref{sec:comparison with existing software}&$l_{\text{max}}$& The maximum triangle edge length&--

\\ \hline \hline
    Appendix~\ref{appendix:bdry pt generation}&$\vec{s}_i$& A boundary approximation point &--
    \\ \hline
    Appendix~\ref{appendix:bdry pt generation}&$\mathcal{S}$& The set of boundary approximation points, $\mathcal{S}=\{\vec{s}_1,\vec{s}_2,\ldots\}$ &--
    \\ \hline
    Appendix~\ref{appendix:bdry pt generation}&$\textit{fac}_{\vec{s}}$&Factor used to calculate $d_{\vec{s}}$, a distance according to which the boundary approximating points should be roughly evenly spaced&0.5
    \\ \hline
    Appendix~\ref{appendix:bdry pt generation}& $d_{\vec{s}}$ & The minimum distance in between adjacent boundary points on the same boundary, $d_{\vec{s}}=\textit{fac}_{\vec{s}}\cdot\min(\Delta x^{\text{geo}},\Delta y^{\text{geo}})$. The boundary approximating points should be roughly evenly spaced according to this distance. &--
    \\ \hline
    Appendix~\ref{appendix:bdry pt generation}&$n_{\text{grid}}$&The initial number of points generated in each boundary grid cell for the generation of boundary approximating points&5

    \\ \hline
    Appendix~\ref{appendix:bdry pt generation}&$\hat{\geps}$&A geometry boundary tolerance to determine whether each generated point in the boundary grid cells is on the geometry boundary, $\hat{\geps}=\textit{fac}_{\geps}\cdot\min(\Delta x^{\text{geo}},\Delta y^{\text{geo}})$ &--
     \\ \hline
    Appendix~\ref{appendix:bdry pt generation}&$\hat{\deps}$&A fixed finite difference step size for approximating derivatives, $\hat{\deps}=\sqrt{\epsilon}\cdot\min(\Delta x^{\text{geo}},\Delta y^{\text{geo}})$&--
     \\ \hline
    Appendix~\ref{appendix:bdry pt generation}&$\vp_i$&A boundary point&--
     \\ \hline
    Appendix~\ref{appendix:bdry pt generation}&$\vp^*$&The closest boundary point to $\vp_i$ &--
    \\ \hline
    Appendix~\ref{appendix:bdry pt generation}&$\vec{m}$& The midpoint of $\vp_i$ and $\vp^*$ &--
\\ \hline
\caption{Symbols used in the paper, as well as control parameters used in the meshing pipeline. The recommended default values for the control parameters are listed in the last column.\label{tbl:meshing_parameters}}
\end{longtable}

\end{appendices}

\bibliography{voro}

\end{document}